# HyperNetWalk: A Unified Framework for Personalized and Cohort-Level Cancer Driver Gene Identification via Reverse Inference on Layered Signaling–Regulatory Network

**Xueqing Xu[1,2], Yonghang Gao[1,2], Duanchen Sun[3*], and Ling-Yun Wu[1,2*]**

[1]State Key Laboratory of Mathematical Sciences, Academy of Mathematics and Systems Science, Chinese Academy of Sciences, Beijing 100190, China

[2]School of Mathematical Sciences, University of Chinese Academy of Sciences, Beijing 100049, China

[3]School of Mathematics, Shandong University, Jinan 250100, China

***Corresponding authors**

dcsun@sdu.edu.cn (D. Sun), and lywu@amss.ac.cn (L.-Y. Wu)

**Keywords: cancer driver gene identification; unified personalized and cohort-level prediction; layered signaling–regulatory network; reverse inference; hypergraph random walk**

## Abstract

Accurate identification of cancer driver genes from the large background of passenger mutations is essential for understanding tumorigenesis and facilitating clinical translation. Existing computational methods usually focus on either personalized or cohort-level driver prediction, and the few approaches that consider both levels often lack a shared inference foundation connecting them. In addition, many methods do not explicitly model the directional information flow from upstream genomic alterations through signaling cascades to downstream transcriptional dysregulation. Here, we present HyperNetWalk, an unsupervised framework that unifies personalized and cohort-level cancer driver gene identification within a single inference architecture. HyperNetWalk constructs a layered signaling–regulatory network that integrates protein–protein interactions, used to approximate upstream signaling, with a gene regulatory network representing downstream transcriptional regulation; transcription factors serve as interface nodes between the two components. Driver gene identification is then formulated as an inverse problem: observed transcriptional changes are traced back to candidate upstream driver genes through reverse random walk on this layered structure. The resulting sample-specific scores provide a shared foundation for both prediction levels: they serve as node weights for cross-sample refinement via hypergraph random walk, using local hypergraphs for personalized prediction and a global hypergraph for cohort-level prediction. Evaluation across 12 TCGA cancer types demonstrated that HyperNetWalk outperformed representative existing methods at both prediction levels. Ablation analyses supported the importance of both the reverse inference formulation and the layered network architecture. Further analyses showed that HyperNetWalk captured cancer-type-specific driver signals, prioritized both recurrent and low-frequency candidate drivers, and yielded predictions supported by drug–gene interaction and clinical actionability annotations.
Code and data are available at: **https://github.com/xqxu921/HyperNetWalk.**

## Introduction

Cancer arises from the accumulation of somatic mutations in the genome. Only a small subset of these alterations, termed driver mutations, confer a selective growth advantage on tumor cells and contribute causally to tumorigenesis, whereas the vast majority are functionally neutral passenger mutations.[1] Genes that harbor driver mutations are referred to as cancer driver genes. Identifying such genes against an overwhelming mutational background remains a central challenge in cancer genomics, with direct implications for understanding tumor biology and guiding therapeutic strategies. This challenge is further complicated by interpatient heterogeneity: even within the same cancer type, tumors from different patients may be driven by distinct genomic events.

A wide range of computational approaches have been developed to address this task. Frequency-based methods, such as MutSigCV[2], identify driver genes by detecting deviations from expected background mutation rates but may miss low-frequency drivers that fail to reach statistical significance despite their functional importance. Functional impact- or consequence-aware approaches, including DriverML[3], OncodriveFML[4] and MutPanning[5], prioritize candidate drivers by incorporating the predicted functional consequences or coding effects of somatic mutations and can improve the detection of rare drivers, although their performance depends on the availability and quality of functional annotations. Network-based methods exploit the interconnected nature of cellular systems. For example, DriverNet[6] links mutated genes to downstream transcriptional changes through network connectivity, HotNet2[7] uses diffusion processes to identify significantly perturbed subnetworks, and DriverMP[8] integrates protein–protein interactions with differential co-expression at the gene-pair level. More recent graph neural network models learn driver-related patterns from biological networks and multi-omics features in supervised or semi-supervised settings.[9–12] Although such models can learn expressive representations from biological networks, their reliance on known driver labels makes them susceptible to annotation incompleteness and limits their ability to discover previously uncharacterized drivers. Overall, most of the methods described above operate primarily at the cohort level, producing a single gene ranking from data aggregated across patients.

To capture sample-specific heterogeneity, personalized network-based methods have also been developed. DawnRank[13] and SCS[14] use patient-specific mutation and expression profiles to prioritize candidate driver genes according to their influence on differentially expressed genes (DEGs). More recent approaches, such as PersonaDrive[15] and PDRWH[16], further exploit cross-sample information through graph- or hypergraph-based modeling. These methods can identify sample-specific drivers, but how personalized predictions should be integrated into coherent cohort-level rankings remains largely unaddressed. Cohort-level methods often obscure sample-specific context, whereas personalized methods are primarily optimized for individual-level prediction and do not explicitly enforce cross-sample consistency. As a result, when cohort-level rankings are required, methods such as DawnRank and PDRWH typically rely on post-hoc rank aggregation, which retains only ordinal information from each sample and therefore provides no guarantee of competitive cohort-level performance. A framework that naturally produces both personalized and cohort-level predictions from a common sample-specific inference process is still lacking.

A separate limitation concerns how biological networks are represented. In many existing methods, signaling interactions and transcriptional regulatory relationships are either used in isolation or collapsed into a single, weakly differentiated graph. Such treatment can obscure the directional information flow from upstream signaling perturbations to downstream transcriptional regulation and blur the functional distinctions among interaction types. Work in related areas suggests that preserving the organizational structure of heterogeneous biological interactions can be more informative than representing them as a homogeneous network. For example, TieDIE[17] identifies intermediate causal pathways that connect genomic perturbations to transcriptional states, while NicheNet[18] explicitly links ligand–receptor interactions to downstream signaling and gene regulatory effects. Although neither method is designed for driver gene identification, they illustrate the value of preserving the structured flow of biological information.

From a mechanistic standpoint, driver mutations can exert their effects through a directional cascade (Fig. 1A). Genomic alterations may perturb upstream signaling processes, leading to dysregulated transcription factor (TF) activity; dysregulated TFs can in turn induce coordinated changes in downstream gene expression, which are observed as DEGs in patient samples. This view has two implications for method design. First, the underlying network should preserve the functional separation between signaling interactions, represented by signaling or protein–protein interaction (PPI) networks, and transcriptional regulation, described by gene regulatory networks (GRNs), with TFs serving as interface nodes between the two layers. Second, driver gene identification can be formulated as an inverse problem: rather than propagating perturbation signals only forward from mutated genes, one can trace observed transcriptional dysregulation upstream to candidate causal drivers.

Here, we present HyperNetWalk, an unsupervised framework for personalized and cohort-level cancer driver gene identification. Building on the reverse-inference principle described above, HyperNetWalk propagates patient-specific transcriptional dysregulation backward through a layered signaling–regulatory network to score candidate mutated genes. These sample-specific scores are then refined in local and global hypergraphs, yielding personalized and cohort-level rankings within the same inference framework. Across 12 TCGA cancer types, HyperNetWalk outperforms representative existing methods at both levels; ablation and downstream analyses support the contributions of the reverse-inference strategy, the layered network architecture, cancer-type-specific signal capture, low-frequency driver prioritization, and therapeutic relevance.

## Results

### Overview of the HyperNetWalk framework

HyperNetWalk is an unsupervised two-stage framework that first infers sample-specific driver scores on a layered signaling–regulatory network and then refines these scores through hypergraph random walk for personalized and cohort-level prediction (Fig. 1).

We first construct a global layered network to represent the mutation-to-expression cascade (Fig. 1B). The network contains a PPI component for upstream signaling and a GRN component for downstream transcriptional regulation. It is organized into three layers: Layer 1 is a PPI layer containing signaling proteins and protein-level representations of TFs; Layer 2 contains TF nodes connected by TF–TF regulatory edges; and Layer 3 contains target genes as transcriptional outputs. Layers 2 and 3 form the regulatory component, and TF nodes in Layer 2 are linked to their corresponding protein nodes in Layer 1 through identity edges.

**Stage I: Sample-specific reverse inference.** For each patient, a sample-specific subnetwork is extracted by connecting mutated genes to DEGs through the layered network (Fig. 1C). A random walk with restart (RWR) is initialized from DEG nodes and run on reversed edges; the stationary probabilities of mutated gene nodes define sample-specific driver scores.

**Stage II: Cross-sample refinement.** Stage II refines these scores by hypergraph random walk (Fig. 1D). Personalized prediction uses a local hypergraph centered on the target sample and neighboring samples defined by shared mutated genes, whereas cohort-level prediction uses a global hypergraph spanning all patients. The outputs are personalized and cohort-level driver rankings, respectively.

**HyperNetWalk achieves superior performance in both personalized and cohort-level driver gene identification**

We benchmarked HyperNetWalk against five personalized methods (DawnRank[13], PDRWH[16], PersonaDrive[15], PITCH[19], and PRODIGY[20]) and five cohort-level methods (dNdScv[21], DriverMP[8], DriverNet[6], DriverRWH[22], and Subdyquency[23]), together with a mutation-frequency ranker and a random ranker. All methods were evaluated across 12 TCGA cancer types using Cancer Gene Census (CGC) Tier 1 driver genes[24] as the reference set. Because known driver genes account for only a small fraction of somatically mutated genes, we focused on top-ranked predictions.

We first compared Precision@$K$ across all cancer types (Fig. 2A). At the personalized level, we determined a cancer-type-specific cutoff $N_c$ using the ranking-evaluation-aggregation (REA) strategy[15] and set $K = \min_c N_c = 6$ for cross-cancer comparison. HyperNetWalk achieved the highest median Precision@$K$ across the 12 cancer types (0.575), outperforming the best baseline (0.454). DawnRank, which also propagates information from downstream expression changes back to candidate mutated genes, was the closest competitor, consistent with the benefit of reverse-propagation strategies.

At the cohort level, we evaluated Precision@100 (Fig. 2A) and additional cutoffs ($K = 50$ and $K = 500$; Supplementary Fig. S1). HyperNetWalk achieved the highest Precision@100 (0.57), outperforming DriverMP (0.54) and DriverRWH (0.415). The comparison with DriverRWH, another hypergraph random-walk method, suggests that Stage I sample-specific scores provide information beyond hypergraph topology alone. Although DriverMP was competitive at $K = 50$, HyperNetWalk achieved higher precision at larger cutoffs, indicating more stable ranking performance beyond the very top candidates.

We further assessed cohort-level performance using AUC-based metrics (Fig. 2B; Supplementary Figs. S2–S4). Given the strong class imbalance between driver and passenger genes, partial AUPRC (pAUPRC) over the low-recall region (recall $\leq 0.1$) is particularly informative for evaluating high-confidence predictions. HyperNetWalk achieved the highest pAUPRC in 10 of 12 cancer types, further supporting its advantage in prioritizing top-ranked driver candidates.

To examine cancer-type-specific behavior, we selected three representative cancer types: BRCA, a well-studied cancer type with moderate mutation rate; LUAD, characterized by high mutational noise and heterogeneity; and THCA, which has low mutational burden. At the personalized level, Precision@$K$ curves showed that HyperNetWalk consistently outperformed all competing methods across all cutoffs in BRCA and LUAD, with relative improvements of 18% and 26%, respectively, at $K = N_c$ (Fig. 2C). In THCA, HyperNetWalk remained the best-performing method overall, although DawnRank showed slightly higher precision at two isolated small cutoffs. The mutation-frequency baseline performed poorly in LUAD, consistent with the high mutational noise of this cancer type. Similar trends were observed across all 12 cancer types and across Recall@$K$ and F1-score@$K$ evaluations (Supplementary Figs. S5–S7).

At the cohort level, we evaluated the top 500 ranked genes for each cancer type (Fig. 2D). HyperNetWalk achieved the best overall performance across BRCA, LUAD, and THCA, with DriverMP as the closest competitor. In the small-cutoff range ($K \leq 50$), HyperNetWalk maintained higher precision than DriverMP, which is particularly relevant for downstream experimental validation. Similar patterns were observed across all 12 cancer types (Supplementary Figs. S8–S10).

A central feature of HyperNetWalk is that personalized and cohort-level predictions are derived from a common inference foundation. We therefore asked whether its cohort-level advantage could be reproduced by aggregating the personalized outputs of existing methods. Even under the most favorable aggregation strategy identified for each competitor (see Methods), HyperNetWalk consistently outperformed all aggregated personalized baselines (Supplementary Figs. S11 and S12). This result indicates that HyperNetWalk's cohort-level performance cannot be explained by post-hoc aggregation alone, but depends on the quality of its Stage I sample-specific scores and their integration during hypergraph refinement. Aggregated personalized methods also remained behind dedicated cohort-level methods such as DriverMP, whereas these cohort-level methods do not provide sample-specific predictions. These results highlight the difficulty of achieving strong performance at both prediction levels simultaneously.

Together, these benchmarks show that HyperNetWalk outperforms existing personalized, cohort level, and aggregated baselines across diverse cancer types.

**Reverse inference and layered architecture contribute to HyperNetWalk performance**

To identify the sources of HyperNetWalk's performance, we designed four model variants, each modifying one design choice while keeping the remaining components unchanged (Fig. 3).

We first asked whether the direction of inference matters. The *Forward* variant replaces reverse propagation from DEG nodes with forward propagation from mutated gene nodes, while retaining the layered network and hypergraph refinement. This variant showed the largest performance decrease among all tested variants. At the personalized level, Forward achieved a median Precision@6 of 0.268 across 12 cancer types, compared with 0.575 for HyperNetWalk (Fig. 3A). At the cohort level, median Precision@100 dropped from 0.570 to 0.105. The pAUPRC heatmap further showed that Forward produced the lowest values among all variants in 11 of 12 cancer types, with pronounced drops in KIRP, LUSC, and THCA (Fig. 3B). The only exception was UCEC, where Forward achieved a higher pAUPRC than HyperNetWalk, possibly reflecting a small number of highly recurrent driver genes that provide strong starting signals for forward propagation. These results indicate that the direction of inference is the dominant contributor among the tested design choices.

We next tested whether the layered architecture contributes beyond specifying a particular biological edge set. The *Merged* variant collapses layer-specific instances of the same gene into a single node, thereby removing TF–protein identity edges, whereas *Merged_Plus* uses a related merged representation with an expanded edge set induced directly from the PPI network and GRN (see Methods). Both variants performed worse than HyperNetWalk at both prediction levels (Fig. 3A), and Merged_Plus did not consistently improve over Merged. In the pAUPRC evaluation, HyperNetWalk achieved the highest value in 9 of 12 cancer types, whereas neither merged variant approached its performance in most cancer types (Fig. 3B). These results indicate that the advantage of the layered architecture is not explained simply by its selected edge set. Rather, preserving layer-specific gene roles, especially for TFs, and constraining transitions between signaling and regulatory components are important for effective reverse propagation.

The *Undirected* variant treats all directed edges as undirected during random walk, while preserving the layered structure. This variant also reduced performance at both prediction levels, although the decrease was generally smaller than that observed for Forward. Thus, edge directionality provides useful constraints for reverse inference, but the direction of inference itself has the strongest effect.

Precision@*K* curves in BRCA, LUAD, and THCA confirmed these patterns: Forward showed the largest decrease, Merged, Merged_Plus, and Undirected showed moderate reductions, and HyperNetWalk maintained the highest precision across most of the evaluated ranking range at both prediction levels (Fig. 3C and D).

Together, these ablation results identify reverse inference as the main driver of HyperNetWalk's performance and show that the layered, directed network architecture provides an additional performance advantage.

**HyperNetWalk captures cancer-type-specific driver and regulatory landscapes**

The benchmarking analyses above used CGC Tier 1 driver genes as a pan-cancer reference set. Because tumorigenesis often reflects tissue- and context-specific mechanisms, we further assessed whether HyperNetWalk recovers cancer-type-specific driver signals. We used IntOGen cancer-type-specific driver sets as an external reference, while recognizing that such reference sets may reflect prior discovery biases.

At the cohort level, we compared the top 100 predicted drivers for each cancer type with IntOGen driver sets from all 12 cancer types. To reduce the influence of broadly shared drivers, we excluded genes annotated in more than three cancer types. After filtering, cancer-type-specific predictions are expected to show the strongest overlap with their matched reference sets, corresponding to the diagonal entries in Fig. 4A. HyperNetWalk approximated this pattern, with the matched cancer type showing the highest overlap in 9 of 12 cancers. COAD- and KIRP-specific reference sets showed little overlap with predictions from other cancer types, suggesting relatively tissue-restricted driver signals that were captured by HyperNetWalk. By contrast, DriverMP, DriverRWH, and DawnRank showed more diffuse off-diagonal overlap patterns (Supplementary Fig. S13). Across the 12 cancer types, HyperNetWalk achieved the highest median overlap with matched cancer-type-specific driver sets (Fig. 4B), indicating that its performance is not driven solely by pan-cancer drivers.

Cancer-type specificity was also preserved at the personalized level. For each cancer type, we measured the proportion of samples with at least one predicted driver matching the corresponding IntOGen reference set among the top $N_c$ predictions. HyperNetWalk outperformed DawnRank in 11 of 12 cancer types and matched it in the remaining cancer type, with improvements in BRCA (0.549 vs. 0.496) and LUAD (0.678 vs. 0.592) (Supplementary Fig. S14A). A similar trend was observed for the mean number of matched hits per sample (Supplementary Fig. S14B). Reference-free cross-cancer overlap analyses further showed that HyperNetWalk predictions remained largely distinct across cancer types (Supplementary Note S3 and Supplementary Fig. S15).

Hallmark enrichment analysis of the top 100 cohort-level predicted drivers revealed both shared and cancer-type-specific pathway patterns (Fig. 4C). Broadly enriched pathways included WNT/$\beta$-catenin signaling, E2F targets, and G2M checkpoint. At the same time, several enrichments reflected known biological differences among cancer types. Predicted drivers in COAD, STAD, and UCEC were enriched for TGF-$\beta$ signaling, consistent with the roles of this pathway in tumor–stroma interactions, microenvironment remodeling, and immune regulation[25,26]. HNSC, KIRC, and LIHC were enriched for PI3K/AKT/mTOR signaling, a pathway linked to proliferative and metabolic reprogramming[27,28]. UCEC also showed enrichment of estrogen response programs, consistent with the established role of hormone signaling in endometrial tumorigenesis[29].

We next examined whether the TF layer, which links signaling perturbations to transcriptional regulation in the layered network, captures cancer-type-associated regulatory variation. Because Stage I random walks traverse this layer during reverse inference, the stationary probabilities assigned to TF nodes provide a sample-level representation of the inferred regulatory state. UMAP projection of these TF-layer probability vectors separated samples by cancer type, as quantified by adjusted Rand index (ARI) and normalized mutual information (NMI) (Fig. 4D). Although raw gene expression produced slightly stronger separation, this pattern may partly reflect broad transcriptomic differences or batch effects across cancer types (Supplementary Fig. S16). In contrast, TF-layer probability vectors provide a network-informed representation that integrates each sample's mutational context with the layered network topology.

To assess the biological content of these TF-layer representations, we identified cancer-type-specific TFs based on their Stage I probabilities and performed Hallmark enrichment analysis of their downstream target-gene programs (Supplementary Notes S4 and S5). In HNSC, the target-gene program associated with cancer-type-specific TFs was enriched for inflammatory response and TNF-$\alpha$ signaling via NF-$\kappa$B (Fig. 4E), consistent with inflammatory and stress-response signaling in head and neck squamous cell carcinoma[30]. In LIHC, the corresponding program was enriched for bile acid metabolism and oncogenic signaling pathways (Fig. 4F), consistent with the liver-associated metabolic context of hepatocellular carcinoma[31]. These examples show that the TF layer captures biologically coherent, cancer-type-specific regulatory programs.

**HyperNetWalk identifies supported candidate drivers missed by competing methods**

We next examined whether HyperNetWalk's predictions were dominated by recurrently mutated genes or also captured low-frequency candidate drivers. Mutated genes were classified into four groups according to their within-cancer-type mutation frequency, using 2% as the threshold, and CGC Tier 1 driver annotation: known high-frequency drivers, known low-frequency drivers, other high-frequency mutated genes, and other low-frequency mutated genes.

At the cohort level, we compared the composition of the top 100 ranked genes predicted by HyperNetWalk and baseline methods in BRCA and LUAD, including a mutation-frequency baseline for reference (Fig. 5A). HyperNetWalk recovered a larger fraction of known low-frequency drivers than competing methods while retaining high precision, whereas most other methods showed a stronger bias toward high-frequency mutated genes. Similar trends were observed across other cancer types (Supplementary Fig. S17).

At the personalized level, we computed the proportions of different gene categories at each ranking position across all samples in BRCA and LUAD (Fig. 5B). Top-ranked positions were enriched for known drivers, as expected and as also observed for DawnRank (Supplementary Fig. S18). However, HyperNetWalk maintained a larger proportion of known low-frequency drivers than DawnRank throughout the evaluated ranking range, indicating that its personalized predictions are not driven solely by recurrent mutation frequency.

We next assessed whether HyperNetWalk produced predictions distinct from existing methods. At the cohort level, comparison of the top 100 ranked genes from HyperNetWalk, DawnRank, DriverMP, and DriverRWH in BRCA and LUAD showed that HyperNetWalk uniquely prioritized 33 genes in BRCA and 37 genes in LUAD that were absent from all three baseline lists (Fig. 5C; Supplementary Fig. S19). At the personalized level, comparison of the union of the top 5 predictions across all samples for each method showed that a substantial fraction of HyperNetWalk predictions was distinct from those of DawnRank and PDRWH (Fig. 5D; Supplementary Fig. S20). Thus, HyperNetWalk nominates additional candidate drivers at both prediction levels rather than merely re-ranking genes already prioritized by other methods.

To evaluate the external support for these HyperNetWalk-specific genes, we classified them into four evidence levels (A–D) according to their support in cancer knowledge resources, including CGC, NCG[32], and CancerMine[33] (Supplementary Note S6). Across both cohort-level and personalized predictions, a substantial fraction of HyperNetWalk-specific genes were assigned to evidence levels B and C (Supplementary Tables S2 and S3), suggesting prior cancer relevance despite their lower rankings by competing methods.

We further examined the local network contexts of representative HyperNetWalk-specific candidates (Supplementary Tables S4 and S5). Among cohort-level predictions uniquely prioritized by HyperNetWalk, two recurrent patterns emerged. First, some candidates are TFs with large regulatory out-degrees in the GRN, even when they lack downstream reachable TF nodes through Layer 1 signaling paths, such as ZNF263 and EBF1 in BRCA and TAF3 in LUAD. Second, some candidates are embedded in compact Layer 1 neighborhoods that connect them to downstream TFs after directed path-based filtering.

TBX3 in BRCA illustrates the second pattern (Fig. 5E). TBX3 is a level A candidate ranked 93rd by HyperNetWalk, but much lower by DawnRank, DriverMP, and DriverRWH. In the induced Layer 1 subnetwork, TBX3 was embedded in a dense signaling neighborhood containing established BRCA-related genes, including CDH1, CCND1, and PIK3CA[34]. All retained directed routes from TBX3 to downstream TFs passed through CDH1, suggesting that CDH1 may relay TBX3-associated signals to the downstream regulatory module. This coherent BRCA-relevant network context may explain why HyperNetWalk prioritized TBX3 despite its lower rankings by other methods.

PHF8 provides a complementary example (Fig. 5E). Although PHF8 is generally regarded as a transcriptional co-factor, it is represented as a TF in the RegNetwork-derived GRN and has substantial regulatory out-degree. Its prioritization, however, does not appear to be driven solely by this property. In the Layer 1 subnetwork, PHF8 connected to downstream TFs through TOPBP1 and was positioned near TP53, BRIP1, E2F1, and CDKN2A, indicating proximity to DNA-damage response and cell-cycle control modules[35]. HyperNetWalk ranked PHF8 82nd, whereas DawnRank ranked it 669th.

For LUAD, GABPA exemplifies a candidate that combines both patterns (Fig. 5F). GABPA is a TF with substantial regulatory out-degree, and TXN served as the only retained intermediate node connecting GABPA to its downstream TFs in the induced Layer 1 subnetwork. These downstream TFs were themselves highly connected in the GRN. HyperNetWalk ranked GABPA 62nd, whereas DawnRank ranked it 2,016th.

Together, these case studies show that HyperNetWalk can prioritize candidates through complementary mechanisms: direct regulatory potential, reflected by TFs with large downstream target programs, and indirect network influence, reflected by compact signaling neighborhoods connected to downstream regulatory modules.

**HyperNetWalk outputs reveal therapeutic and prognostic relevance**

We next evaluated whether HyperNetWalk-derived outputs have therapeutic relevance and prognostic associations. Using the Drug–Gene Interaction Database (DGIdb)[36] as a reference, we assessed druggability from three perspectives: genes with any annotated drug–gene interaction, genes associated with anti-neoplastic drugs, and genes with interactions involving approved drugs.

At the cohort level, more than 70% of the top 100 predictions had at least one annotated drug–gene interaction across all 12 cancer types, and more than 50% were associated with anti-neoplastic drugs (Fig. 6A), indicating substantial enrichment for therapeutically tractable genes.

At the personalized level, the union of top $N_c$ predictions across all samples was larger and more heterogeneous than the cohort-level top 100 list, which can lower the apparent overlap proportion with DGIdb. Nevertheless, DGIdb-supported genes remained broadly represented among personalized predictions (Supplementary Fig. S21), and most individual samples contained multiple DGIdb-supported genes among their top $N_c$ predictions under all three annotation criteria (Fig. 6B).

We also examined OncoKB annotations[37,38] as a stricter reference for clinical actionability. Restricting the analysis to genes annotated with evidence levels 1–4, HyperNetWalk recovered OncoKB-supported genes across multiple cancer types (Supplementary Fig. S22). Together, the DGIdb and OncoKB analyses indicate that HyperNetWalk predictions overlap with genes of established or emerging therapeutic relevance.

Finally, we examined whether TF-layer representations from Stage I were associated with patient survival. For each cancer type, we performed consensus clustering on sample-level TF-layer probability vectors, with the number of clusters determined by the delta area criterion without using survival information. Kaplan–Meier analysis and log-rank tests showed nominally significant survival differences among the inferred subtypes in six cancer types (Supplementary Fig. S23). Three cancer types with clear survival separation — KIRP, LIHC, and UCEC — are highlighted in Fig. 6C, suggesting that TF-layer-derived regulatory states capture clinically relevant interpatient heterogeneity.

Comparison with available molecular subtype annotations revealed partial concordance (Supplementary Fig. S24). For example, in UCEC, TF-layer-derived subtypes S1 and S4 were more frequently associated with TCGA-defined subtype 1. This partial overlap suggests that TF-layer representations captured known subtype-related signals as well as additional regulatory variation not fully described by existing molecular classifications.

Together, these analyses show that HyperNetWalk-derived outputs are associated with both therapeutic annotations and survival-relevant patient stratification.

## Discussion

In this study, we presented HyperNetWalk, an unsupervised framework that unifies personalized and cohort-level cancer driver gene identification through a shared sample-specific inference process. HyperNetWalk combines reverse random walk on a layered signaling–regulatory network with hypergraph-based cross-sample refinement, producing both patient-specific and cohort-level driver rankings. Across 12 TCGA cancer types, HyperNetWalk outperformed representative existing methods, while additional analyses supported the cancer-type specificity, low-frequency driver prioritization, therapeutic relevance, and survival-associated regulatory information captured by its outputs.

The ablation analysis identified the direction of inference as the most important design choice. This result may reflect an inherent asymmetry in driver gene identification. In a typical cancer sample, the number of DEGs far exceeds the number of true driver mutations. Forward propagation from mutated genes therefore starts from a sparse and noisy initial signal, because most mutated genes are passengers, and the signal may become diluted as it spreads through the network. Reverse propagation instead starts from a dense and phenotypically grounded signal, namely DEGs, and traces this signal upstream toward mutated genes that are connected to dysregulated downstream programs. This formulation naturally concentrates probability mass on candidate causal perturbations. The strong performance of DawnRank, the closest personalized competitor and another reverse-propagation method, is consistent with this interpretation. HyperNetWalk extends this idea by embedding reverse inference in a layered architecture that provides additional structural constraints.

The layered architecture also contributed beyond the biological edge set itself. Collapsing layer-specific gene representations into a single graph reduced performance even when many underlying PPI and regulatory edges were preserved, indicating that network organization carries information not captured by a homogeneous adjacency matrix. A key feature of the layered design is the separation of TFs into signaling-level and regulatory-level representations. Together with directed TF–protein identity edges, this separation constrains transitions between the signaling and regulatory components during reverse propagation and helps focus on the walk toward upstream candidate drivers. More broadly, these findings suggest that network-based biological inference can benefit from preserving the functional roles of heterogeneous interaction types rather than merging them into a single undifferentiated graph. The directed path-based filtering used in Stage I further controls the search space by restricting sample-specific subnetworks to plausible propagation routes, reducing noise from the dense global PPI topology.

A second implication concerns the relationship between personalized and cohort-level prediction. Cohort-level methods gain robustness by aggregating patients but often lose sample-specific context, whereas personalized methods preserve individual mutational profiles but do not necessarily yield coherent cohort-level rankings. Post-hoc aggregation methods connect these two levels only after independent personalized rankings have already been produced. HyperNetWalk takes a different approach: Stage I generates mechanistically informed scores that are comparable across samples, and Stage II refines them through local or global hypergraph structures depending on the prediction level. The post-hoc aggregation experiment showed that applying aggregation or hypergraph refinement to other personalized methods did not reproduce HyperNetWalk's cohort-level performance, suggesting that the informativeness of the sample-specific scores is a major determinant of success. This supports a general design principle for multi-level genomic prediction: when sample-level inference is grounded in a biologically structured model, personalized and cohort-level predictions can be derived from a common inference foundation rather than from separate procedures.

HyperNetWalk is unsupervised in the sense that it does not require known driver-gene labels for training. This distinguishes it from supervised and semi-supervised graph learning approaches, which can achieve strong performance when high-quality labels are available but may inherit biases from incomplete annotations. Genes that have not yet been recognized as drivers may be treated as negatives or non-driver examples during training, potentially limiting discovery of novel or context-dependent drivers. This issue is particularly relevant for personalized driver identification, where the functional relevance of a mutated gene may depend on the patient's molecular context. HyperNetWalk mitigates this limitation by using each sample's observed transcriptional phenotype and the layered network structure to score mutated genes without training on driver labels. At the same time, HyperNetWalk is not free of prior knowledge: it relies on curated PPI networks, GRNs, and TF annotations. Thus, "unsupervised" refers specifically to the absence of driver-gene labels, not to the absence of biological priors.

The ability to prioritize low-frequency and method-specific candidate drivers further illustrates the value of this design. Many cancer-relevant genes are not recurrently mutated enough to be detected by frequency-dominated approaches, yet they may influence downstream regulatory programs in specific samples or cancer types. By integrating mutation profiles with transcriptional dysregulation through a structured network, HyperNetWalk can prioritize candidates supported by either direct regulatory potential or indirect network influence. The case studies in BRCA and LUAD illustrate these complementary mechanisms and suggest that HyperNetWalk-specific predictions include biologically plausible candidates rather than merely algorithm-specific artifacts.

Several limitations should be noted. First, HyperNetWalk depends on prior biological networks. The PPI networks used to approximate signaling are incomplete, may contain false-positive interactions, and are largely static rather than tumor-context-specific. Although sample-specific subnetworks and expression-informed edge weights partially mitigate this issue, the underlying topology may still bias predictions toward well-studied genes and pathways. The GRN introduces a related limitation: the current regulatory network is context-independent and may not fully capture cancer-type-specific or cell-state-specific regulation. Incorporating cancer-type-specific or sample-level regulatory maps, for example from chromatin accessibility, single-cell regulatory inference, or perturbation data, could improve the specificity of the TF layer and downstream driver predictions.

Second, evaluation remains constrained by incomplete reference sets. CGC Tier 1 provides a high-confidence driver reference, but genes absent from CGC may still be bona fide cancer drivers. Therefore, reported precision values should be interpreted as reference-based estimates that may underestimate the true precision of all methods. The evidence-level analysis of HyperNetWalk-specific predictions partially addresses this concern by showing that many additional candidates are supported by other cancer gene resources, but experimental validation will be needed to establish their causal roles.

Third, the current framework simplifies several aspects of biological regulation. Molecular relationships are represented primarily as pairwise edges, which do not capture combinatorial TF regulation, protein complex formation, or pathway-level logic. Higher-order representations could model these multi-molecule interactions more faithfully. HyperNetWalk also does not distinguish the sign of regulation: regulatory and signaling edges are treated as propagation channels without specifying whether a perturbation activates or represses downstream expression. Incorporating regulatory signs together with the direction of expression changes could help distinguish activating from repressing effects and better separate oncogene-like from tumor-suppressor-like patterns. Integrating additional genomic and epigenomic data types, such as copy-number alterations and DNA methylation, may further refine driver scoring beyond sequence-level mutations.

From a computational perspective, HyperNetWalk is scalable because Stage I and the personalized component of Stage II can be processed independently across samples, making them straightforward to parallelize. The cohort-level hypergraph refinement scales mainly with the number of samples and the union of mutated genes within each cancer type. Detailed runtime and hardware information are reported in the Supplementary Note S7.

In summary, HyperNetWalk demonstrates that structuring biological networks according to functional roles and formulating driver gene identification as a reverse inference problem can improve both personalized and cohort-level prediction. Its layered network and hypergraph refinement strategy produce biologically informed rankings while also yielding interpretable TF-layer representations associated with regulatory heterogeneity and patient survival. These principles may extend to other genomic inference settings in which upstream perturbations must be inferred from downstream molecular phenotypes.

## Methods

### Datasets

**TCGA Multi-omics Data.** Somatic mutation and gene expression data were obtained from TCGA via the UCSC Xena platform[39], which provides uniformly processed data across cancer types.

We analyzed 12 cancer types with at least 30 normal samples to ensure reliable differential expression estimation: BRCA, COAD, HNSC, KIRC, KIRP, LIHC, LUAD, LUSC, PRAD, STAD, THCA, and UCEC (Supplementary Table S1).

Somatic mutations were represented as a binary matrix, where an entry of 1 indicates at least one non-silent mutation in the corresponding gene. Gene expression values were obtained as $\log_2(\text{TPM} + 1)$.

**Protein–Protein Interaction (PPI) Network.** The PPI network was constructed by integrating data from STRING (version 12.0)[40] and OmniPath[41]. From STRING, we retained interactions with a combined confidence score $\geq 0.4$, yielding 19,488 proteins and 929,472 interactions. OmniPath contributed 11,682 proteins and 139,054 directed, literature-curated interactions, of which 77,318 overlap with STRING. The integrated network combines the broad interaction coverage of STRING with the directed signaling information from OmniPath.

**Gene Regulatory Network (GRN).** The GRN was obtained from RegNetwork[42], a curated repository of transcriptional and post-transcriptional regulatory relationships. We specifically extracted directed regulatory interactions between human TFs and their target genes. This process yielded a GRN consisting of 32,919 genes and 1,111,313 directed edges, explicitly connecting 1,717 TFs to 31,202 target genes. Among these TFs, 53 have no corresponding node in the above integrated PPI network.

**Problem Formulation and Network Construction**

**Notation and Input Representation.** For a specific cancer type $c$, the somatic mutation profile is denoted as a binary matrix $\mathbf{M}^c \in \{0,1\}^{m_c \times n_c}$, where $m_c$ is the number of genes mutated in at least one sample and $n_c$ is the number of tumor samples. An entry $[\mathbf{M}^c]_{gi} = 1$ indicates that gene $g$ harbors at least one non-silent mutation in sample $i$, and 0 otherwise. Patients with fewer than three mutated genes are excluded to reduce noise.

The gene expression profiles for tumor and normal samples are represented as matrices $\mathbf{X}_{\mathrm{T}}^c \in \mathbb{R}^{l_c \times n_c}$ and $\mathbf{X}_{\mathrm{N}}^c \in \mathbb{R}^{l_c \times n'_c}$, respectively, where $l_c$ is the number of expression-profile genes and $n'_c$ is the number of normal samples. All tumor samples are required to have matched mutation and expression data.

The PPI network is represented as a mixed graph $G_{\mathrm{PPI}} = (V_{\mathrm{PPI}}, E_{\mathrm{PPI}})$, where $V_{\mathrm{PPI}}$ denotes proteins. The edge set $E_{\mathrm{PPI}}$ contains directed edges (from OmniPath consensus interactions) and undirected edges (from OmniPath non-consensus interactions or STRING). Undirected edges are treated as bidirectional during propagation.

The GRN is represented as a directed graph $G_{\text{GRN}} = (V_{\text{GRN}}, E_{\text{GRN}})$. Based on its topology, genes are partitioned into TFs ($V_{\text{TF}}$: nodes with out-degree $> 0$) and target genes ($V_{\text{TG}}$: nodes with in-degree $> 0$ and out-degree $= 0$).

**Construction of the Layered Biological Network.** To model the mutation-to-expression cascade, we construct a global layered network $G_{\text{GLOBAL}} = (V_{\text{GLOBAL}}, E_{\text{GLOBAL}})$ that integrates a PPI network for upstream signaling with a GRN for downstream transcriptional regulation. The network is organized into three functional layers for computational implementation: a PPI layer (Layer 1, signaling), a TF layer (Layer 2, regulation), and a target-gene layer (Layer 3, expression output). The GRN is embedded across Layers 2 and 3.

**Layer Definitions:**

Layer 1 (PPI Layer): This layer captures signaling propagation. The node set $V_{\text{Layer1}}$ includes all proteins in $V_{\text{PPI}}$ as well as the protein counterparts of TFs in $V_{\text{TF}}$. For TFs absent from the integrated PPI network, the counterpart node is added to Layer 1 without PPI intra-layer edges. Including TF proteins in this layer ensures that TFs which are themselves mutated can be captured as perturbation origins, so that candidate driver genes represented in the network reside in a single search space within Layer 1.

Layer 2 (TF Layer): This layer encodes the regulatory mediators. The node set $V_{\text{Layer2}}$ corresponds to $V_{\text{TF}}$.

Layer 3 (Target-Gene Layer): This layer defines the downstream transcriptional outputs. The node set $V_{\text{Layer3}}$ corresponds to $V_{\text{TG}}$.

The overall node set is $V_{\text{GLOBAL}} = V_{\text{Layer1}} \cup V_{\text{Layer2}} \cup V_{\text{Layer3}}$. The GRN is embedded as a directed subnetwork spanning Layers 2 and 3, covering both TF–TF and TF–target regulatory edges. TFs link the PPI layer to the GRN through identity mapping between their protein nodes in Layer 1 and their regulatory nodes in Layer 2, enabling signal propagation from signaling to transcriptional regulation.

**Edge Definitions and Weighting Scheme:**

1. Layer 1 Intra-connections (PPI): Edges represent protein interactions and include both directed and undirected edges. Undirected edges are treated as bidirectional during propagation. To incorporate context-specific information, we refine PPI edge weights using gene co-expression patterns. Specifically, for cancer type $c$, we compute the Pearson correlation matrix $\mathbf{S}^c$ from tumor expression data $\mathbf{X}_{\mathrm{T}}^{c}$, following a commonly used strategy in co-expression analysis (e.g., WGCNA[43]). Unlike WGCNA, we use correlation magnitudes without additional power transformations, as edge weighting is applied to a predefined interaction network rather than used to construct one. To avoid vanishing influence from weakly correlated pairs, a lower bound of 0.1 is imposed. Genes lacking expression data are assigned a default weight of 0.1. Thus, the weights are defined as:

$$\left[\mathbf{W}_{\mathrm{Layer1}}\right]_{uv} = \begin{cases} \max(|[\mathbf{S}^c]_{uv}|, 0.1) & \text{if } (u, v) \in E_{\mathrm{PPI}} \\ 0 & \text{otherwise} \end{cases}.$$

2. Layer 2 Intra-connections (TF–TF): Directed edges represent regulatory interactions among TFs, extracted from $G_{\mathrm{GRN}}$. These edges are assigned a uniform weight of 1, because the GRN is treated as a curated directional prior rather than a quantitative map of condition-specific regulatory strengths.
3. Layer 3 Intra-connections: No intra-layer edges are defined for target genes, as they serve as endpoints of the regulatory cascade.
4. Inter-layer Connections:
   - Layer 1 → Layer 2: Directed edges connect TF protein nodes in Layer 1 to their corresponding TF regulatory nodes in Layer 2, implementing identity mapping from the PPI layer to the TF layer. For TF complexes, all component proteins are connected to the corresponding TF node. The weights of these edges are defined during sample-specific subnetwork construction (see below) to enable adaptive transition probabilities between the two layers.
   - Layer 2 → Layer 3: Directed edges represent transcriptional regulation from TFs to target genes, derived from $E_{\mathrm{GRN}}$, and are likewise assigned a uniform weight of 1.

**Matrix Representation.** The global weighted adjacency matrix $\mathbf{W}_{\text{GLOBAL}}$ of the layered network is formulated as a block matrix structure:

$$\mathbf{W}_{\text{GLOBAL}} = \begin{bmatrix} \mathbf{W}_{\text{Layer1}} & \mathbf{W}_{1\to2} & \mathbf{0} \\ \mathbf{0} & \mathbf{W}_{\text{Layer2}} & \mathbf{W}_{2\to3} \\ \mathbf{0} & \mathbf{0} & \mathbf{0} \end{bmatrix}. \quad (1)$$

Here $\mathbf{W}_{\text{Layer1}} \in \mathbb{R}^{|V_{\text{Layer1}}|\times|V_{\text{Layer1}}|}$ and $\mathbf{W}_{\text{Layer2}} \in \mathbb{R}^{|V_{\text{Layer2}}|\times|V_{\text{Layer2}}|}$ are the intra-layer adjacency matrices for the PPI and TF layers, respectively. $\mathbf{W}_{1\to2} \in \mathbb{R}^{|V_{\text{Layer1}}|\times|V_{\text{Layer2}}|}$ encodes the connections from signaling proteins to TFs, and $\mathbf{W}_{2\to3} \in \mathbb{R}^{|V_{\text{Layer2}}|\times|V_{\text{Layer3}}|}$ encodes the regulation of target genes by TFs. Note that the sub-block $\begin{bmatrix} \mathbf{W}_{\text{Layer2}} & \mathbf{W}_{2\to3} \\ \mathbf{0} & \mathbf{0} \end{bmatrix}$ corresponds to the topology of the gene regulatory network $G_{\text{GRN}}$.

**Problem Formulation.** Given the global layered network $G_{\text{GLOBAL}}$ and multi-omics data for a cancer type $c$, the goal is to identify cancer driver genes at both personalized and cohort levels.

For each patient $i$ in cancer type $c$, let $\mathcal{M}_i \subseteq V_{\text{Layer1}}$ denote the set of mutated genes. The personalized driver gene identification task aims to assign a score $s_i(g)$ to each gene $g \in \mathcal{M}_i$, reflecting its likelihood of being a driver gene for patient $i$.

At the cohort level, the objective is to identify genes that exhibit consistent driver signals across patients within the same cancer type. This is formulated as assigning a cohort-level score $s^c(g)$ to each gene $g$, based on aggregated evidence from multiple patients.

We formulate driver gene identification as an inverse problem: upstream driver genes are inferred from downstream transcriptional dysregulation by reversing the direction of information flow in the mutation-to-expression cascade.

In this study, we consider driver gene identification within a given cancer type. For notational simplicity, we omit the index $c$ for sample-specific variables indexed by patient $i$ (e.g., mutated-gene sets $\mathcal{M}_i$), as each sample is associated with a fixed cancer type. Quantities defined across multiple samples (e.g., cohort-level variables) retain the index $c$ when necessary.

**Stage I: Sample-Specific Reverse Inference**

In this stage, HyperNetWalk performs a reverse random walk with restart on a sample-specific layered network for each patient, operating independently and relying solely on the sample-specific network topology and omics profiles. The goal is to score mutated genes by their ability to explain the observed downstream transcriptional dysregulation.

**Construction of Sample-Specific Layered Network.** For a patient $i$ of cancer type $c$, we construct a sample-specific layered network $G_i = (V_i, E_i)$ as a subgraph of the global network $G_{\text{GLOBAL}}$. To capture mutation-to-expression propagation while reducing noise introduced by the dense topology of the global PPI network, we define the sample-specific layered network through a directed path-based filtering strategy.

1. **Signaling propagation in Layer 1.** Starting from the mutated-gene set $\mathcal{M}_i$, we identify TFs whose protein counterparts in Layer 1 are reachable within two hops. STRING interactions are largely undirected and form a small-world topology in which almost all genes are mutually reachable within a few hops. To preserve signaling directionality and ensure specificity in TF selection, the reachability search is therefore restricted to directed interactions from OmniPath. We retain the mutated genes, the reachable TF protein counterparts, and the intermediate proteins located on these directed paths. Although undirected interactions are excluded from the reachability search, they are retained in the final subnetwork as a structural backbone to maintain network connectivity during the random walk.
2. **Regulatory cascade expansion in Layers 2 and 3.** The TFs identified in Layer 1 are mapped to their corresponding nodes in Layer 2. Starting from these TF nodes, we traverse the GRN component of the layered network $G_{\text{GLOBAL}}$ along directed regulatory edges to identify downstream nodes within three hops. This expansion includes both TF–TF regulatory cascades in Layer 2 and TF–target regulatory relationships from Layer 2 to Layer 3.

3. **Phenotypic pruning in Layer 3.** To focus on transcriptional outcomes that are most perturbed in the patient, we retain only DEGs among reachable targets. For each reachable target gene $g$, we compute a sample-specific z-score:

$$z_{gi} = \frac{[\mathbf{X}_{\mathrm{T}}^{c}]_{gi} - \mu_{\mathrm{N}}^{cg}}{\sigma_{\mathrm{N}}^{cg}},$$

where $[\mathbf{X}_{\mathrm{T}}^{c}]_{gi}$ denotes the expression value of gene $g$ in tumor sample $i$, $\mu_{\mathrm{N}}^{cg}$ is the mean expression of gene $g$ across all normal samples of cancer type $c$, and $\sigma_{\mathrm{N}}^{cg}$ is the corresponding standard deviation. We then rank the reachable target genes by $|z_{gi}|$ and retain the top 500 genes as the sample-specific DEG set, denoted by $\mathcal{D}_i$.

Let $V_i$ denote the union of all selected nodes, including mutated and pathway-associated proteins in Layer 1, reachable TFs in Layer 2, and the selected DEGs in Layer 3. The sample-specific layered network is then defined as the induced subgraph of the global layered network on $V_i$, formally expressed as $G_i = G_{\mathrm{GLOBAL}}[V_i]$. This construction preserves all intra-layer and inter-layer edges among the selected nodes.

**Reverse Random Walk with Restart.** To infer upstream drivers from downstream transcriptional changes, we perform a random walk with restart on the sample-specific layered network $G_i$ with reversed edge directions.

Let $\mathbf{W}_i \in \mathbb{R}^{|V_i| \times |V_i|}$ denote the weighted adjacency matrix of $G_i$, obtained as the corresponding submatrix of $\mathbf{W}_{\mathrm{GLOBAL}}$ defined in Equation 1. Edge weights for intra-layer connections in Layers 1–2 and inter-layer connections from Layer 2 to Layer 3 follow the definitions above. It remains to specify weights for the inter-layer edges from Layer 1 to Layer 2 in the sample-specific network.

For each TF node $v \in V_{\mathrm{Layer2}}$ that has at least one upstream TF neighbor in Layer 2 and a corresponding protein node $\hat{v} \in V_{\mathrm{Layer1}}$, we define the weight of the edge from $\hat{v}$ to $v$ as

$$[\mathbf{W}_i]_{\hat{v}v} = \frac{\alpha}{1-\alpha} \sum_{u \in V_{\mathrm{Layer2}}} [\mathbf{W}_i]_{uv},$$

where $\alpha \in (0,1)$ is a structural parameter controlling the probability of transitioning from Layer 2 to Layer 1 during reverse propagation. This construction ensures that, after normalization, the transition probability from a TF node to its Layer 1 counterpart is $\alpha$, with the remaining probability $1 - \alpha$ distributed over upstream TF neighbors in Layer 2. We set $\alpha$ to 0.5 as a balanced default, so that no prior preference is imposed between the two routes when both are available. For TF nodes without upstream TF neighbors, the corresponding inter-layer edge weight is set to 1.

Based on the fully specified adjacency matrix $\mathbf{W}_i$, we construct the row-stochastic transition matrix $\widetilde{\mathbf{P}}_i$ for the reverse random walk as

$$\widetilde{\mathbf{P}}_i = \mathbf{D}_i^{-1}\mathbf{W}_i^T,$$

where $\mathbf{D}_i$ is a diagonal matrix with

$$[\mathbf{D}_i]_{vv} = \sum_u [\mathbf{W}_i]_{uv}.$$

Under this definition, $[\widetilde{\mathbf{P}}_i]_{uv}$ represents the probability of transitioning from node $u$ to node $v$ along reversed edges.

The random walk is initialized from the sample-specific DEG set $\mathcal{D}_i$, serving as the downstream transcriptional evidence from which upstream drivers are inferred. The initial probability vector $\widetilde{\mathbf{q}}_{i,0} \in [0,1]^{|V_i|}$ is defined as:

$$[\widetilde{\mathbf{q}}_{i,0}]_v = \begin{cases} \frac{1}{|\mathcal{D}_i|}, & v \in \mathcal{D}_i, \\ 0, & \text{otherwise}. \end{cases}$$

Let $\widetilde{\mathbf{q}}_{i,t} \in [0,1]^{|V_i|}$ denote the probability vector at iteration $t$, where $[\widetilde{\mathbf{q}}_{i,t}]_v$ represents the probability of the random walker being at node $v$. Starting from this initialization, the reverse random walk with restart is iteratively updated as

$$\widetilde{\mathbf{q}}_{i,t+1} = (1 - r)\widetilde{\mathbf{P}}_i^T \widetilde{\mathbf{q}}_{i,t} + r\widetilde{\mathbf{q}}_{i,0}\,, \tag{2}$$

where $r \in (0,1)$ is the restart probability, set to 0.15 by default, following the conventional setting in PageRank-style random walk algorithms. The restart term ensures that the walk remains anchored to the observed transcriptional dysregulation. The iteration continues until convergence, yielding the stationary distribution $\tilde{\mathbf{q}}_i$, which quantifies the influence of each node in explaining the observed transcriptional changes.

Finally, the Stage I driver score for each mutated gene $g \in \mathcal{M}_i$ is defined as

$$\tilde{s}_i(g) = [\tilde{\mathbf{q}}_i]_g. \tag{3}$$

**Stage II: Cross-Sample Refinement**

The Stage I scores encode the topological and transcriptional evidence linking each sample's mutations to its expression phenotype. Because cancer samples within the same type often share driver genes and mutational patterns, cross-sample information can refine these initial estimates. We use hypergraph-based random walk, motivated by PDRWH[16], to integrate information across related samples with modifications tailored to the Stage I reverse-inference scores. We then construct two hypergraph structures for personalized and cohort-level refinement, respectively.

**Random Walk on Hypergraph.** We adopt a two-step random walk formulation on hypergraphs[44]. Let $\mathcal{H} = (\mathcal{V}, \mathcal{E})$ denote a hypergraph. The incidence matrix $\mathbf{H} \in \{0,1\}^{|\mathcal{V}|\times|\mathcal{E}|}$ is defined as:

$$[\mathbf{H}]_{ve} = \begin{cases} 1, & v \in e, \\ 0, & \text{otherwise.} \end{cases}$$

Let $\mathbf{W}_V \in \mathbb{R}^{|\mathcal{V}|\times|\mathcal{E}|}$ denote the weighted incidence matrix where $[\mathbf{W}_V]_{ve}$ is the weight of node $v$ in hyperedge $e$, and $\mathbf{W}_E \in \mathbb{R}^{|\mathcal{E}|\times|\mathcal{E}|}$ denote the diagonal matrix of hyperedge weights where $[\mathbf{W}_E]_{ee} = w_e$ is the weight of hyperedge $e$.

From a node $u \in \mathcal{V}$, the walker first selects a hyperedge $e \in \mathcal{E}$ from all hyperedges incident to $u$ with probability proportional to the hyperedge weight $w_e$. Then, the walker moves to a node $v \in e$ with probability proportional to the node weight $[\mathbf{W}_V]_{ve}$ in that hyperedge. Formally, the transition probability from node $u$ to node $v$ is:

$$[\mathbf{P}]_{uv} = \sum_{e\in\mathcal{E}} \frac{[\mathbf{H}]_{ue} w_e}{\sum_{\hat{e}\in\mathcal{E}}[\mathbf{H}]_{u\hat{e}} w_{\hat{e}}} \cdot \frac{[\mathbf{W}_V]_{ve}}{\sum_{\hat{v}\in\mathcal{V}}[\mathbf{W}_V]_{\hat{v}e}}.$$

The row-stochastic transition matrix can be written as:

$$\mathbf{P} = \mathbf{D}_V^{-1}\mathbf{H}\mathbf{W}_E\mathbf{D}_E^{-1}\mathbf{W}_V^T,$$

where $\mathbf{D}_V$ is the diagonal matrix of vertex weighted degrees as $[\mathbf{D}_V]_{vv} = \sum_e [\mathbf{H}]_{ve} w_e$ and $\mathbf{D}_E$ is the diagonal matrix of hyperedge weighted degrees as $[\mathbf{D}_E]_{ee} = \sum_v [\mathbf{W}_V]_{ve}$.

We perform the random walk with restart on $\mathcal{H}$ with a uniform initial distribution $[\mathbf{q}_0]_v = 1/|\mathcal{V}|$. The iterative update follows:

$$\mathbf{q}_{t+1} = (1-\gamma)\mathbf{P}^{\mathrm{T}}\mathbf{q}_t + \gamma\mathbf{q}_0\,, \tag{4}$$

where $\mathbf{q}_t \in [0,1]^{|\mathcal{V}|}$ is the probability distribution at iteration $t$, and $\gamma$ is the restart probability parameter set to 0.15 in this study. This relatively small restart probability ensures that the Stage I-informed hypergraph transition structure primarily governs score propagation; Stage II serves as a refinement step rather than a de novo ranking procedure. After convergence, we obtain the stationary distribution $\mathbf{q}$.

**Personalized Driver Gene Identification.** For a given sample $i$ of cancer type $c$ (referred to as the target sample), we construct a local hypergraph $\mathcal{H} = (\mathcal{V}, \mathcal{E})$ to identify personalized driver genes by integrating information from the target sample and its neighboring samples. The set of neighboring samples is defined as: $\mathcal{N}_i = \{j \mid |\mathcal{M}_j \cap \mathcal{M}_i| \geq 2,\ j \neq i\}$, i.e., samples of cancer type $c$ sharing at least two mutated genes with the target sample.

The node set of the local hypergraph is $\mathcal{V} = \mathcal{M}_i$, i.e., the set of mutated genes in the target sample. The hyperedge set is $\mathcal{E} = \{e_j \mid j \in \{i\} \cup \mathcal{N}_i\}$, where $e_j = \mathcal{M}_i \cap \mathcal{M}_j$ is the set of common mutated genes between sample $j$ and the target sample $i$.

To incorporate the Stage I driver signals, we define the weighted incidence matrix $\mathbf{W}_V$ as:

$$[\mathbf{W}_V]_{ge_j} = \begin{cases} \tilde{s}_j(g), & g \in e_j, \\ 0, & \text{otherwise,} \end{cases}$$

where $\tilde{s}_j(g)$ denotes the Stage I driver score of gene $g$ in sample $j$.

When assigning weights to hyperedges, relying solely on transcriptomic similarity or treating all neighboring samples equally may introduce bias due to recurrent but functionally irrelevant mutations. To address this, we further incorporate the driver relevance of shared mutated genes based on Stage I scores, so that a neighboring sample contributes more strongly only when it is both transcriptionally similar and informative with respect to shared candidate drivers. Expression similarity between samples $i$ and $j$ is:

$$\mathrm{sim}(i,j) = \exp\left(-\frac{\left(1-\rho_{ij}\right)^2}{2\sigma^2}\right),$$

where $\rho_{ij}$ is the Pearson correlation coefficient between the expression profiles of samples $i$ and $j$, and $\sigma$ is a bandwidth parameter set to 0.1 in this study. Driver relevance is computed as:

$$\mathrm{rel}_i(j) = \sum_{g \in \mathcal{S}_i(j,k)} \frac{1}{\log_2\left(r_{gj}\right)+1},$$

where $\mathcal{S}_i(j,k) = \text{top-}k\left(\mathcal{M}_i \cap \mathcal{M}_j; \tilde{s}_j\right)$ denotes the top-$k$ shared mutations ranked by Stage I scores in sample $j$, $r_{gj}$ is the rank of gene $g$ in sample $j$ by Stage I scores, and $k = 3$ by default. Thus, the hyperedge weight is

$$w_{e_j} = \mathrm{sim}(i,j) \cdot \mathrm{rel}_i(j).$$

The stationary distribution $\mathbf{q}$ of random walk on this local hypergraph provides the final personalized driver gene scores for sample $i$. That is, for each mutated gene $g \in \mathcal{M}_i$, the refined driver score is defined as $s_i(g) = [\mathbf{q}]_g$.

**Cohort-Level Driver Gene Identification.** For cancer type $c$ with $n_c$ samples, we construct a global hypergraph $\mathcal{H} = (\mathcal{V}, \mathcal{E})$, where:

- $\mathcal{V}$: the set of all genes mutated in at least one sample of cancer type $c$,
- $\mathcal{E} = \{e_1, e_2, \dots, e_{n_c}\}$: the set of hyperedges, with each hyperedge $e_j$ here corresponding to the set of mutated genes in sample $j$ of cancer type $c$.

The incidence matrix $\mathbf{H} \in \{0,1\}^{|\mathcal{V}|\times|\mathcal{E}|}$ and the weighted incidence matrix $\mathbf{W}_V \in \mathbb{R}^{|\mathcal{V}|\times|\mathcal{E}|}$ are defined analogously to the personalized case. For cohort-level analysis, we assign equal weight to all hyperedges since we do not prioritize any particular sample. Thus, $w_e = 1$ for all $e \in \mathcal{E}$, and the hyperedge weight matrix $\mathbf{W}_E$ is simply the identity matrix.

The stationary distribution $\mathbf{q}$ of random walk on this global hypergraph provides cohort-level driver gene scores for cancer type $c$. Similarly, for each gene $g \in \mathcal{V}$, the cohort-level driver score is defined as $s^c(g) = [\mathbf{q}]_g$. Genes are ranked by their scores $s^c(g)$ in descending order to identify cohort-level driver genes across all samples of the cancer type.

**Ablation Variants**

To evaluate the contribution of individual design choices, we constructed four structural variants of HyperNetWalk, each modifying one component while keeping others unchanged. All variants use the same Stage II hypergraph refinement as the full model; differences are restricted to Stage I.

**Forward.** The reverse random walk in Stage I is replaced with a forward random walk. The walk is initialized with a uniform distribution over mutated gene nodes (instead of DEG nodes), and propagation follows the original (non-reversed) edge directions in the sample-specific layered network. The stationary probabilities at DEG nodes are discarded; instead, the probabilities at mutated gene nodes after forward propagation are used as Stage I scores.

**Merged.** The layered architecture is removed by collapsing the sample-specific layered network into a single graph. Specifically, layer-specific instances of the same gene are merged into a single node when that gene appears in multiple layers, namely the PPI layer, TF layer, and target-gene layer. The identity mapping edges between Layer 1 and Layer 2 are then removed. The biological edges (PPI edges, TF–TF regulatory edges, and TF–target regulatory edges) are preserved, but the merged node is now simultaneously incident to both PPI and regulatory edges. Reverse random walk is performed on this merged graph.

**Merged_Plus.** The layered architecture is also removed with nodes merged in the same way as Merged, but the graph is constructed by directly inducing a subgraph from the global PPI and GRN on the set of selected nodes. Because the layered construction imposes a block structure that excludes certain edges (e.g., PPI edges among target genes), the edge set of Merged_Plus is a superset of that of Merged, potentially restoring edges that the layered formulation removes.

**Undirected.** All directed edges in the sample-specific layered network (from OmniPath and GRN) are treated as undirected during the random walk, while the layered structure is preserved. This variant tests the contribution of edge directionality to reverse inference on the layered network.

**Evaluation**

Following standard practice in driver gene identification studies, we evaluate performance by comparing the top-ranked predictions against CGC Tier 1 driver genes[24], a curated reference set of 584 high-confidence cancer driver genes. For both personalized and cohort-level settings, reference positives are defined as the intersection between the mutated gene set and CGC Tier 1. Genes without predicted scores are assigned the lowest ranks to ensure fair comparison across methods.

Because driver gene identification methods are primarily used to prioritize candidates, we focus on evaluating performance at the top of the ranking. We report top-$K$ metrics, including Precision@$K$, Recall@$K$, and F1@$K$, together with AUC-based metrics, including AUPRC, AUROC, pAUPRC, and pAUROC. For personalized prediction, top-$K$ metrics are computed for each sample and then averaged across samples within the same cancer type. The largest $K$ for personalized evaluation, denoted as $N_c$, is adaptively determined for each cancer type following the REA strategy[15]. For cohort-level prediction, top-$K$ metrics are computed directly from the ranked gene list for each cancer type. AUC-based metrics are computed by treating each ranked list as a binary classifier for reference positives, with partial metrics focused on the early-retrieval regions Recall $\leq 0.1$ for pAUPRC and FPR $\leq 0.1$ for pAUROC. Detailed metric definitions and evaluation procedures are provided in the Supplementary Note S1.

To assess robustness to parameter choices, we further perform one-factor-at-a-time sensitivity analyses for the key parameters of HyperNetWalk (Supplementary Note S2 and Supplementary Fig. S25).

**Cohort-Level Aggregation of Personalized Methods**

To test whether the cohort-level performance of HyperNetWalk can be attributed to post-hoc aggregation of personalized predictions, we constructed cohort-level rankings for each competing personalized method. Methods with built-in aggregation procedures were evaluated using their original strategies: DawnRank derives cohort-level rankings from its sample-specific scores using a modified Condorcet rank aggregation, and PDRWH constructs a pairwise voting-based graph for cohort-level ranking. DawnRank's aggregation requires scores for all genes (not only mutated genes) and is therefore not directly applicable to other personalized methods. For methods without predefined aggregation schemes (PersonaDrive, PITCH, and PRODIGY), we applied two general strategies: (1) the pairwise voting-based graph method from PDRWH, and (2) the cohort-level hypergraph random walk used in HyperNetWalk Stage II, with node weights set to the respective method's personalized scores. For each method, we retained the aggregation result that yielded higher performance, providing a favorable comparison baseline.

## Data Availability Statement

The results presented in this study are in part based upon data generated by the TCGA Research Network (https://www.cancer.gov/tcga). TCGA somatic mutation and gene expression data were accessed through the UCSC Xena platform. All data used in this study are publicly available. The source code and analysis scripts for HyperNetWalk are available at: https://github.com/xqxu921/HyperNetWalk. Large supporting data files required to reproduce the main analyses, including processed input data and reference network files, are archived on Zenodo at https://doi.org/10.5281/zenodo.20627042.

## Conflict of Interest

None declared.

## Acknowledgments

This work was supported by the National Key Research and Development Program of China (No. 2020YFA0712400); National Natural Science Foundation of China (Nos. 12231018 and 62202269). The results published here are in whole or part based upon data generated by The Cancer Genome Atlas (TCGA) Research Network (https://www.cancer.gov/tcga). We are grateful to the TCGA

Research Network and to all patients and their families for their invaluable contributions to cancer research.

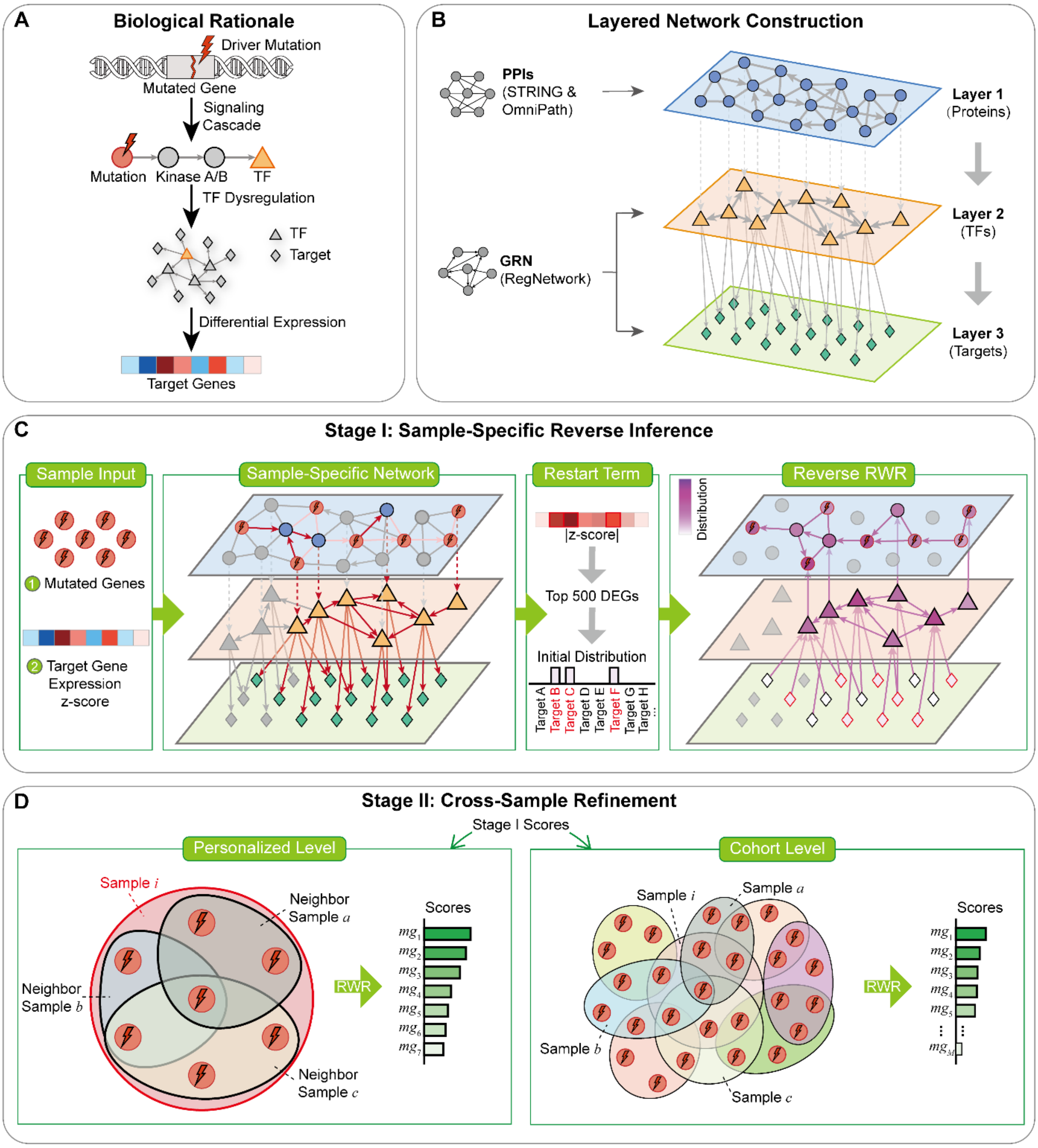

**Figure 1. Overview of the HyperNetWalk framework. A.** Biological rationale. Driver mutations may perturb upstream signaling, dysregulate TF activity, and induce coordinated downstream expression changes observed as DEGs. **B.** Layered network construction. The global network integrates a PPI network and a GRN into three layers: a PPI layer (Layer 1, blue circles), a TF layer (Layer 2, orange triangles), and a target-gene layer (Layer 3, green diamonds). The GRN component spans Layers 2 and 3 through TF–TF and TF–target regulatory edges. TF nodes in Layer 2 are connected to their corresponding protein nodes in Layer 1 by identity edges (grey dashed lines). Layer 1 contains directed OmniPath consensus interactions and undirected STRING or OmniPath non-consensus interactions. **C.** Stage I: sample-specific reverse inference. For each patient, a sample-specific subnetwork is constructed by directed path-based filtering from mutated genes to reachable TFs (red directed paths), while undirected backbone edges are retained (pink edges). RWR is initialized uniformly over the top 500 DEG nodes (red-outlined diamonds) and propagates upstream by traversing directed edges in reverse and treating undirected edges as bidirectional. Stationary probabilities at mutated gene nodes define Stage I sample-specific driver scores. **D.** Stage II: cross-sample refinement. For personalized prediction,

a local hypergraph is constructed around target sample $i$, with candidate nodes representing its mutated genes and hyperedges representing the target and neighboring samples. RWR on this hypergraph refines personalized driver rankings using Stage I scores as transition weights. For cohort-level prediction, a global hypergraph spanning all patients produces cohort-level driver gene rankings.

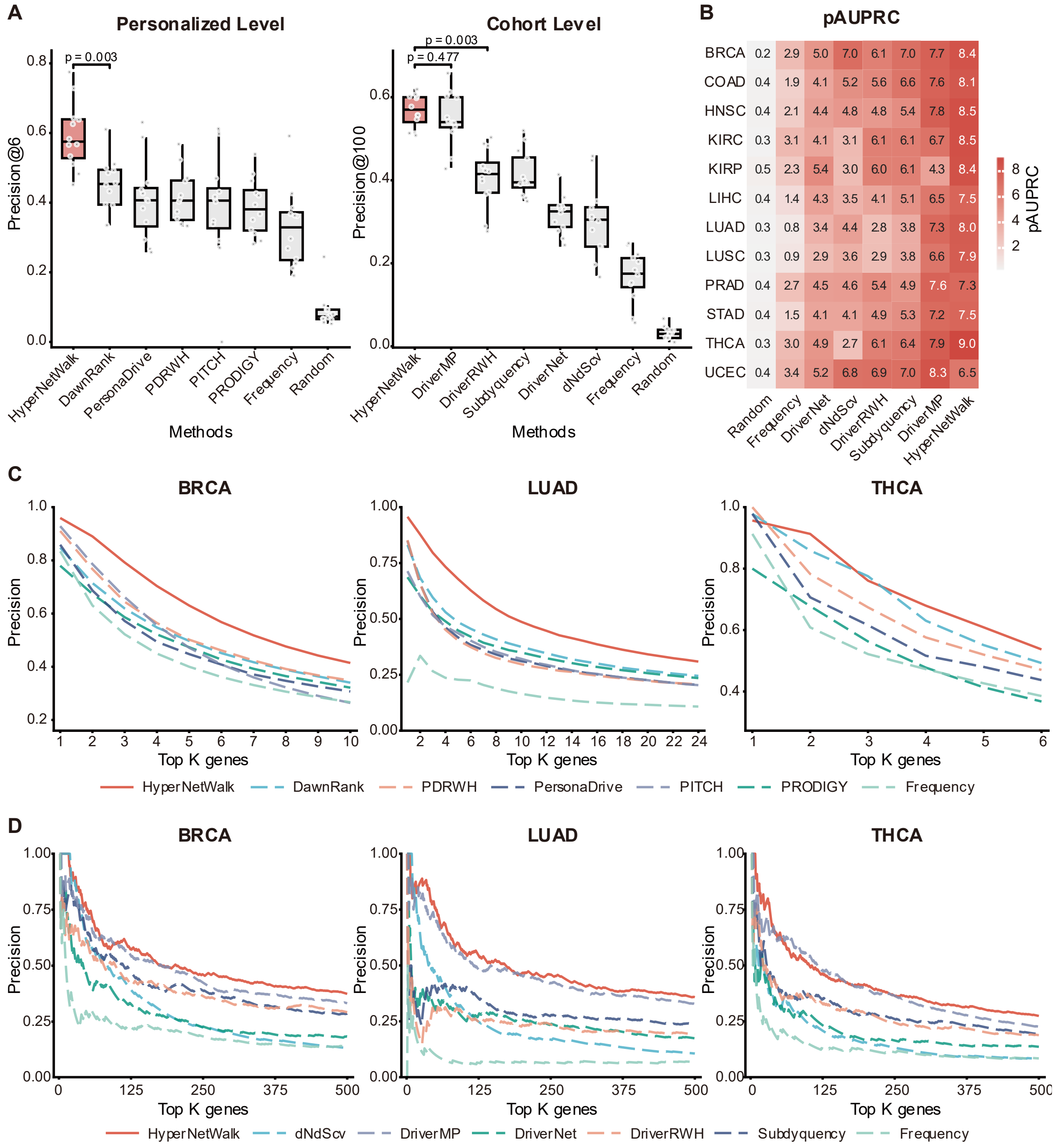

**Figure 2. Comprehensive benchmarking of HyperNetWalk. A.** Boxplots of Precision@$K$ across 12 TCGA cancer types for personalized predictions (left, $K = 6$) and cohort-level predictions (right, $K = 100$). Each point represents one cancer type. P-values were calculated using two-sided paired Wilcoxon signed-rank tests, with cancer type as the paired unit. **B.** Heatmap of cohort-level pAUPRC values in the low-recall region (recall $\leq 0.1$) across 12 cancer types; the best-performing method for each cancer type is highlighted. **C.** Precision@$K$ curves for personalized predictions in BRCA, LUAD, and THCA. The random baseline is omitted for clarity because of its substantially lower performance. **D.** Precision@$K$ curves for cohort-level predictions in the same three cancer types.

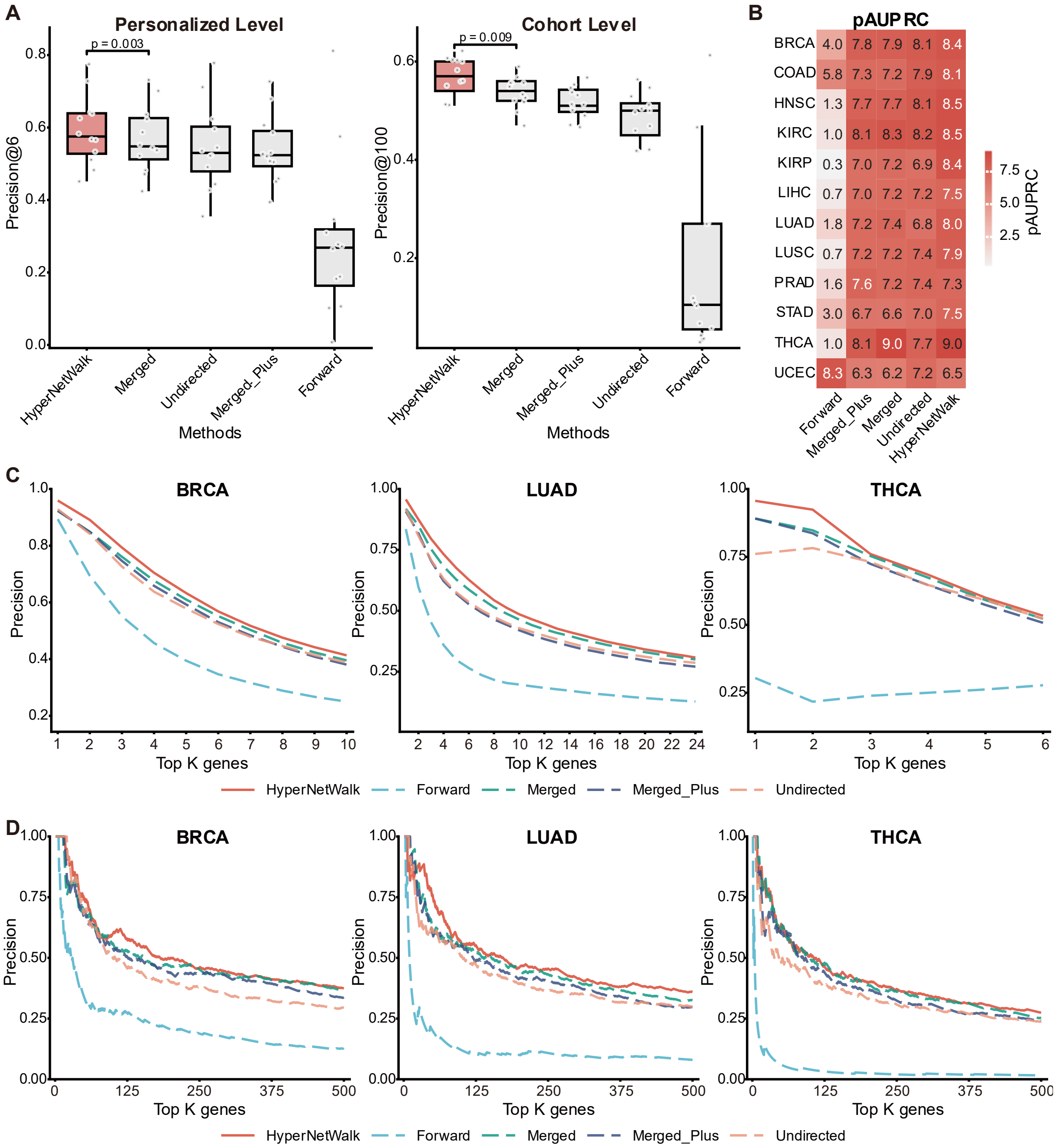

**Figure 3. Ablation analysis of HyperNetWalk design components.** Four model variants were compared with the full model: *Forward* replaces Stage I reverse propagation with forward propagation from mutated genes; *Merged* collapses layer-specific instances of the same gene into a single node, removing TF–protein identity edges; *Merged_Plus* also uses merged gene nodes but induces edges directly from the PPI network and GRN over the selected nodes; and *Undirected* treats directed OmniPath and GRN edges as undirected during random walk. **A.** Boxplots of Precision@$K$ across 12 TCGA cancer types for personalized predictions (left, $K = 6$) and cohort-level predictions (right, $K = 100$). Each point represents one cancer type. P-values indicate two-sided paired Wilcoxon signed-rank tests, with cancer type as the paired unit. **B.** Heatmap of cohort-level pAUPRC values in the low-recall region (recall $\leq 0.1$) across 12 cancer types; the best-performing model or variant for each cancer type is highlighted. pAUPRC values are scaled by 100 for readability. **C.** Precision@$K$ curves for personalized predictions in BRCA, LUAD, and THCA. **D.** Precision@$K$ curves for cohort-level predictions in the same three cancer types.

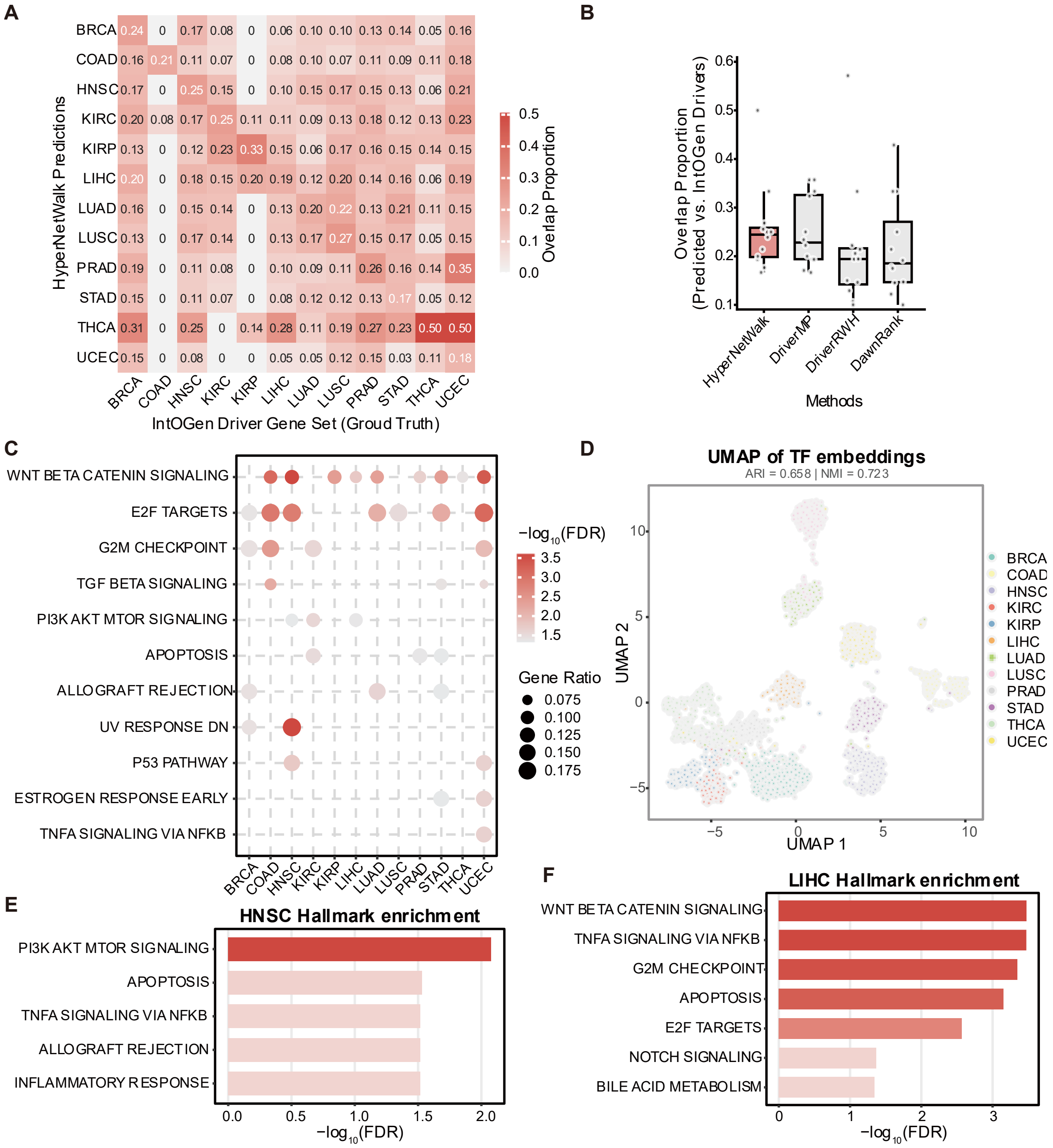

**Figure 4. HyperNetWalk captures cancer-type-specific driver patterns at gene, pathway, and regulatory levels. A.** Heatmap of the overlap proportion between the top 100 cohort-level predictions for each cancer type and IntOGen cancer-type-specific driver sets. Rows denote prediction cancer types, and columns denote reference cancer types. For each row, the largest value is highlighted. **B.** Boxplots of the overlap proportion with the matched cancer-type-specific driver set, corresponding to the diagonal elements in panel A. Each dot represents one cancer type. **C.** Dot plot of Hallmark enrichment results for the top 100 cohort-level predictions in each cancer type. Rows represent the union of significantly enriched pathways across all cancer types, and columns represent cancer types. Dot size indicates the gene ratio, and dot color indicates enrichment significance, measured as $-\log_{10}(\text{FDR})$. **D.** UMAP visualization of sample-level TF-layer probability vectors derived from Stage I reverse inference. Each point represents one sample and is colored by cancer type. Adjusted Rand index (ARI) and normalized mutual information (NMI) values are shown. **E.** Representative Hallmark enrichment results for the downstream target-gene program associated with cancer-type-specific TFs in HNSC. **F.**

Representative Hallmark enrichment results for the downstream target-gene program associated with cancer-type-specific TFs in LIHC.

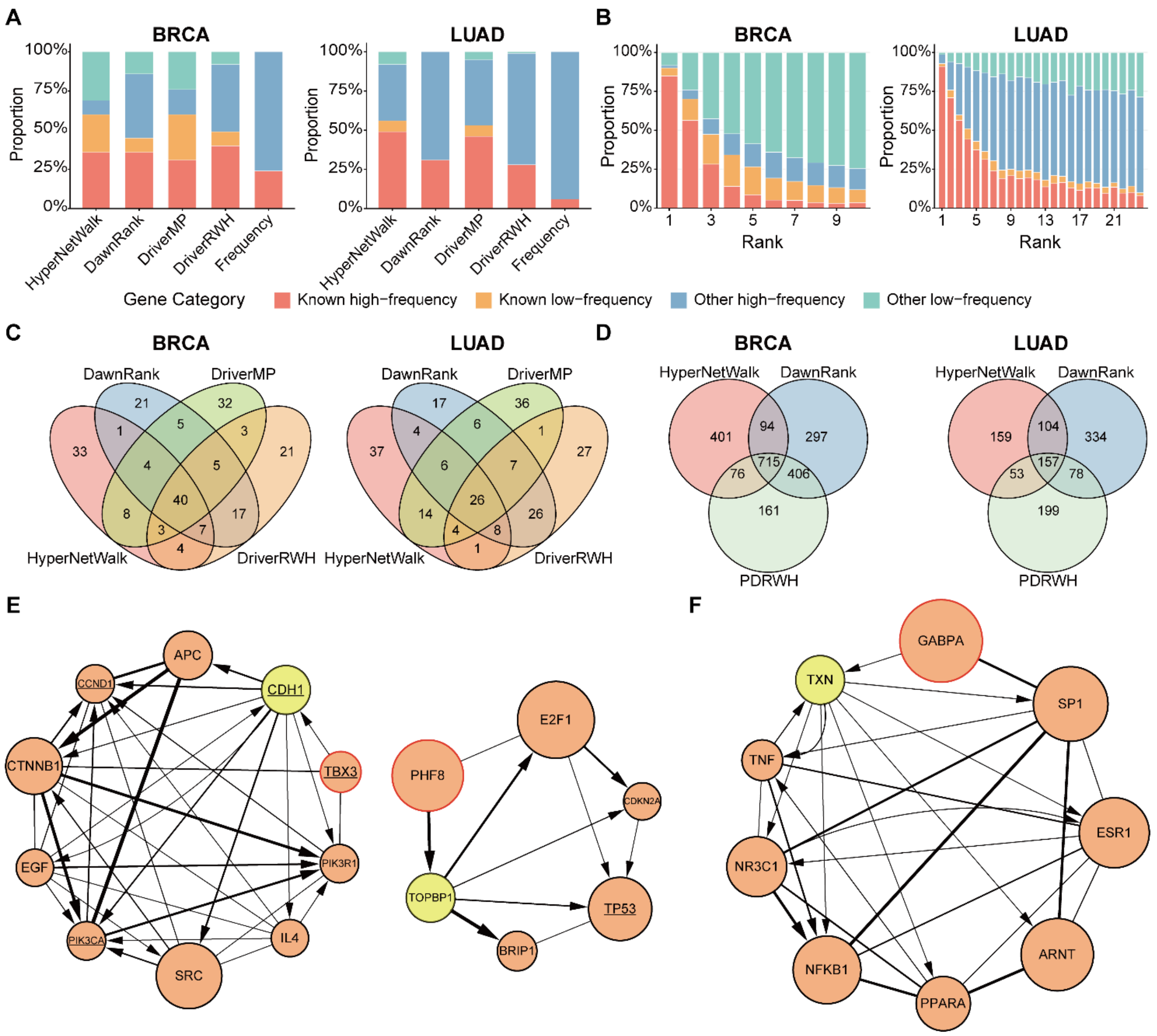

**Figure 5. HyperNetWalk identifies supported candidate drivers missed by competing methods. A.** Stacked bar plots showing the composition of top 100 cohort-level predictions in BRCA and LUAD. Predicted genes were grouped into four categories using a 2% mutation-frequency threshold and CGC Tier 1 annotation: known high-frequency drivers, known low-frequency drivers, other high-frequency mutated genes, and other low-frequency mutated genes. **B.** Rank-wise composition of personalized predictions in BRCA and LUAD. For each rank position, proportions of the same four gene categories were calculated across all samples. The maximum rank shown for each cancer type corresponds to $N_c$, as determined by the REA strategy. **C.** Venn diagrams showing overlap among the top 100 cohort-level predictions from HyperNetWalk and baseline methods in BRCA and LUAD. **D.** Venn diagrams showing overlap among personalized predictions in BRCA and LUAD, where each set denotes the union of the top 5 predicted genes across all samples for the corresponding method. **E.** Local network contexts of representative HyperNetWalk-specific candidates in BRCA, TBX3 and PHF8. Subnetworks were induced using the Stage I directed path-based filtering strategy in Layer 1. TF nodes are colored orange, with node size proportional to their GRN out-degree; intermediate non-TF nodes are shown in yellow; candidate genes are highlighted in red; and CGC Tier 1 BRCA-related drivers are underlined. Edge width is proportional to the random-walk edge weight. **F.** Local

network context of the HyperNetWalk-specific candidate GABPA in LUAD, shown using the same visualization scheme as in panel E.

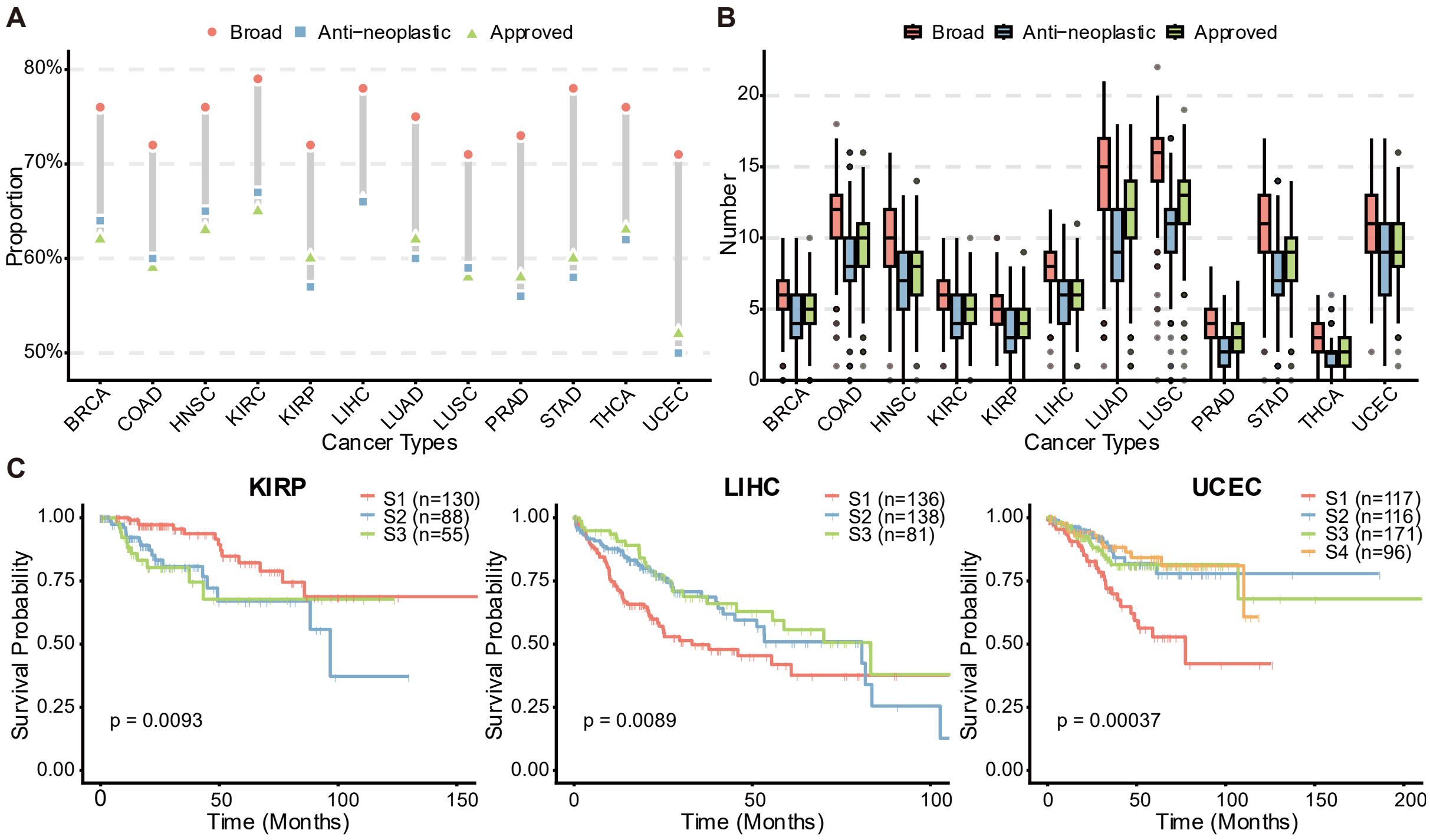

**Figure 6. HyperNetWalk outputs reveal therapeutic relevance and survival-associated patient subtypes. A.** Cleveland dot plot showing the proportion of the top 100 cohort-level predictions with DGIdb annotations across 12 cancer types. Three DGIdb-supported categories were evaluated: genes with any annotated drug–gene interaction (broad), genes associated with anti-neoplastic drugs, and genes with interactions involving approved drugs. **B.** Boxplots showing the number of DGIdb-supported genes among personalized predictions across samples for each cancer type. The same three DGIdb categories as in panel A were evaluated. **C.** Kaplan–Meier survival curves for KIRP, LIHC, and UCEC stratified by TF-layer-derived subtypes. Patient subtypes were inferred by consensus clustering on Stage I TF-layer probability vectors, with the number of clusters determined by the delta area criterion without using survival information. Log-rank p-values are shown.

**Supplementary Materials for**

# HyperNetWalk: A Unified Framework for Personalized and Cohort-Level Cancer Driver Gene Identification via Reverse Inference on Layered Signaling–Regulatory Network

Xueqing Xu[1,2], Yonghang Gao[1,2], Duanchen Sun[3,*], Ling-Yun Wu[1,2,*]

[*]Corresponding authors: dcsun@sdu.edu.cn (D. Sun); lywu@amss.ac.cn (L.-Y. Wu)

## Supplementary Notes

### Note S1: Evaluation Metrics and Strategies

We use top-$K$ metrics and area under the curve (AUC)-based metrics to quantify prediction performance. Unless otherwise specified, CGC Tier 1 genes are used as the reference driver set. Reference positives are defined as the intersection between the mutated gene set and CGC Tier 1.

**Top-$K$ Metrics.** For a given sample or cohort, let $\mathrm{TP}_K$ denote the number of true driver genes in the top $K$ predictions, P denote the total number of true driver genes in the evaluation set, and $K$ denote the number of predictions considered. We compute:

- **Precision@$K$**: the proportion of true driver genes among the top $K$ predictions,

$$\text{Precision}@K = \frac{\mathrm{TP}_K}{K}.$$

- **Recall@$K$**: the proportion of true driver genes recovered in the top $K$ predictions,

$$\text{Recall}@K = \frac{\mathrm{TP}_K}{\mathrm{P}}.$$

- **F1-score@$K$**: the harmonic mean of Precision@$K$ and Recall@$K$,

$$\text{F1}@K = 2 \cdot \frac{\text{Precision}@K \cdot \text{Recall}@K}{\text{Precision}@K + \text{Recall}@K}.$$

**AUC-Based Metrics.** To provide a comprehensive assessment across all possible thresholds, we compute:

- **Area Under the Precision-Recall Curve (AUPRC)** summarizes the precision–recall trade-off. It is particularly informative under class imbalance, where driver genes are rare.
- **Area Under the Receiver Operating Characteristic Curve (AUROC)** measures the ability to distinguish driver and non-driver genes.
- **Partial AUPRC (pAUPRC)** restricts the area under the precision-recall curve to the low-recall region (specifically, Recall $\leq 0.1$). This metric emphasizes precision among the most highly ranked, high-confidence candidate genes. To compute this precisely, we use linear interpolation to determine the precision at the recall cutoff and calculate the area using trapezoidal integration.

- **Partial AUROC (pAUROC)** represents the area under the ROC curve constrained to a low false positive rate (FPR). Specifically, we restrict the specificity to $\geq 0.90$ (which equates to FPR $\leq 0.1$). This enforces a stringent evaluation, focusing strictly on the model's ability to limit false alarms among the top-ranked predictions.

For the partial metrics, rather than evaluating the entire prediction spectrum, we focus strictly on these conservative thresholds (Recall $\leq 0.1$ and FPR $\leq 0.1$). This approach rigorously assesses the model's ability to prioritize the most likely driver genes with minimal false positives—aligning with the practical use case where only a limited number of top candidates are selected for experimental validation.

**Personalized Evaluation Strategy.** For personalized driver gene identification, we employ the REA strategy. Following this approach, we evaluate only samples containing at least three reference positives (i.e., $|\mathcal{M}_i \cap \text{CGC Tier 1}| \geq 3$). For each cancer type $c$, we evaluate the top $N_c$ predictions for each sample, where:

$$N_c = 2 \cdot \text{median}\{|\mathcal{M}_i \cap \text{CGC Tier 1}| : i \in \text{samples of type } c\}.$$

If a sample has fewer than $N_c$ mutated genes, i.e., $|\mathcal{M}_i| < N_c$, it is excluded when computing metrics for $K > |\mathcal{M}_i|$. We aggregate the evaluation metrics (Precision@$K$, Recall@$K$, and F1@$K$ for various $K$ values) across all evaluated samples by computing their mean to obtain the overall performance for personalized driver gene identification.

**Cohort-Level Evaluation Strategy.** For cohort-level evaluation, we rank all mutated genes in each cancer type $c$. The reference set is defined as the intersection of all mutated genes in that cancer type and CGC Tier 1. We compute Precision@$K$, Recall@$K$, and F1@$K$ for multiple values of $K$. We also report AUPRC, AUROC, pAUPRC, and pAUROC.

**Note S2: Parameter Sensitivity Analysis**

To assess whether the performance of HyperNetWalk depended sensitively on specific parameter choices, we performed a one-factor-at-a-time sensitivity analysis for five key parameters: the number of DEGs used for Stage I initialization, the structure parameter $\alpha$ controlling the transition probability from Layer 2 to Layer 1 during reverse propagation, the similarity bandwidth parameter $\sigma$, the Stage I restart probability $r$, and the Stage II hypergraph diffusion restart probability $\gamma$. For each analysis, one parameter was varied while all other parameters were fixed at their default values used in the main analysis.

The tested values were selected to cover a broad but interpretable range around the default settings. For the number of DEGs, we evaluated 100, 250, 500, 750, and 1000 DEGs, with 500 used as the default. For $\alpha$, we evaluated 0.2, 0.5, and 0.8 with 0.5 as the default, covering a range from more conservative (favoring Layer 2) to more aggressive (favoring Layer 1) transition probabilities. For $\sigma$, we evaluated 0.05, 0.1, 0.3, 0.5, 1, and 2, with 0.1 used as the default. For the restart probabilities $r$ and $\gamma$, we evaluated 0.05, 0.15, 0.3, 0.5, and 0.8, covering a range from low to high restart rates, with 0.15 as the default.

Performance was evaluated using the same primary metrics as in the main benchmark: personalized Precision@$K$ with $K = 6$ and cohort-level Precision@$K$ with $K = 100$. Across the tested ranges, HyperNetWalk maintained stable performance for both personalized and cohort-level prediction (Supplementary Fig. S25). The default parameter values generally lay within high-performing regions rather than isolated optima. In particular, performance was relatively stable with respect to the number of DEGs, $\alpha$, and $\sigma$, indicating that the method is not strongly dependent on a narrowly tuned choice of these parameters. For the restart probabilities, particularly for Stage I restart $r$, there was a more noticeable decline in performance at the highest tested values (0.5 and 0.8), suggesting that excessive restart can hinder effective propagation of information through the network. However, the default value of 0.15 was well within a stable performance region, and moderate increases to 0.3 still maintained reasonable performance. Overall, these results indicate that the benchmark performance of HyperNetWalk is robust to moderate parameter perturbations and is not driven by fine-tuning of individual parameters.

### Note S3: Reference-Free Assessment of Cancer-Type Specificity

Because IntOGen annotations may still reflect discovery bias, we further evaluated cancer-type specificity without relying on external driver references. If a method's predictions are truly cancer-type-specific, we would expect low overlap between the top-ranked predictions for different cancer types. Conversely, high overlap may indicate that the method is primarily capturing general properties of mutated genes rather than cancer-specific drivers.

At the cohort level, we computed Jaccard indices between the top 100 predicted driver genes for each pair of cancer types. At the personalized level, we first took the union of the top $N_c$ predictions across all samples within each cancer type $c$, and then computed Jaccard indices between these cancer-level unions for different cancer types. Since DriverMP and DriverRWH do not produce personalized predictions, we only evaluated their cohort-level Jaccard indices. In both cases, HyperNetWalk achieved low cross-cancer overlap for most cancer pairs, indicating that its predictions remain distinct across cancer types (Supplementary Fig. S15). The highest similarity was observed between LUAD and LUSC, a pattern shared by baseline methods and likely reflecting their common tissue origin and partially overlapping molecular landscapes.

### Note S4: Identification of Cancer-Type-Specific TFs and Downstream Target-Gene Programs

To assess whether TF-layer representations captured interpretable cancer-type-specific regulatory programs, we use the Stage I stationary probabilities assigned to TF nodes as sample-level TF scores. For each TF $g$ and cancer type $c$, we compute a one-versus-rest specificity score:

$$\widetilde{z}_{gc} = \frac{\widetilde{\mu}_{gc} - \widetilde{\mu}_{g\neg c}}{\widetilde{\sigma}_g},$$

where $\widetilde{\mu}_{gc}$ is the mean Stage I probability of TF $g$ across samples of cancer type $c$, $\widetilde{\mu}_{g\neg c}$ is the corresponding mean across samples from all other cancer types, and $\widetilde{\sigma}_g$ is the standard deviation across all samples. TFs with $\widetilde{z}_{gc} > 1$ are defined as cancer-type-specific TFs for cancer type $c$, indicating that their Stage I probabilities are positively enriched in that cancer type compared to others. Then, for each cancer type, downstream target-gene programs are constructed as the union of direct GRN targets of the corresponding cancer-type-specific TFs.

### Note S5: Functional Enrichment Analysis

Functional enrichment analysis is performed for a given query gene set against a specified background universe of genes, using MSigDB Hallmark gene sets (msigdbr, *Homo Sapiens*) and over-representation analysis (ORA) with clusterProfiler::enricher. Statistical significance is assessed by the Benjamini-Hochberg FDR. Pathways with FDR $< 0.05$ are considered significantly enriched.

For cohort-level driver predictions, enrichment analysis is performed separately for each cancer type, with the top 100 HyperNetWalk-ranked genes as the query set and all genes ranked by HyperNetWalk for the corresponding cancer type as the background. Enrichment results are summarized by gene ratio and $-\log_{10}(\mathrm{FDR})$. For the TF-layer regulatory analysis, the query set is the downstream target program of each cancer type's cancer-type-specific TFs, with all GRN target genes as the background.

### Note S6: Evidence-Level Annotation of HyperNetWalk-Specific Predictions

To assess whether genes uniquely prioritized by HyperNetWalk have prior cancer-related support, we annotated HyperNetWalk-specific predictions using external cancer knowledge resources. HyperNetWalk-specific genes were defined separately for the cohort-level and personalized analyses. At the cohort level, a HyperNetWalk-specific gene was defined as a gene present in the HyperNetWalk top 100 predictions but absent from the corresponding baseline prediction sets. At the personalized level, a HyperNetWalk-specific gene was defined as a gene present in the top 5 predictions for at least one sample of cancer type $c$ by HyperNetWalk, but absent from the corresponding baseline prediction sets for all samples of that cancer type. The resulting HyperNetWalk-specific genes were then annotated using CGC, NCG, and CancerMine.

Genes were assigned to four mutually exclusive evidence levels according to the strength of their external support. Level A included genes annotated in CGC Tier 1 for the matched cancer type and therefore represented the strongest evidence category. Level B included genes annotated in CGC Tier 1 for other cancer types, as well as genes annotated for the matched cancer type in CGC Tier 2, NCG, or CancerMine. Level C included genes present in CGC Tier 2, NCG, or CancerMine but annotated only for cancer types other than the matched cancer type. Level D included genes without support from any of these resources.

When a gene satisfied multiple criteria, the highest applicable evidence level was assigned according to the hierarchy Level A (>) Level B (>) Level C (>) Level D. This annotation scheme was used only to summarize the degree of prior support for HyperNetWalk-specific predictions; it was not used during model training or driver ranking. Across both cohort-level and personalized predictions, a substantial fraction of HyperNetWalk-specific predictions fell into levels B and C (Supplementary Tables S2 and S3), indicating that many of these additional candidates have prior cancer relevance despite not being highly ranked by competing methods.

### Note S7: Scalability and Runtime Analysis

The computational cost of HyperNetWalk scales favorably with the number of samples because the most computationally intensive sample-specific steps are directly parallelizable. In Stage I and the personalized component of Stage II, samples are processed independently, allowing these steps to be distributed across parallel workers. For cohort-level hypergraph refinement, the number of vertices is determined by the union of mutated genes within each cancer type, whereas the number of hyperedges corresponds to the number of samples. The cost of constructing the hypergraph transition matrix therefore depends on the number of samples, the size of the mutated-gene union, and the sparsity of the sample–gene incidence structure. Once the transition matrix has been constructed, the stationary distribution computation is governed primarily by the size and sparsity of this matrix.

In our implementation, a complete run across the 12 TCGA cancer types analyzed in this study required 2,447 seconds of wall-clock time, approximately 41 minutes, using 50 parallel CPU workers on a Linux server equipped with two AMD EPYC 7763 64-core processors and 2.0 TiB of RAM.

# Supplementary Figures

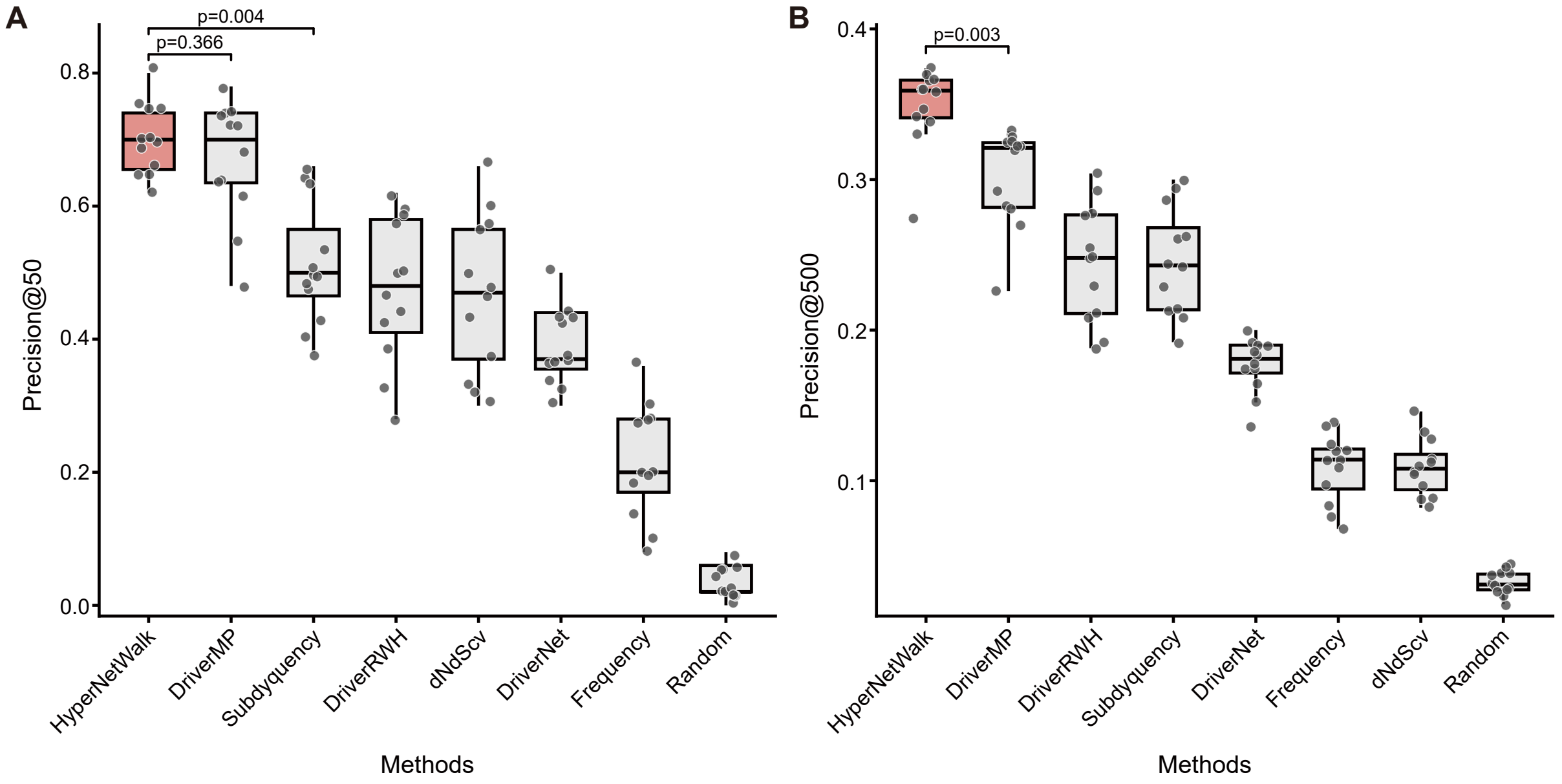

**Figure S1. Cohort-level Precision@$K$ ($K$ = 50 and $K$ = 500) of HyperNetWalk and competing baseline methods. A.** Boxplots of Precision@50 across 12 TCGA cancer types for cohort-level predictions. **B.** Boxplots of Precision@500 across 12 TCGA cancer types for cohort-level predictions. Each point represents one cancer type. P-values were calculated using two-sided paired Wilcoxon signed-rank tests, with cancer type as the paired unit.

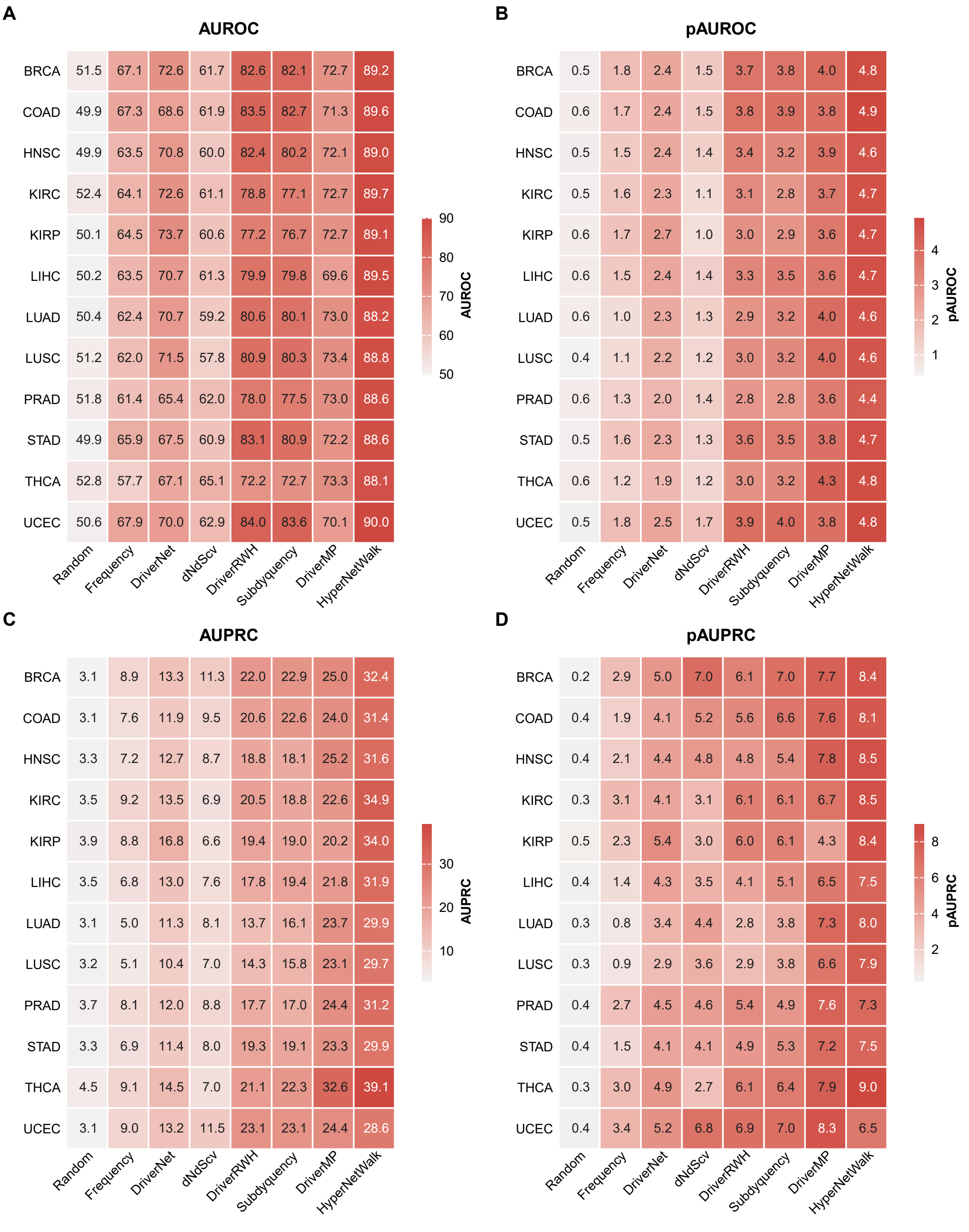

**Figure S2. Cohort-level AUC-based metrics across 12 cancer types.** Heatmaps of **A.** AUROC, **B.** pAUROC, **C.** AUPRC, and **D.** pAUPRC values. Columns represent cohort-level methods, and rows represent cancer types. For each cancer type, the best-performing method is highlighted.

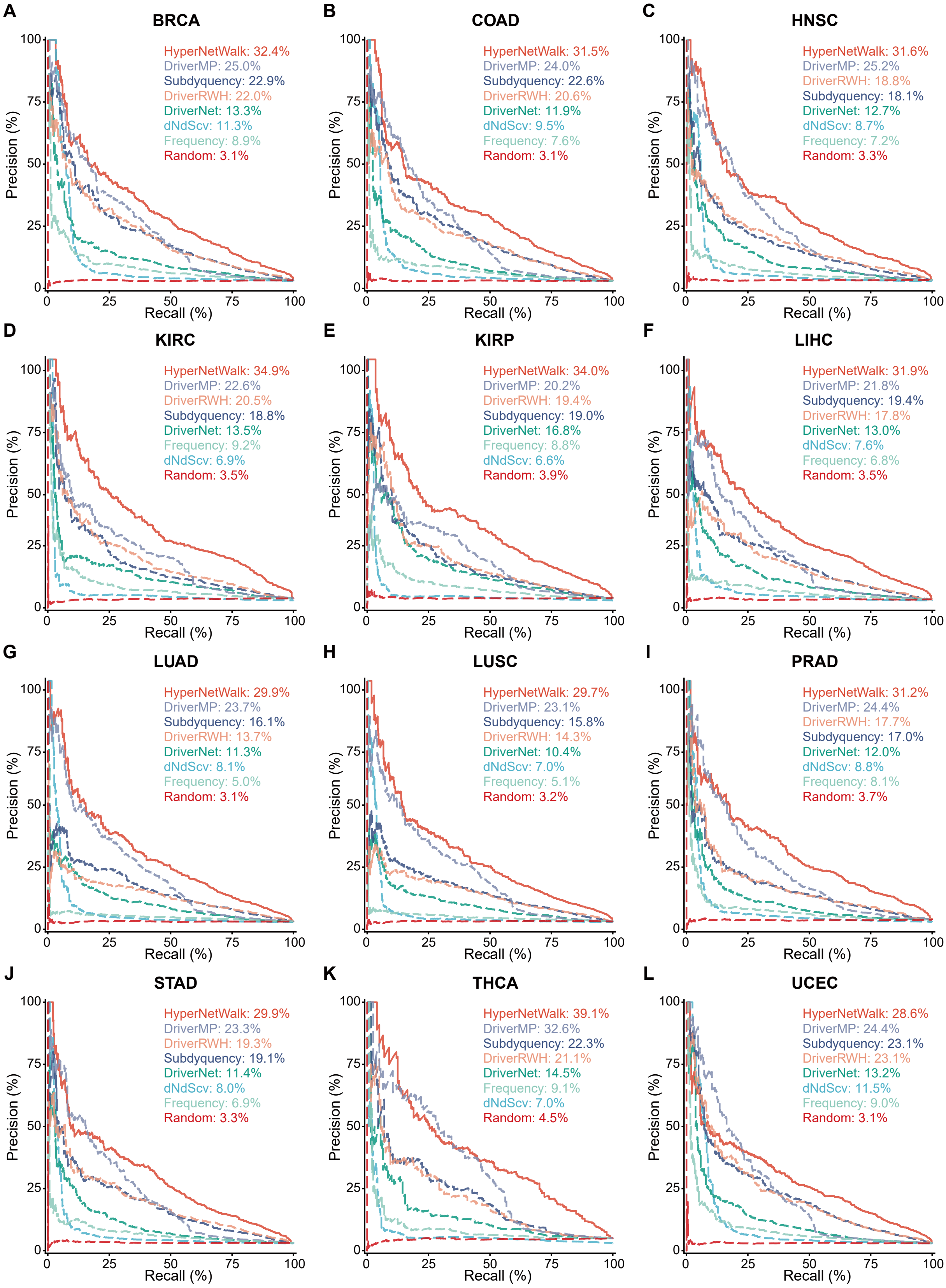

**Figure S3. Cohort-level PR curves across 12 cancer types.** Precision-recall (PR) curves comparing HyperNetWalk with baseline methods in cohort-level driver gene identification across 12 TCGA cancer types. Panels (**A–L**) correspond to the 12 cancer types. AUPRC values are shown in the legends.

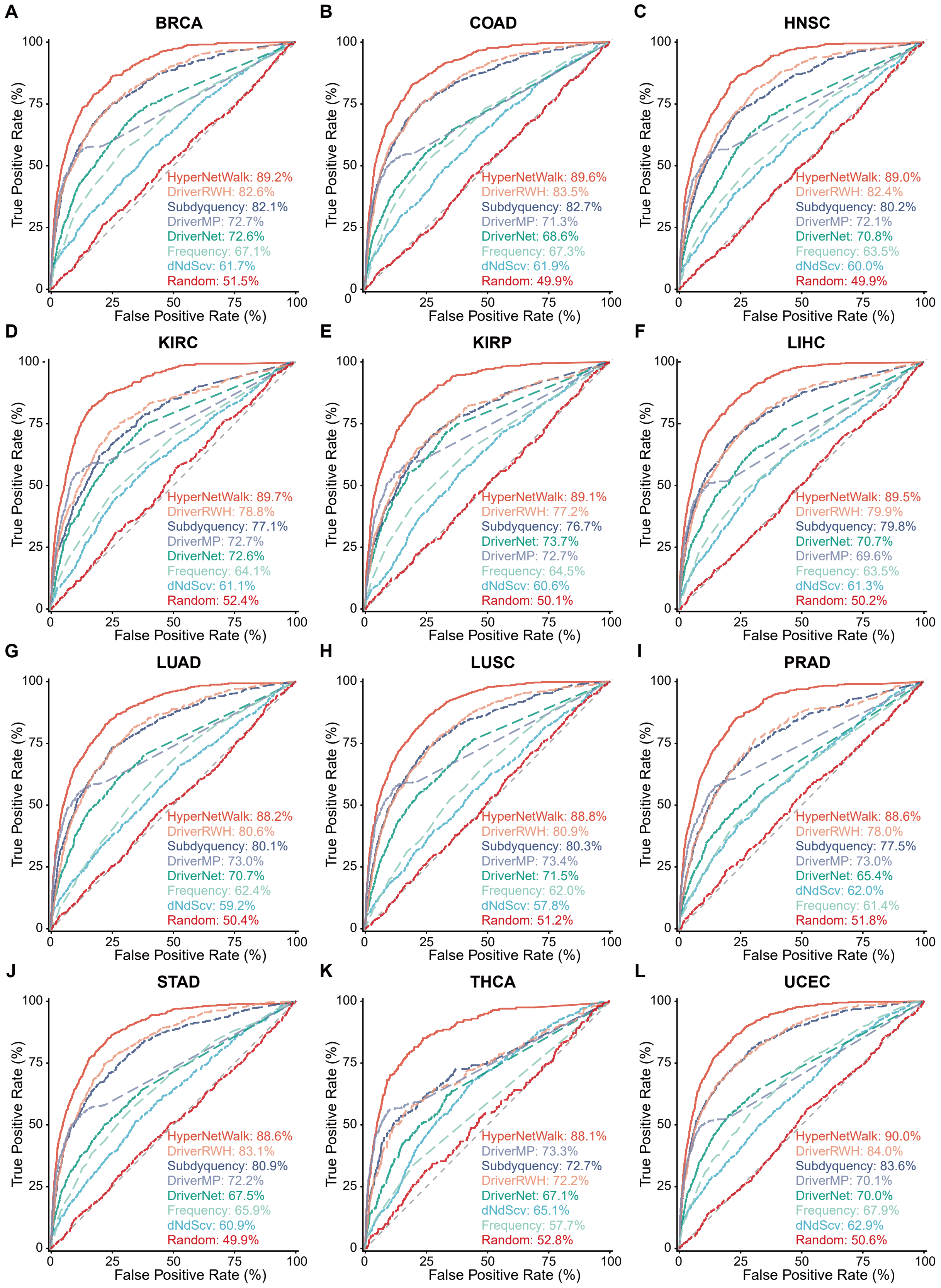

**Figure S4. Cohort-level ROC curves across 12 cancer types.** Receiver operating characteristic (ROC) curves comparing HyperNetWalk with baseline methods in cohort-level driver gene identification across 12 TCGA cancer types. Panels (**A–L**) correspond to the 12 cancer types. AUROC values are shown in the legends.

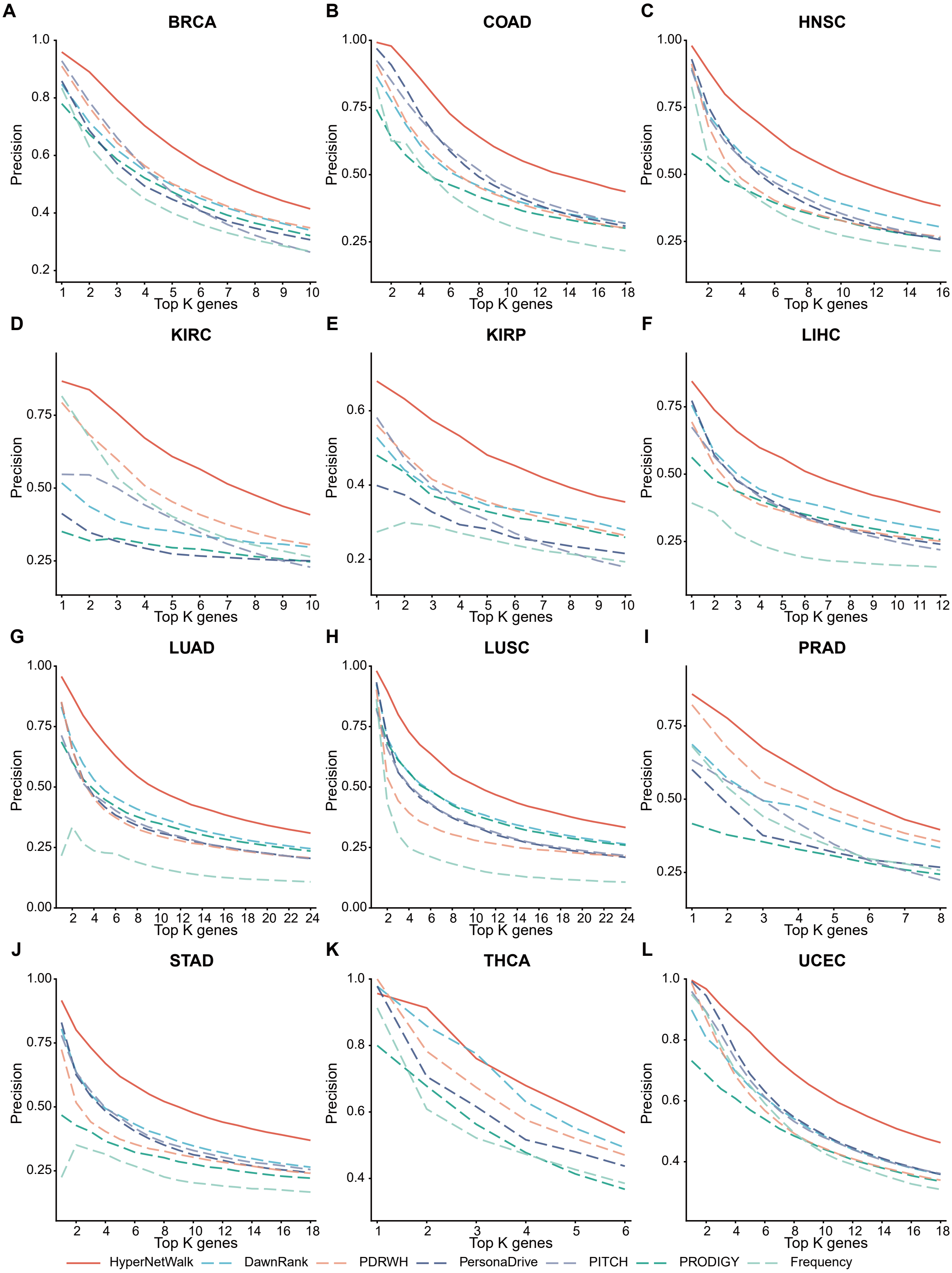

**Figure S5. Personalized Precision@$K$ curves across 12 cancer types.** Precision@$K$ curves comparing HyperNetWalk with baseline methods in personalized driver gene identification across 12 TCGA cancer types. Panels (**A–L**) correspond to the 12 cancer types.

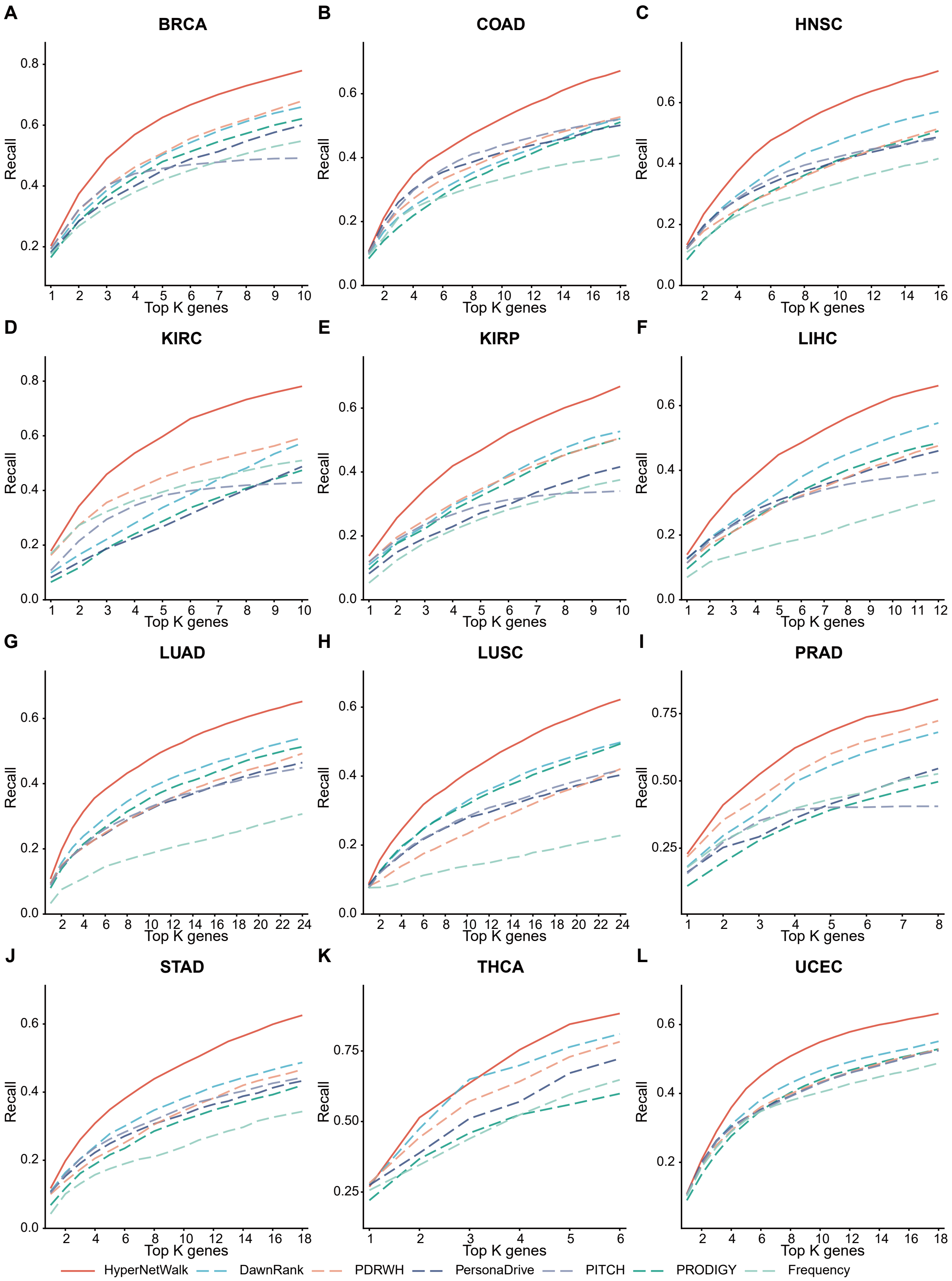

**Figure S6. Personalized Recall@$K$ curves across 12 cancer types.** Recall@$K$ curves comparing HyperNetWalk with baseline methods in personalized driver gene identification across 12 TCGA cancer types. Panels (**A–L**) correspond to the 12 cancer types.

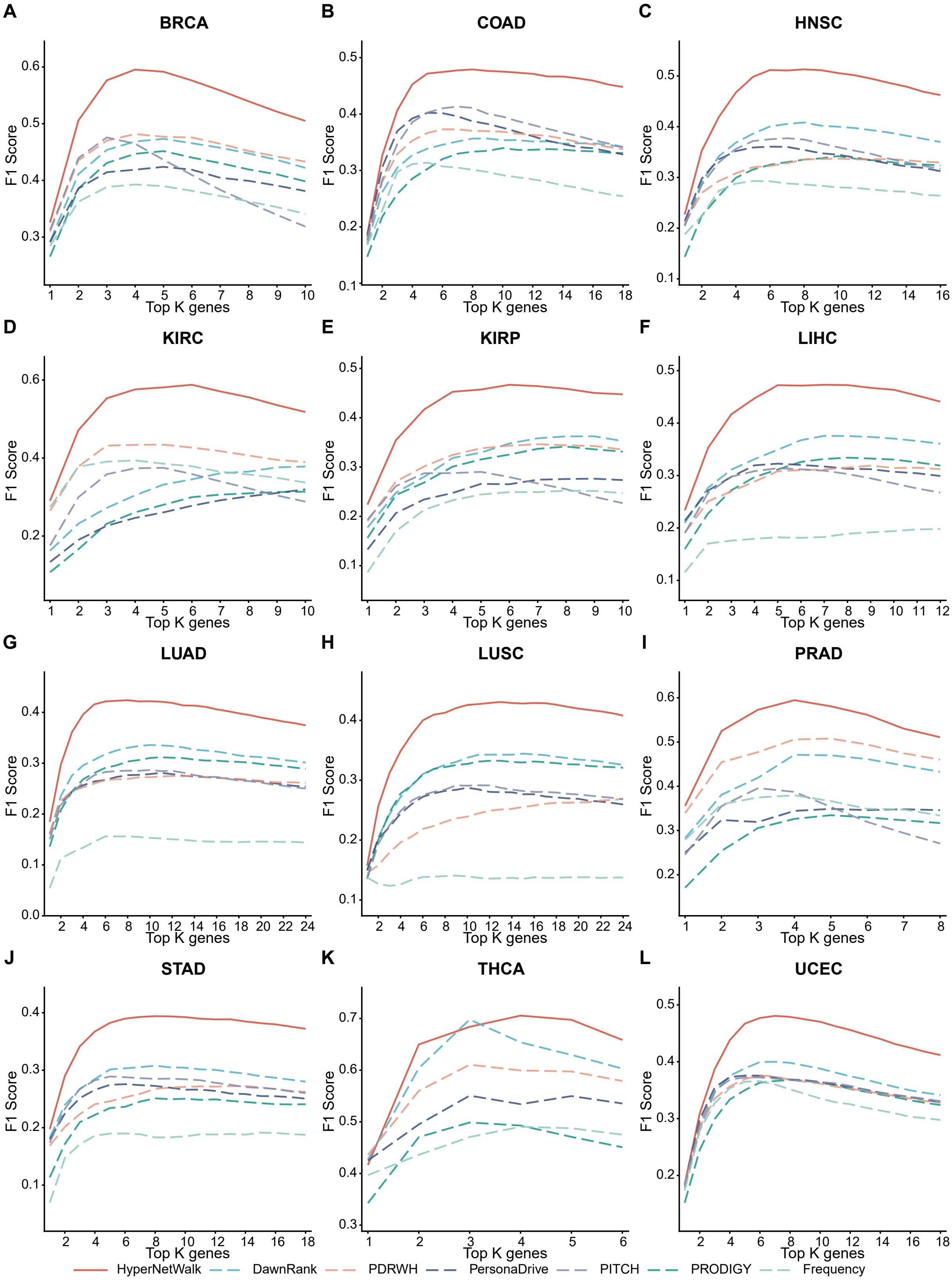

**Figure S7. Personalized F1-score@$K$ curves across 12 cancer types.** F1-score@$K$ curves comparing HyperNetWalk with baseline methods in personalized driver gene identification across 12 TCGA cancer types. Panels (**A–L**) correspond to the 12 cancer types.

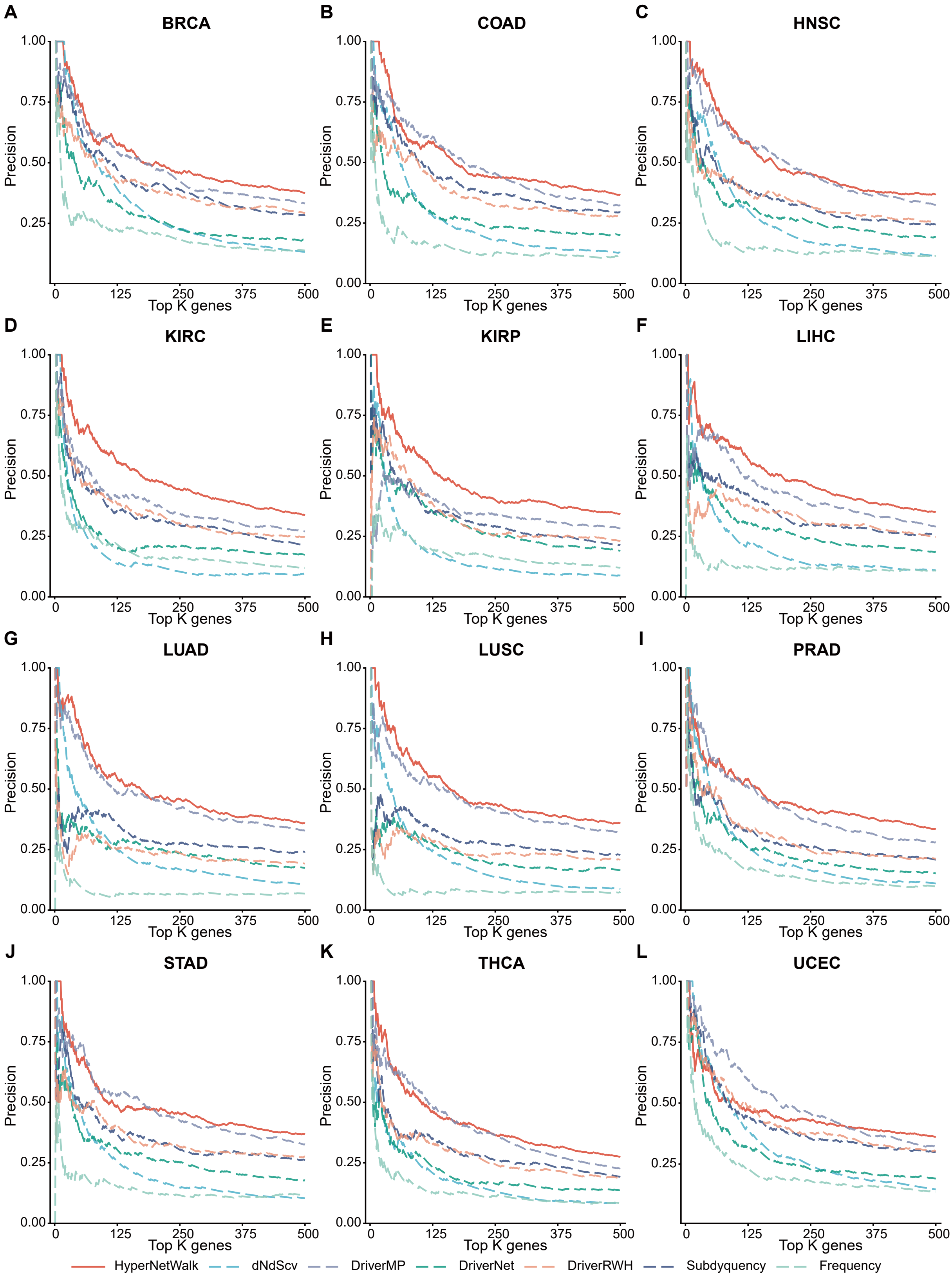

**Figure S8. Cohort-level Precision@$K$ curves across 12 cancer types.** Precision@$K$ curves comparing HyperNetWalk with baseline methods in cohort-level driver gene identification across 12 TCGA cancer types. Panels (**A–L**) correspond to the 12 cancer types.

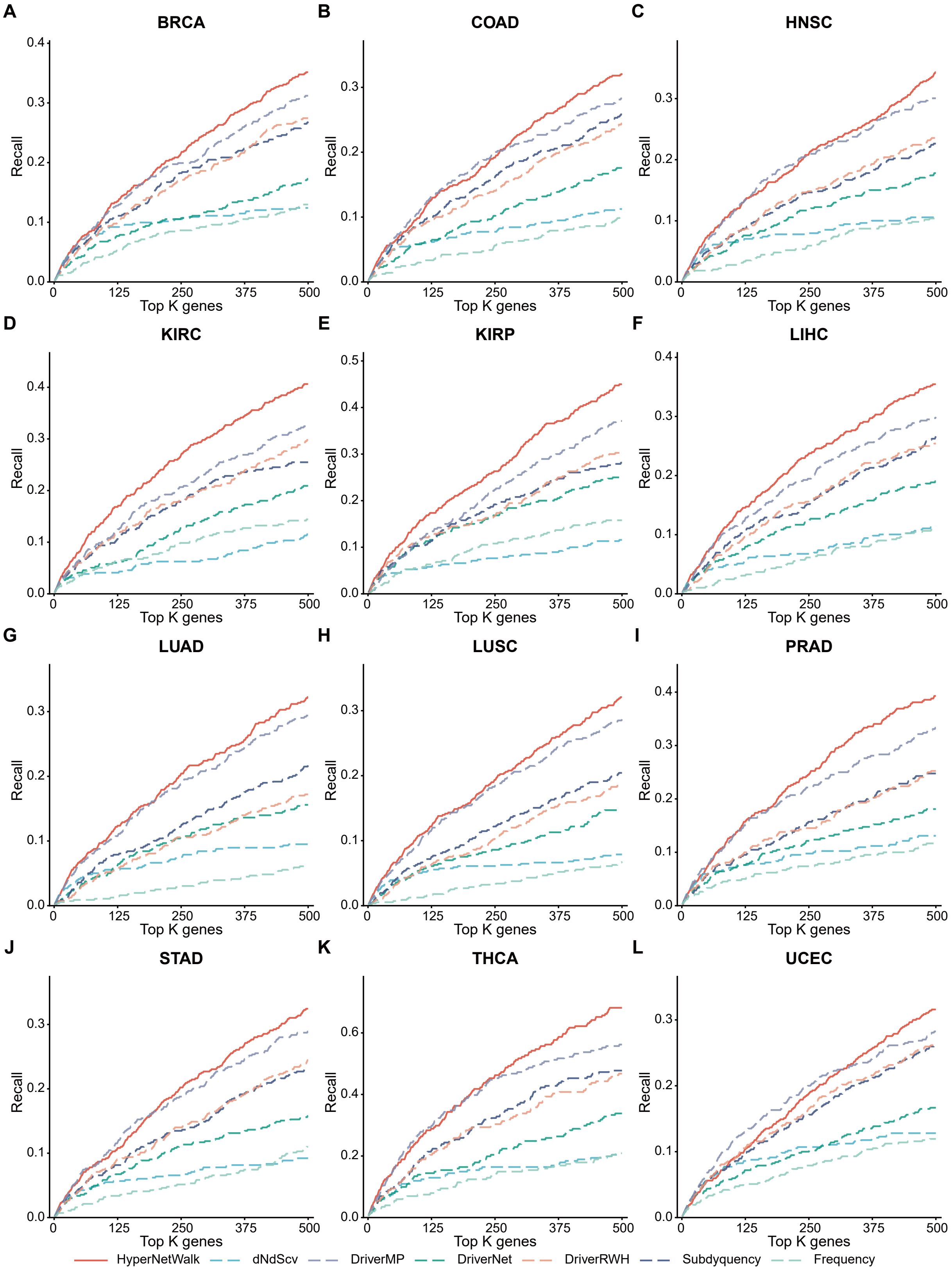

**Figure S9. Cohort-level Recall@$K$ curves across 12 cancer types.** Recall@$K$ curves comparing HyperNetWalk with baseline methods in cohort-level driver gene identification across 12 TCGA cancer types. Panels (**A–L**) correspond to the 12 cancer types.

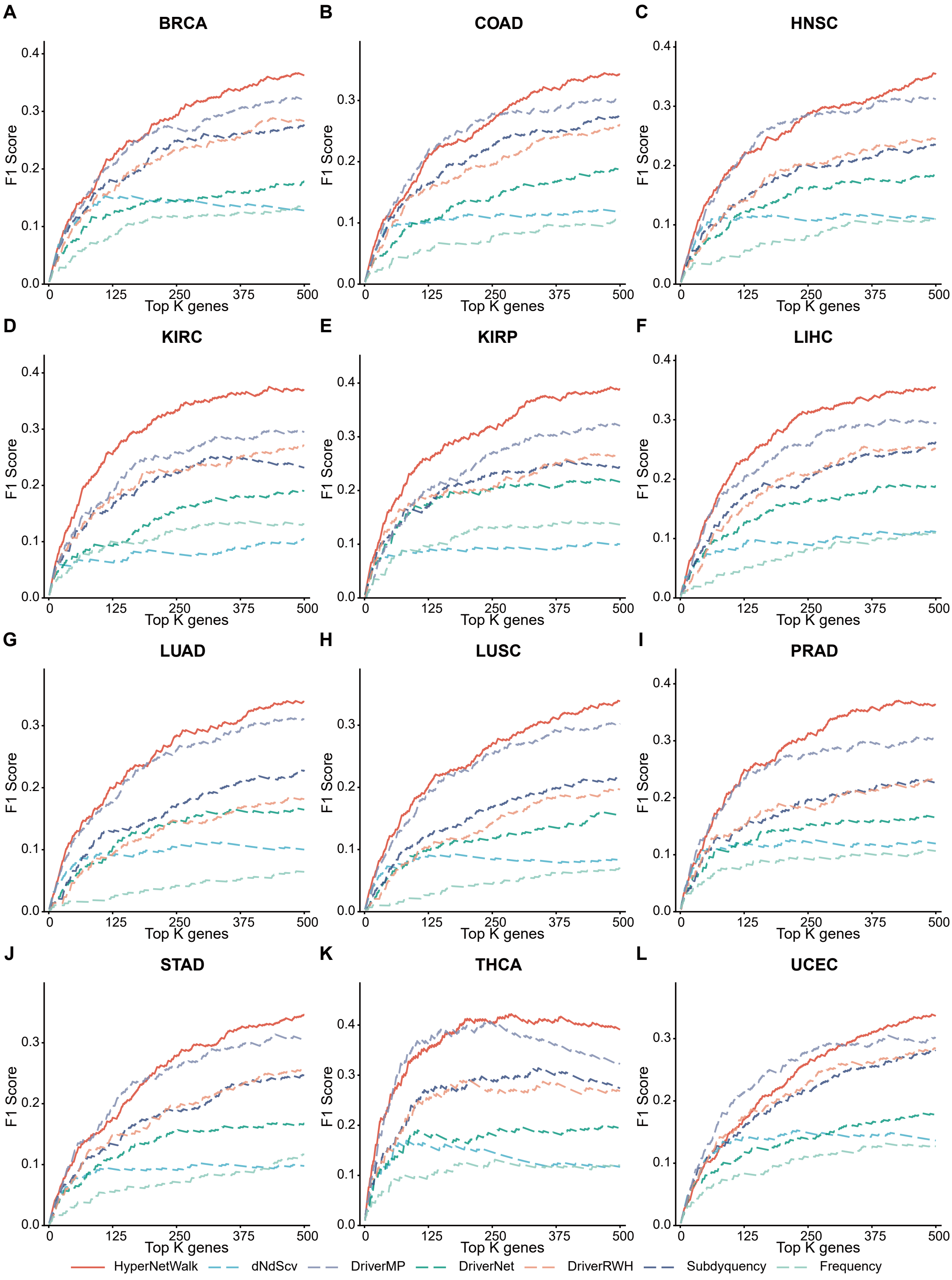

**Figure S10. Cohort-level F1-score@$K$ curves across 12 cancer types.** F1-score@$K$ curves comparing HyperNetWalk with baseline methods in cohort-level driver gene identification across 12 TCGA cancer types. Panels (**A–L**) correspond to the 12 cancer types.

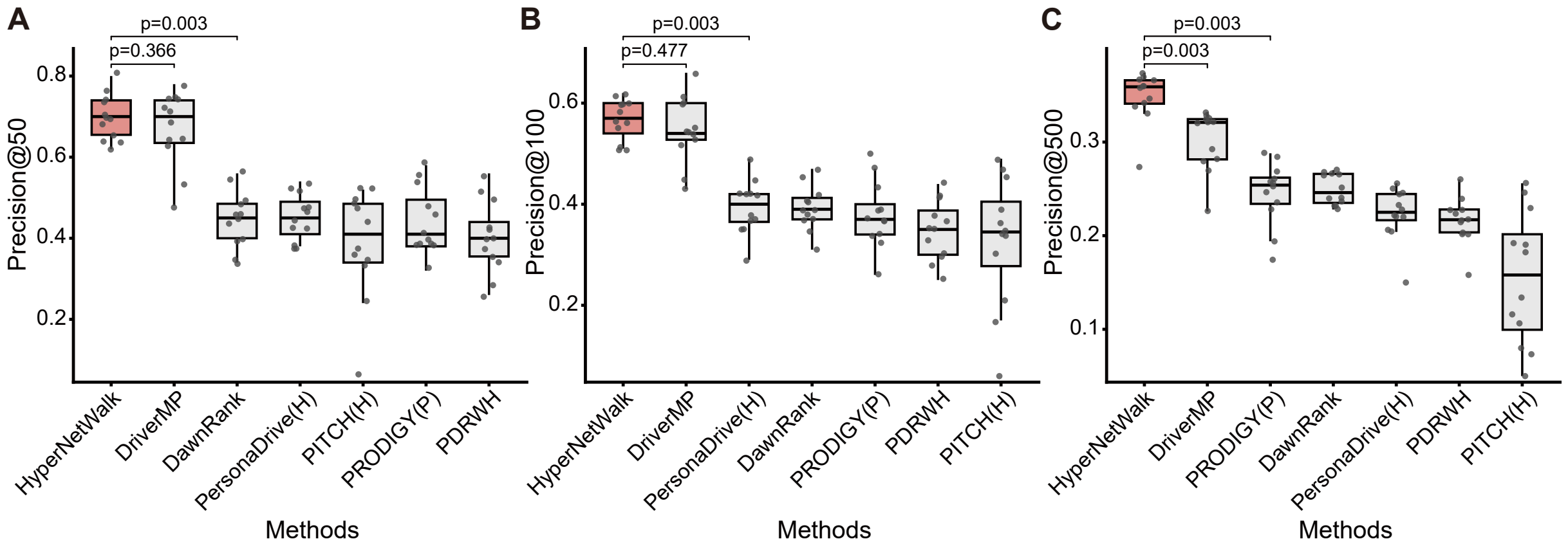

**Figure S11. Cohort-level Precision@$K$ of aggregated personalized methods across 12 cancer types.** Cohort-level rankings are obtained by aggregating personalized predictions from existing methods. For methods with pre-defined aggregation procedures (e.g., DawnRank and PDRWH), the original strategies are used. For other methods, two general aggregation strategies are applied: a pairwise voting-based approach proposed in PDRWH (suffix (P)) and the hypergraph-based aggregation used in HyperNetWalk (suffix (H)). For each personalized method, the aggregation strategy yielding better performance is retained. HyperNetWalk is evaluated using its native cohort-level predictions without additional aggregation. DriverMP is included as a representative strong cohort-level baseline for reference. Each point represents one cancer type. P-values were calculated using two-sided paired Wilcoxon signed-rank tests, with cancer type as the paired unit. Because the paired Wilcoxon signed-rank test is rank-based, identical p-values can occur when HyperNetWalk shows consistently higher Precision@$K$ across all cancer types, even if the magnitude of improvement differs among baselines. **A.** $K = 50$. **B.** $K = 100$. **C.** $K = 500$.

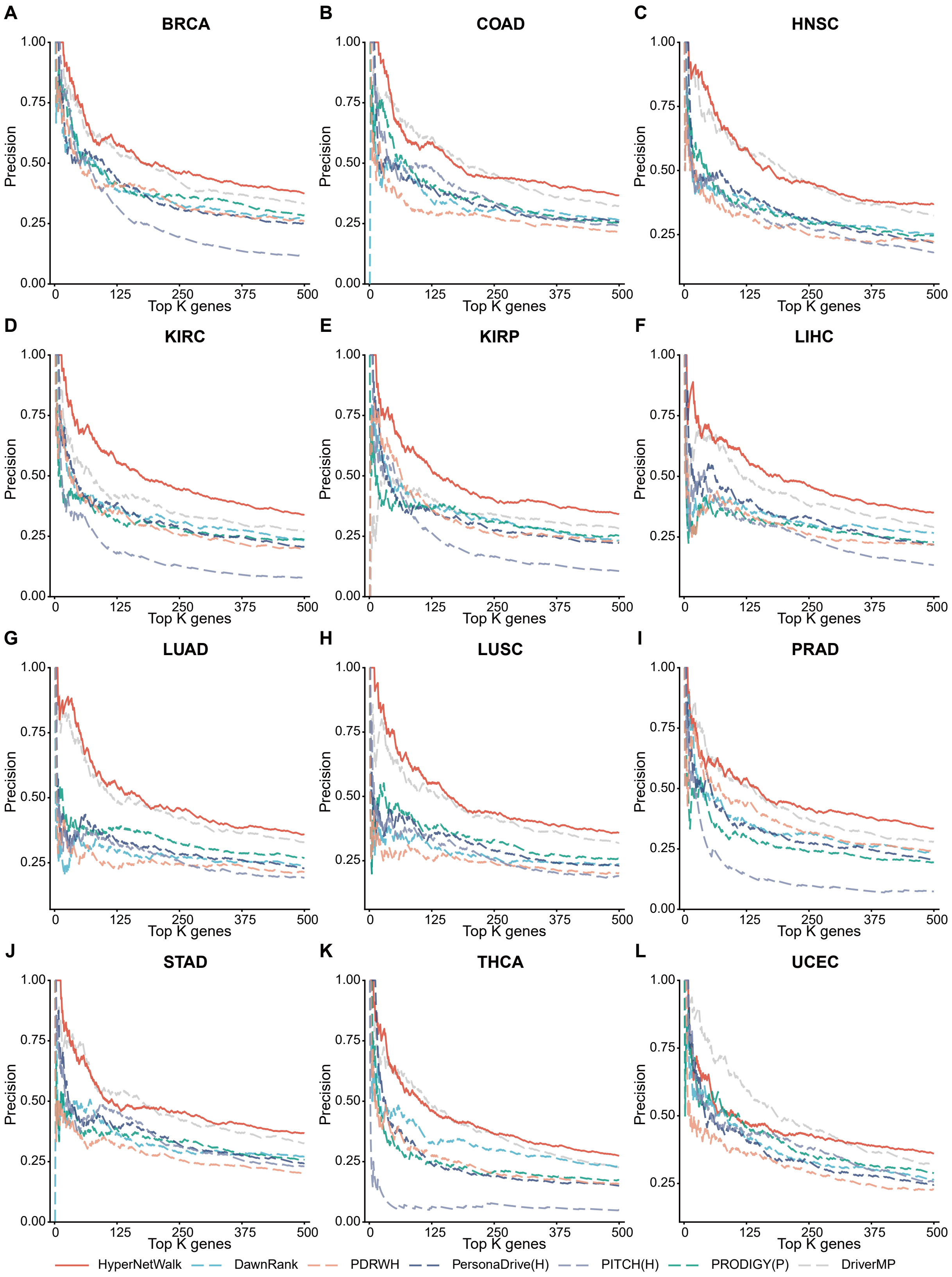

**Figure S12. Precision@$K$ curves of aggregated personalized methods across 12 cancer types.** Cohort-level rankings are obtained by aggregating personalized predictions as described in Supplementary Fig. S11. Panels **A–L** correspond to the 12 cancer types. HyperNetWalk is evaluated using its native cohort-level predictions, and DriverMP is included as a representative strong cohort-level baseline for reference.

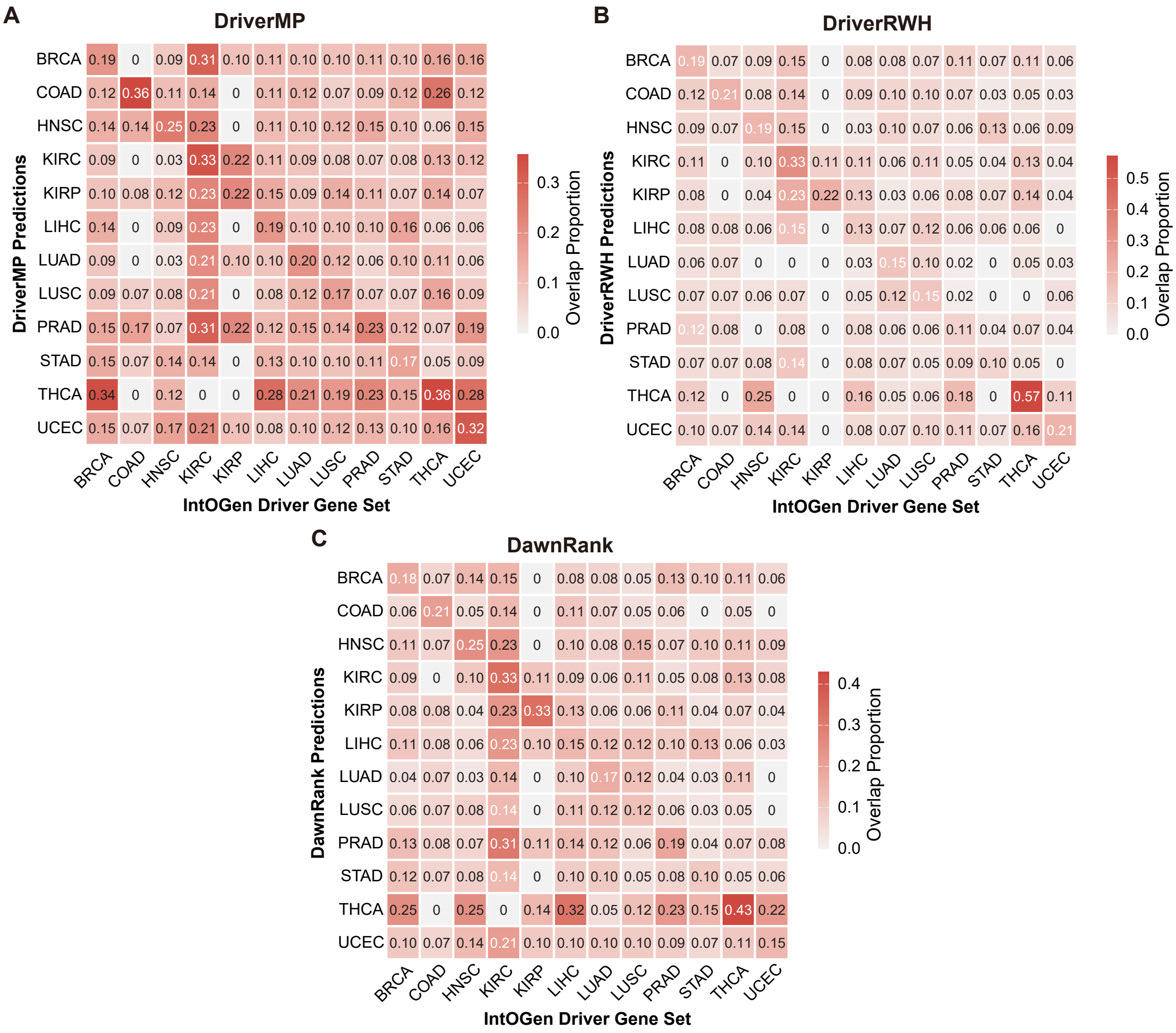

**Figure S13. Cross-cancer specificity of predicted driver genes.** Heatmap showing the overlap proportion between the top 100 cohort-level predictions from DriverMP (**A**), DriverRWH (**B**), DawnRank (**C**) and cancer-type-specific driver gene sets derived from IntOGen. Rows indicate the predicted cancer type and columns indicate the reference cancer type. Each value represents the proportion of reference drivers recovered among the top-ranked predictions.

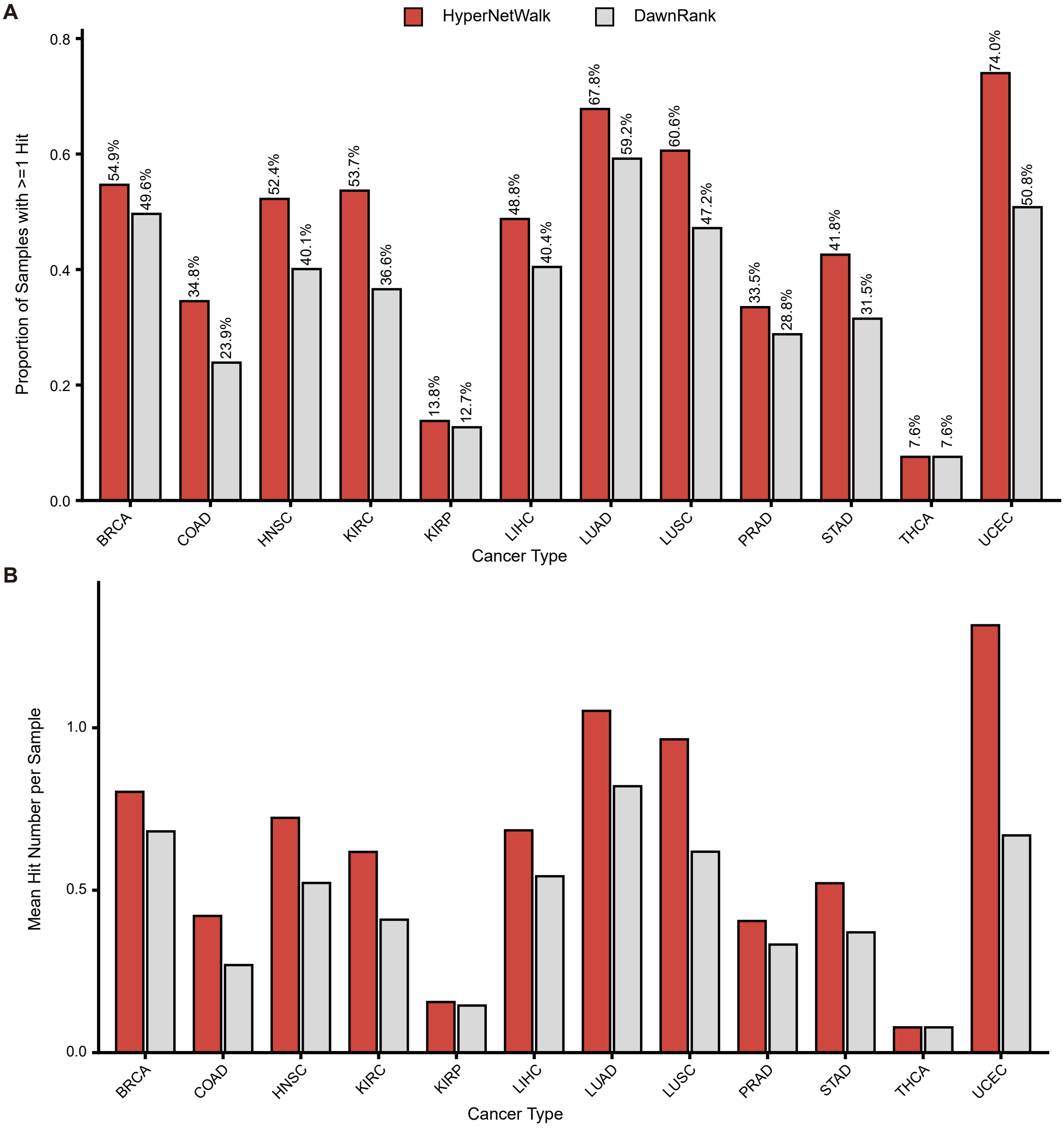

**Figure S14. Cross-cancer specificity of predicted personalized driver genes. A.** Bar plot showing the proportion of samples with at least one cancer-type-specific driver gene (from IntOGen) in the top $N_c$ personalized predictions from HyperNetWalk and DawnRank across 12 cancer types. **B.** Bar plot showing the mean number of cancer-type-specific driver genes (from IntOGen) in the top $N_c$ personalized predictions from HyperNetWalk and DawnRank across 12 cancer types.

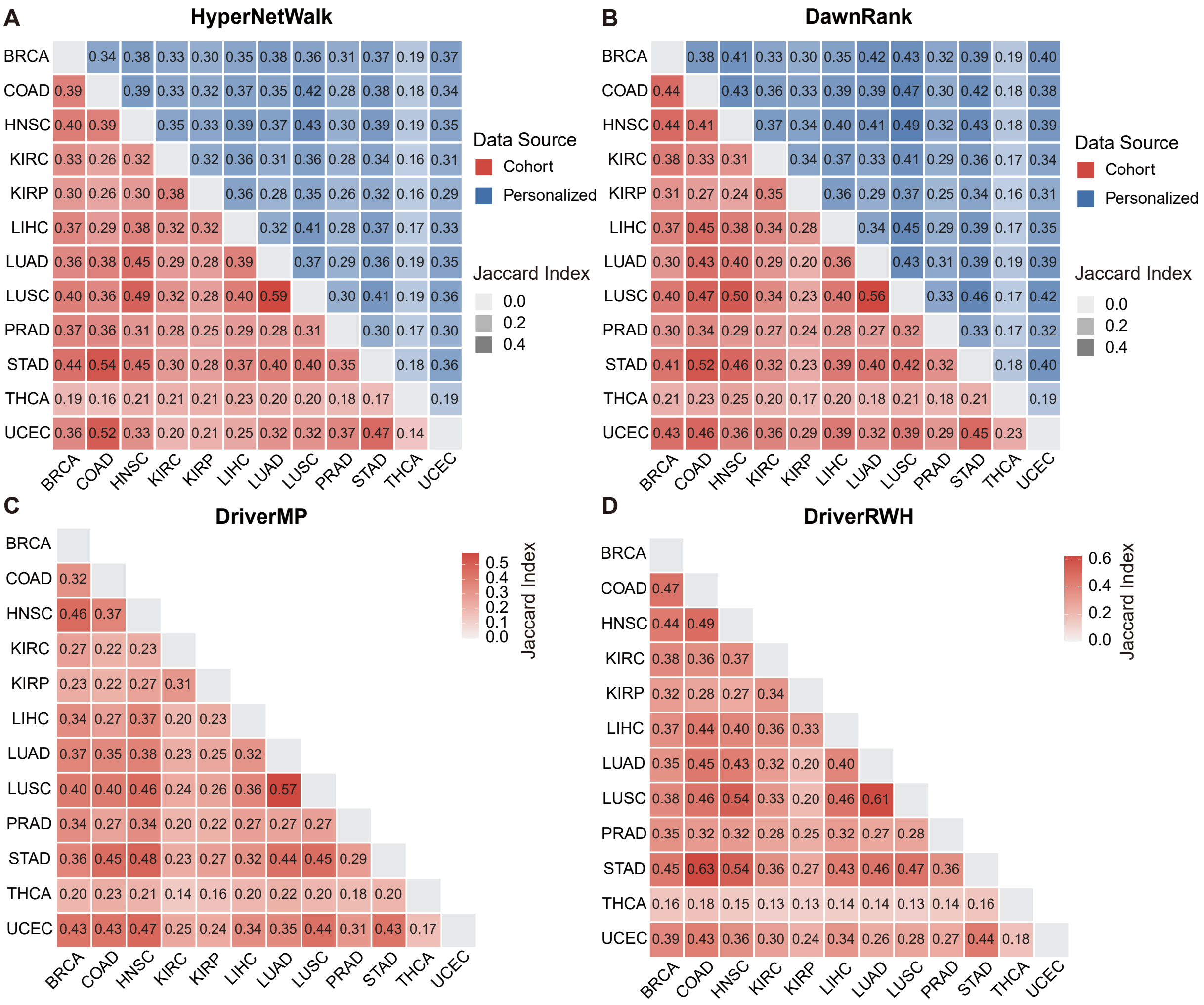

**Figure S15. Cross-cancer Jaccard similarity of predicted driver genes.** Heatmap showing the Jaccard similarity between the top 100 cohort-level predictions of each cancer type (red) and the Jaccard similarity between the union of top $N_c$ personalized predictions across samples of each cancer type (blue) for HyperNetWalk (**A**), DawnRank (**B**), DriverMP (**C**), and DriverRWH (**D**), where $N_c$ denotes the cancer-type-specific cutoff used for personalized evaluation. The diagonal elements are masked to focus on cross-cancer similarity. Since DriverMP and DriverRWH do not produce personalized predictions, only the cohort-level Jaccard similarity is shown for these methods.

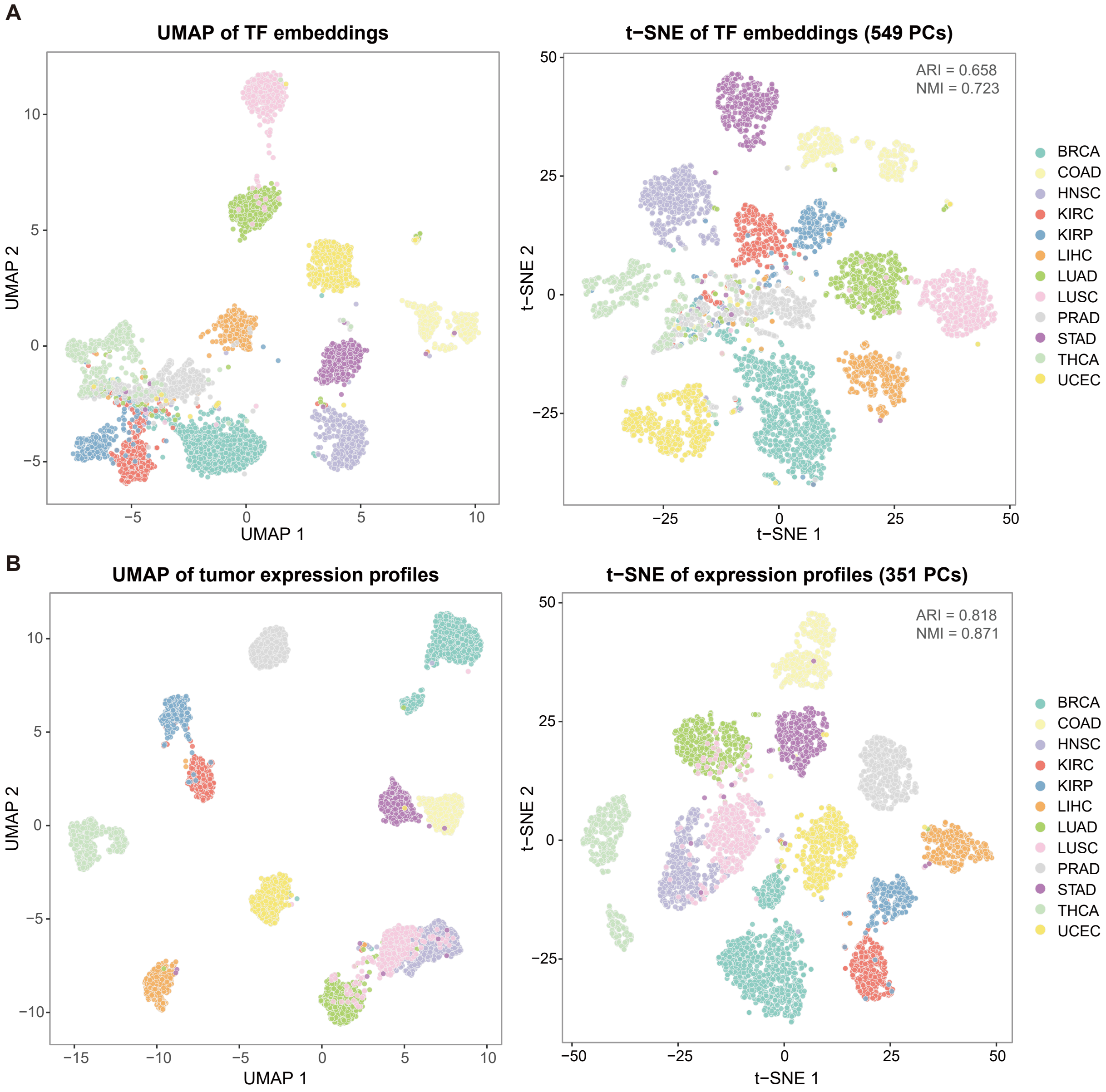

**Figure S16. Visualization of TF-layer representations and expression profiles. A.** UMAP (left) and t-SNE (right) visualizations of sample-level TF-layer probability vectors derived from HyperNetWalk Stage I reverse inference. **B.** UMAP (left) and t-SNE (right) visualizations of tumor expression profiles across 12 cancer types. Each point represents one sample and is colored by cancer type; ARI and NMI values are shown in the legends.

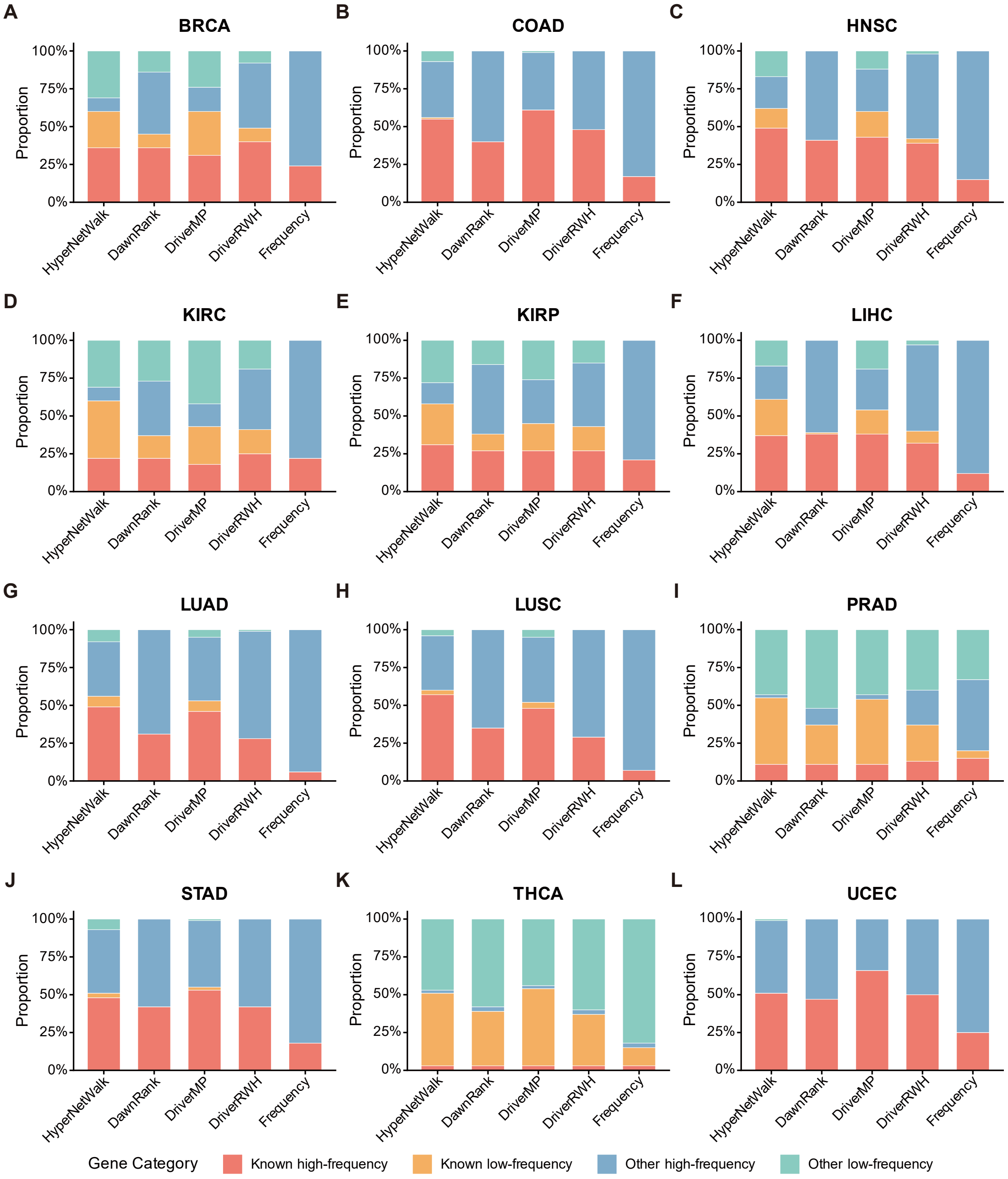

**Figure S17. Composition of cohort-level top 100 predictions across 12 cancer types.** Stacked bar plots of the cohort-level top 100 predictions in 12 cancer types (**A–L**). Predicted genes are grouped into four categories: known high-frequency drivers, known low-frequency drivers, other high-frequency mutated genes, and other low-frequency mutated genes.

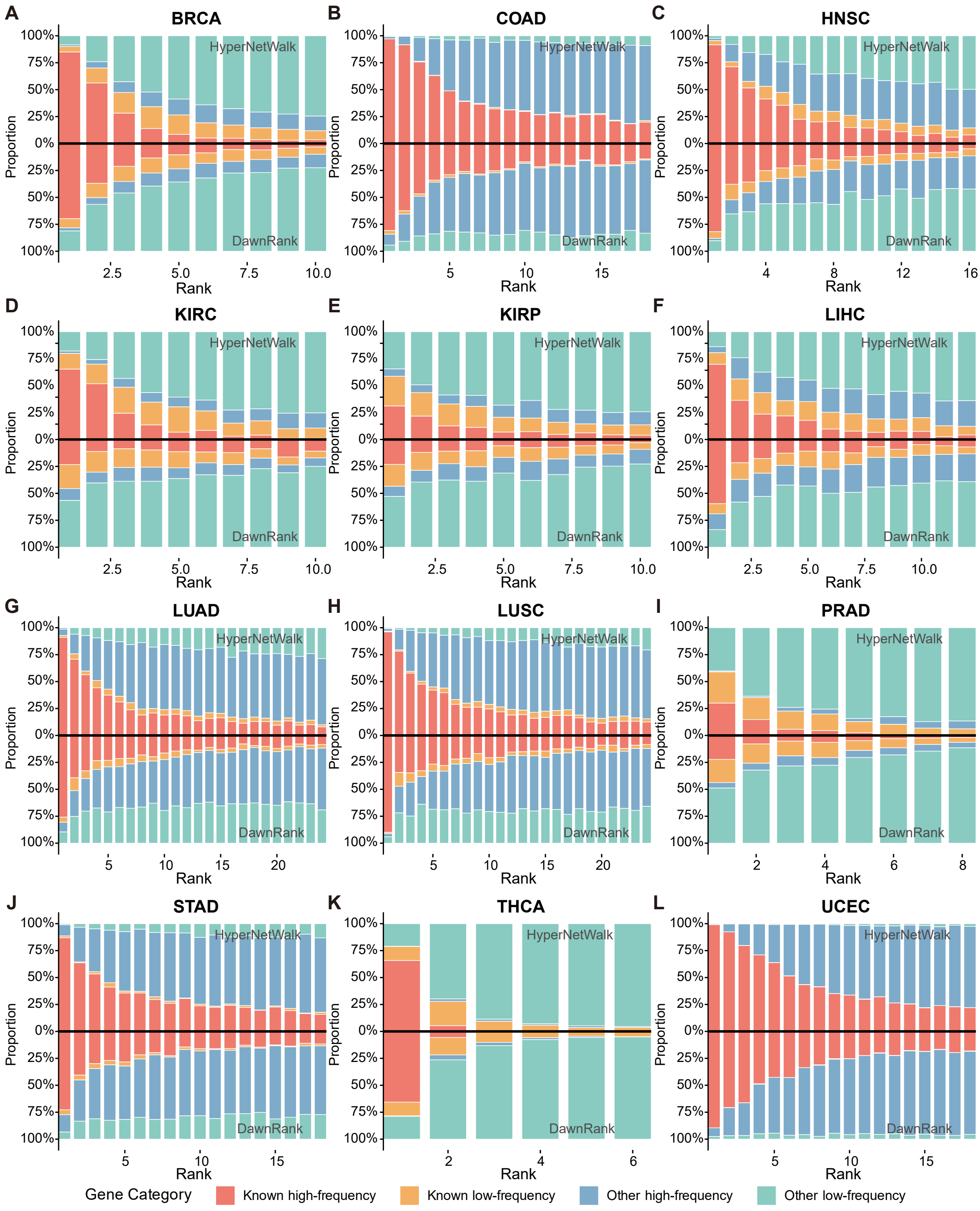

**Figure S18. Rank-wise composition of personalized predictions across 12 cancer types.** Mirrored stacked bar plots of the personalized predictions in 12 cancer types (**A–L**). For each rank position, predicted genes are grouped into four categories: known high-frequency drivers, known low-frequency drivers, other high-frequency mutated genes, and other low-frequency mutated genes. The upper and lower parts of each mirrored plot represent HyperNetWalk and DawnRank, respectively.

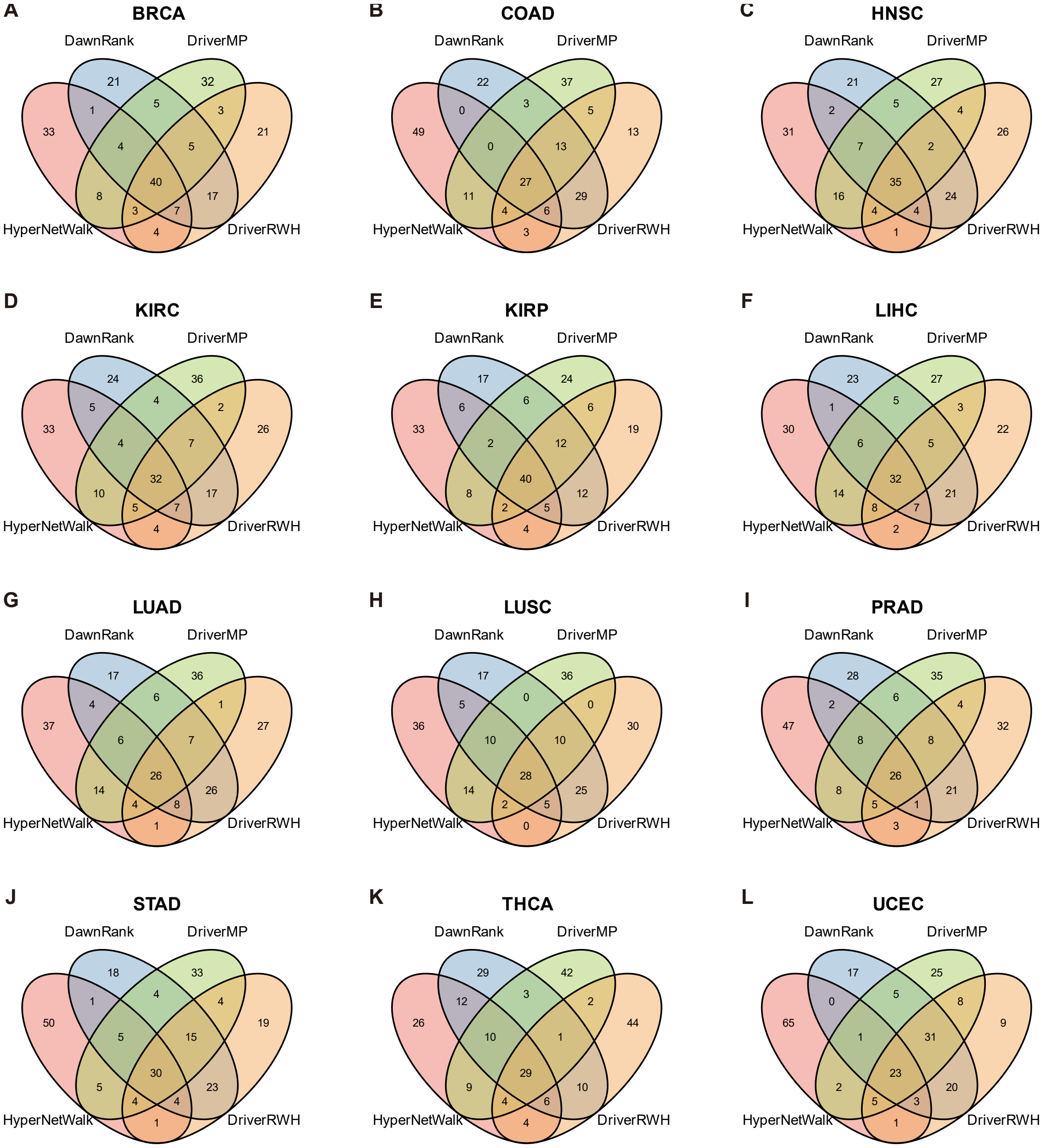

**Figure S19. Overlap among cohort-level top 100 predictions from HyperNetWalk and baseline methods across 12 cancer types.** Venn diagrams for 12 cancer types (**A–L**) showing the unique and shared cohort-level top 100 predictions of HyperNetWalk, DawnRank, DriverMP, and DriverRWH.

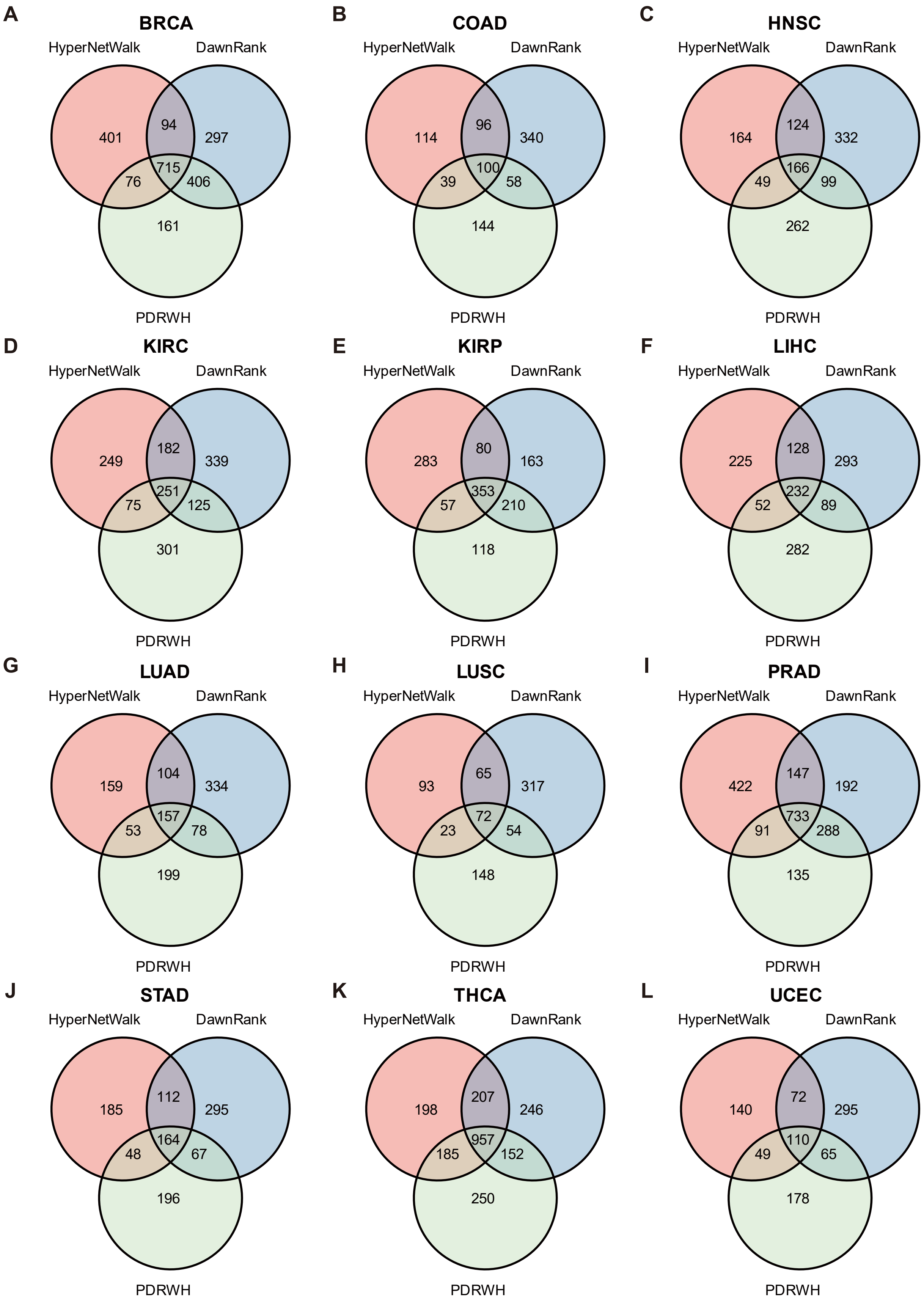

**Figure S20. Overlap among personalized predictions across all cancer types.** Venn diagrams for 12 cancer types (**A–L**) showing the unique and shared genes in the union of top 5 personalized predictions from HyperNetWalk, DawnRank, and PDRWH across all samples.

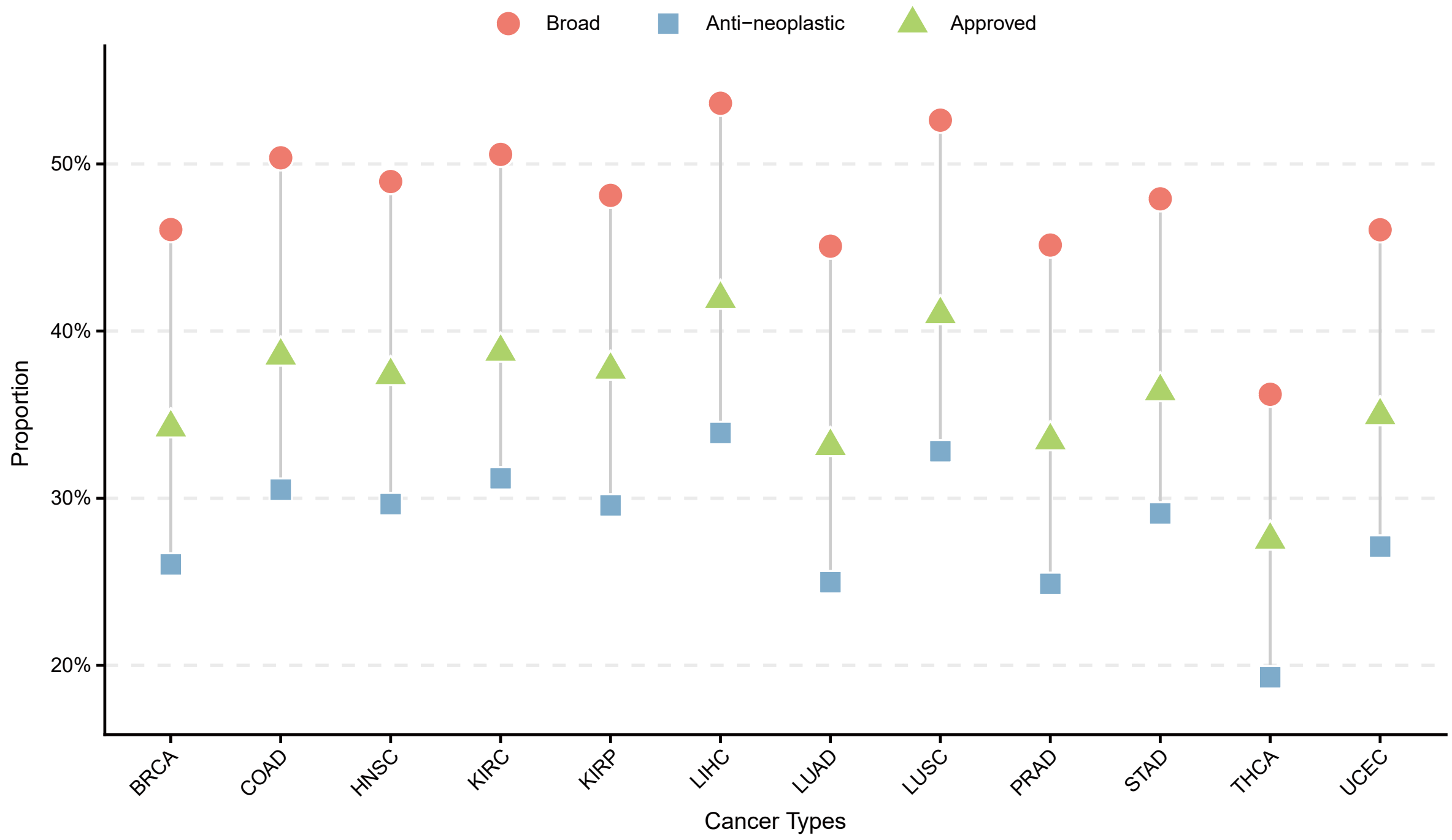

**Figure S21. DGIdb annotations of HyperNetWalk personalized predictions.** Cleveland dot plot showing the proportion of genes in the union of personalized top $N_c$ predictions from HyperNetWalk that are annotated in DGIdb across 12 cancer types. Three DGIdb-supported categories are evaluated: genes with any drug–gene interaction annotation (broad), genes associated with anti-neoplastic drugs, and genes with approved drug interactions.

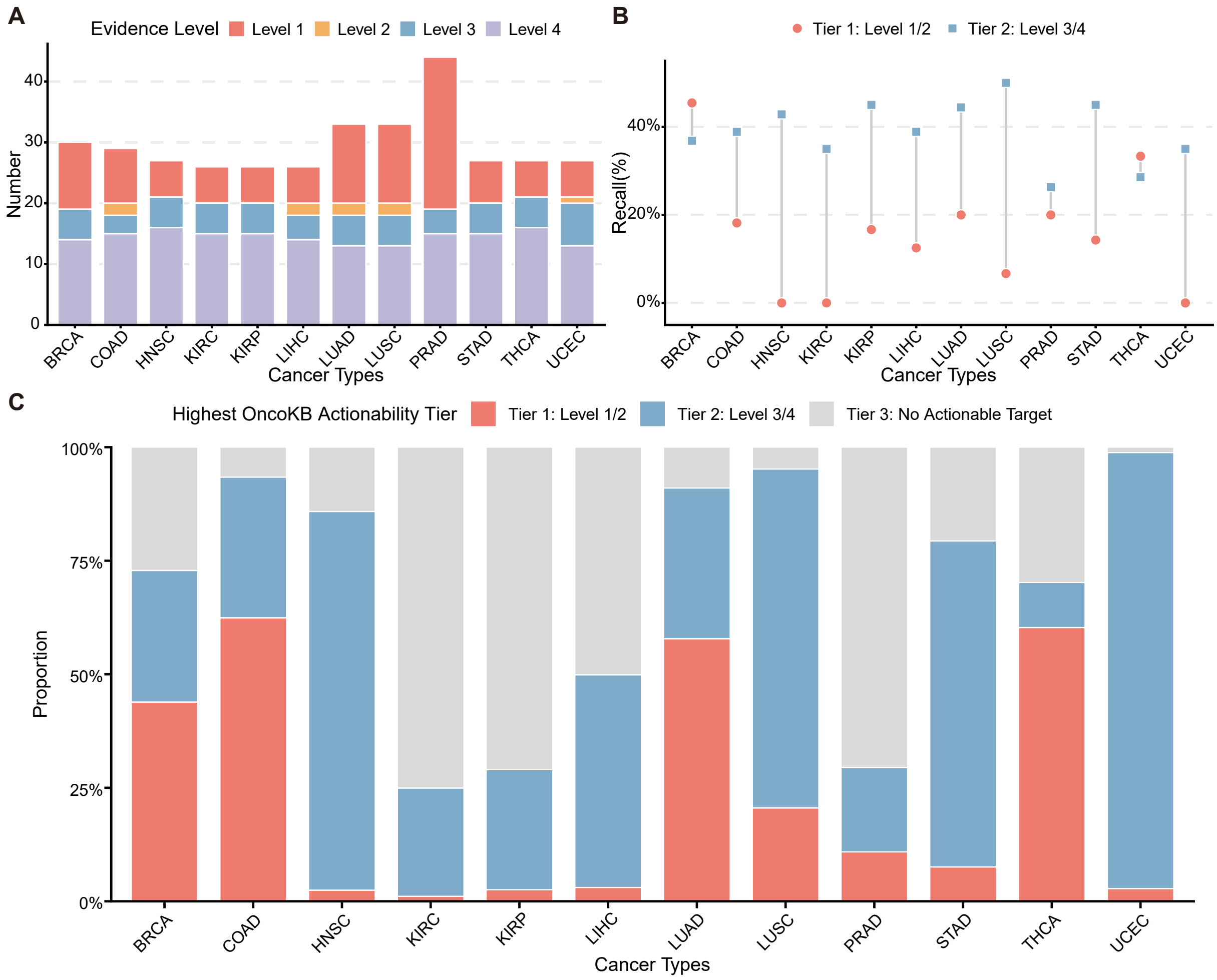

**Figure S22. OncoKB-based actionability annotations of HyperNetWalk predictions. A.** Stacked bar plot showing the number of genes annotated to each OncoKB evidence level across 12 cancer types. For genes with multiple OncoKB annotations, the highest evidence level within each cancer type is retained, and only evidence Levels 1–4 are considered. **B.** Cleveland dot plot showing the proportion of cohort-level top 100 predictions from HyperNetWalk annotated in OncoKB across 12 cancer types. Proportions are shown separately by the retained highest OncoKB evidence level. **C.** Stacked bar plot showing the distribution of the highest OncoKB evidence level detected among the personalized top $N_c$ predictions for each sample. For each cancer type, bars represent the proportion of samples assigned to each highest evidence level category. Samples without any OncoKB-supported gene among the top $N_c$ predictions are assigned to the Tier 3 category.

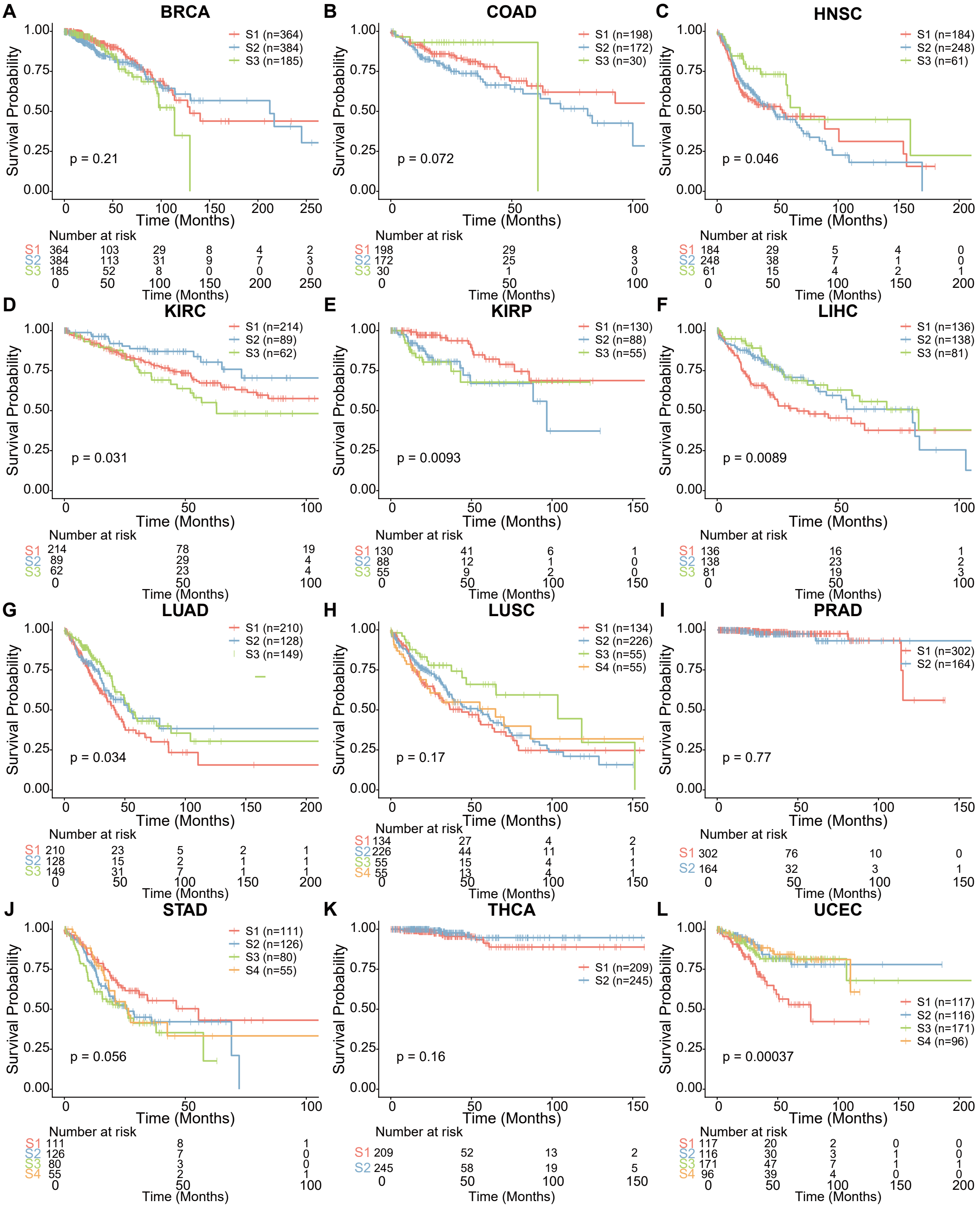

**Figure S23. Kaplan–Meier survival curves for 12 cancer types stratified by TF-layer-derived subtypes.** For each cancer type, patient subtypes were inferred by consensus clustering on Stage I TF-layer probability vectors, with the number of clusters determined by the delta area criterion without using survival information. Log-rank p-values are shown for all 12 cancer types (**A–L**). Risk tables are shown below the curves, and log-rank test p-values are reported in each panel.

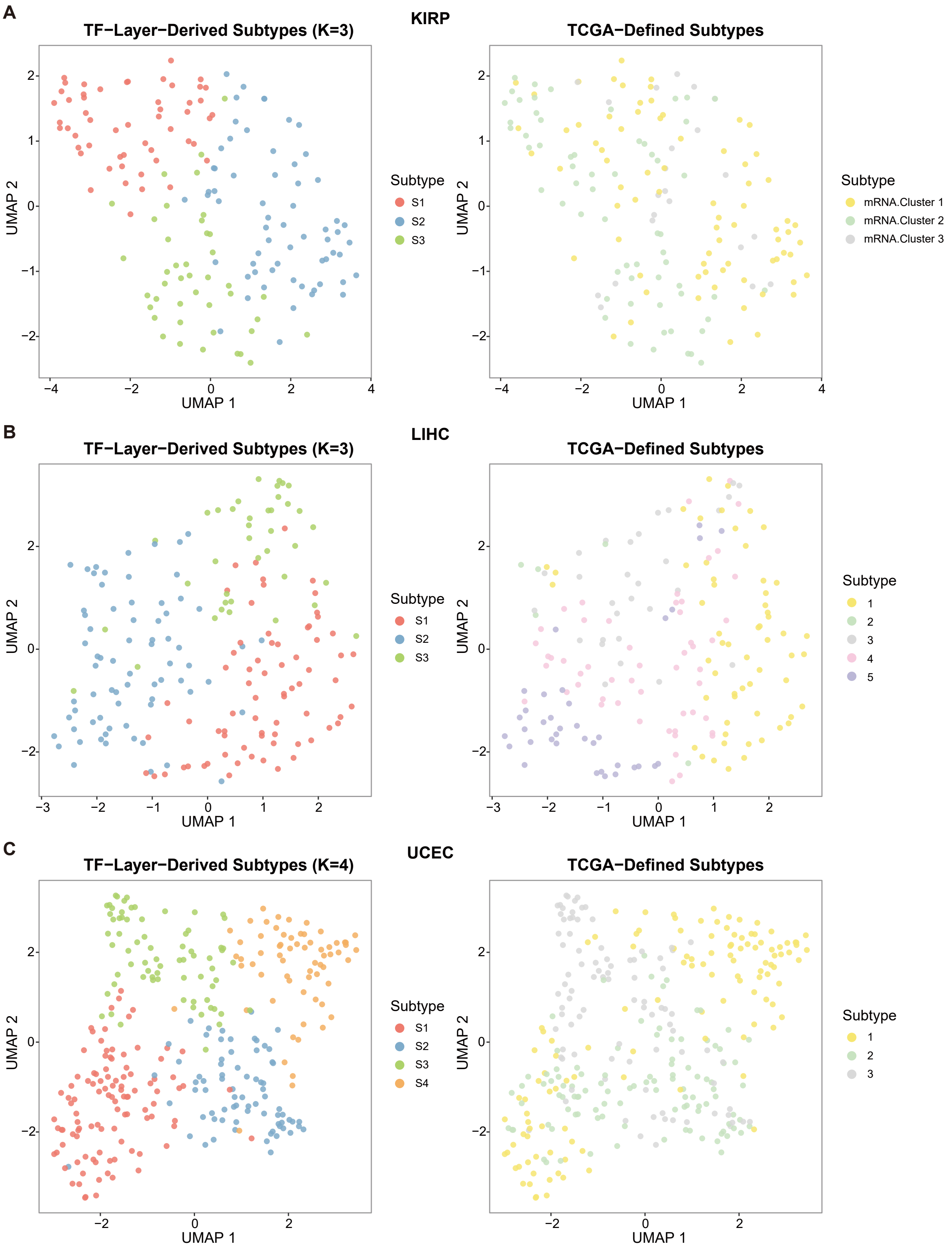

**Figure S24. Comparison between TF-layer-derived subtypes and TCGA molecular subtype annotations.** UMAP projections of TF-layer probability vectors are shown for KIRP (**A**), LIHC (**B**), and UCEC (**C**). For each cancer type, each point represents one patient sample and is colored either by TF-layer-derived subtypes (left) or by available TCGA molecular subtype annotations (right). TF-layer-derived subtype labels were obtained from HyperNetWalk-analyzed samples, whereas the visual comparison with TCGA molecular subtypes was restricted to patients with available subtype annotations; therefore, the left and right panels for each cancer type show the same subset of overlapping patients.

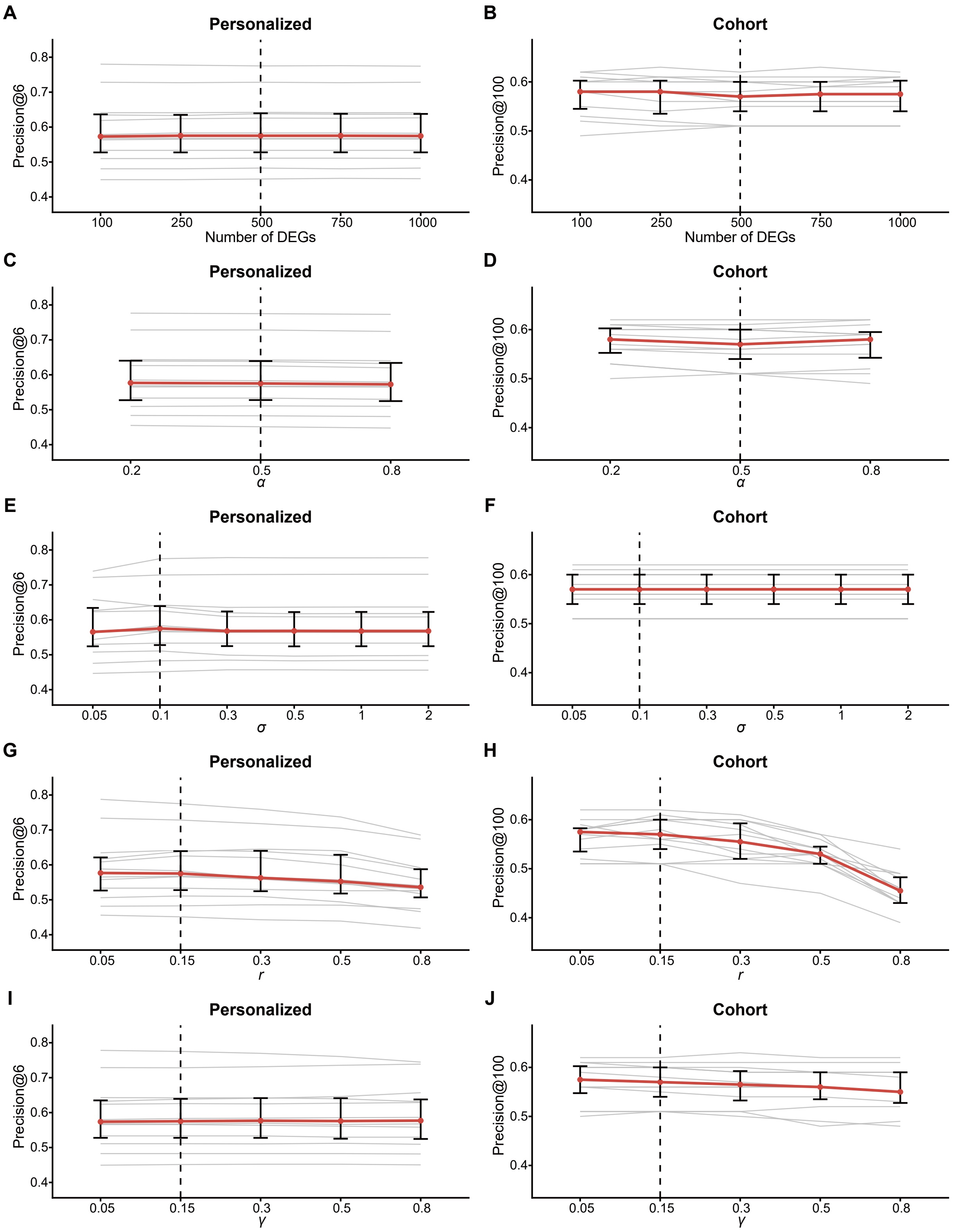

**Figure S25. Parameter sensitivity analysis of HyperNetWalk.** One-factor-at-a-time sensitivity analysis was performed for five key parameters: the number of DEGs used for Stage I initialization (**A–B**), the structural parameter $\alpha$ controlling the probability of transitioning from Layer 2 to Layer 1 during reverse propagation (**C–D**), the similarity bandwidth parameter $\sigma$ for personalized hyperedge construction (**E–F**), the Stage I restart probability $r$ (**G–H**), and the Stage II hypergraph diffusion restart probability $\gamma$ (**I–J**). For each analysis, one parameter is varied while all other parameters are fixed at their default values. Performance is evaluated across 12 TCGA cancer types using personalized Precision@$K$ ($K$ = 6) and cohort-level Precision@$K$ ($K$ = 100). Parameter values are shown as ordered discrete settings tested in the sensitivity analysis. Thin gray lines represent individual cancer types, the red line represents the median across cancer types, and error bars indicate the interquartile range. Vertical dashed lines indicate the default parameter values used in this study. Because $\sigma$ only affects personalized hyperedge construction, cohort-level performance remains unchanged across different $\sigma$ values.

# Supplementary Tables

**Table S1. Summary of TCGA datasets analyzed in this study.**

| Cancer type | Mut. samples | Mut. genes | Expr. tumor | Expr. normal | Expr. genes |
|---|---|---|---|---|---|
| BRCA | 985 | 16,662 | 1,081 | 99 | 59,427 |
| COAD | 440 | 18,654 | 455 | 41 | 59,427 |
| HNSC | 510 | 16,509 | 515 | 44 | 59,427 |
| KIRC | 379 | 10,929 | 529 | 72 | 59,427 |
| KIRP | 279 | 9,920 | 289 | 32 | 59,427 |
| LIHC | 371 | 14,340 | 369 | 50 | 59,427 |
| LUAD | 575 | 17,705 | 513 | 58 | 59,427 |
| LUSC | 491 | 17,623 | 496 | 51 | 59,427 |
| PRAD | 495 | 11,075 | 483 | 51 | 59,427 |
| STAD | 431 | 18,003 | 410 | 36 | 59,427 |
| THCA | 496 | 4,163 | 500 | 57 | 59,427 |
| UCEC | 515 | 18,993 | 539 | 35 | 59,427 |

**Note:** Mut., mutation; Expr., expression. Mutation samples indicate tumor samples with available somatic mutation profiles. Expression tumor and normal indicate the numbers of tumor and normal samples with available gene expression profiles.

**Table S2. Evidence-level summary of HyperNetWalk-specific candidate genes prioritized at the cohort level across 12 cancer types.**

| Cancer type | Total candidates | Level A | Level B | Level C | Level D |
|---|---|---|---|---|---|
| BRCA | 33 | 1 | 24 | 7 | 1 |
| COAD | 49 | 2 | 27 | 19 | 1 |
| HNSC | 31 | 0 | 18 | 13 | 0 |
| KIRC | 33 | 0 | 24 | 8 | 1 |
| KIRP | 33 | 0 | 15 | 18 | 0 |
| LIHC | 30 | 2 | 24 | 4 | 0 |
| LUAD | 37 | 0 | 29 | 7 | 1 |
| LUSC | 36 | 0 | 21 | 15 | 0 |
| PRAD | 47 | 0 | 30 | 16 | 1 |
| STAD | 50 | 0 | 33 | 16 | 1 |
| THCA | 26 | 0 | 12 | 14 | 0 |
| UCEC | 65 | 3 | 30 | 29 | 3 |

**Note:** HyperNetWalk-specific candidate genes at the cohort level are defined as genes present in the HyperNetWalk top 100 predictions but absent from the corresponding baseline prediction sets for each cancer type. Candidates are categorized into four mutually exclusive evidence levels based on cancer-related databases (CGC, NCG, and CancerMine):

**Level A (Matched high-confidence):** Genes in CGC Tier 1 annotated for the matched cancer type.

**Level B (Moderately supported):** Genes in CGC Tier 1 for other cancer types, or genes annotated for the matched cancer type in CGC Tier 2, NCG, or CancerMine.

**Level C (Indirect/Weak support):** Genes present in CGC Tier 2, NCG, or CancerMine but annotated for other cancer types.

**Level D (No external database support):** Genes without support from any of these resources, retained as exploratory candidates for further evaluation.

**Table S3. Evidence-level summary of HyperNetWalk-specific candidate genes prioritized at the personalized level across 12 cancer types.**

| Cancer type | Total candidates | Level A | Level B | Level C | Level D |
|---|---|---|---|---|---|
| BRCA | 401 | 6 | 167 | 151 | 77 |
| COAD | 114 | 5 | 51 | 41 | 17 |
| HNSC | 164 | 2 | 62 | 84 | 16 |
| KIRC | 249 | 1 | 74 | 139 | 35 |
| KIRP | 283 | 0 | 63 | 185 | 35 |
| LIHC | 225 | 4 | 122 | 80 | 19 |
| LUAD | 159 | 2 | 85 | 57 | 15 |
| LUSC | 93 | 0 | 37 | 51 | 5 |
| PRAD | 422 | 5 | 99 | 225 | 93 |
| STAD | 185 | 2 | 77 | 91 | 15 |
| THCA | 198 | 1 | 15 | 97 | 85 |
| UCEC | 140 | 3 | 48 | 80 | 9 |

**Note:** HyperNetWalk-specific candidate genes at the personalized level are defined as genes present in the top 5 predictions for at least one sample of a cancer type by HyperNetWalk, but absent from the corresponding baseline prediction sets for all samples of that cancer type. Candidates are categorized into four mutually exclusive evidence levels based on cancer-related databases (CGC, NCG, and CancerMine):

**Level A (Matched high-confidence):** Genes in CGC Tier 1 annotated for the matched cancer type.

**Level B (Moderately supported):** Genes in CGC Tier 1 for other cancer types, or genes annotated for the matched cancer type in CGC Tier 2, NCG, or CancerMine.

**Level C (Indirect/Weak support):** Genes present in CGC Tier 2, NCG, or CancerMine but annotated for other cancer types.

**Level D (No external database support):** Genes without support from any of these resources, retained as exploratory candidates for further evaluation.

**Table S4. HyperNetWalk-specific candidate genes prioritized at the cohort level in BRCA and their evidence levels.**

| Gene | GRN out-degree | Mut. freq. | HNW rank | Best baseline rank | Evidence level |
|---|---|---|---|---|---|
| TBX3 | 6 | 3.64% | 93 | 146 | A |
| MYH11 | 9,177 | 1.87% | 29 | 130 | B |
| MAZ | 17,162 | 0.73% | 30 | 3,219 | B |
| KDM5B | 7,631 | 1.56% | 31 | 208 | B |
| ETS1 | 13,272 | 0.83% | 35 | 189 | B |
| BRD4 | 14,508 | 0.42% | 37 | 402 | B |
| BCOR | 8,466 | 1.46% | 47 | 518 | B |
| SP1 | 8,414 | 0.62% | 52 | 787 | B |
| ERG | 12,812 | 0.83% | 53 | 3,895 | B |
| STAG1 | 7,419 | 1.66% | 55 | 665 | B |
| E2F1 | 11,361 | 0.62% | 58 | 171 | B |
| NCOA3 | 89 | 1.25% | 66 | 206 | B |
| KDM4C | 6,686 | 1.35% | 71 | 1,602 | B |
| SETDB1 | 981 | 1.66% | 72 | 133 | B |
| SMAD4 | 233 | 0.94% | 74 | 132 | B |
| ZEB1 | 4,034 | 1.14% | 75 | 331 | B |
| ATF2 | 3,589 | 0.73% | 78 | 410 | B |
| TCF12 | 4,801 | 0.94% | 81 | 829 | B |
| PHF8 | 5,441 | 1.35% | 82 | 669 | B |
| STAT5A | 1,323 | 1.14% | 88 | 119 | B |
| JUN | 12,891 | 0.42% | 90 | 181 | B |
| EBF1 | 8,362 | 0.62% | 91 | 1,272 | B |
| JAK2 | 0 | 1.04% | 92 | 138 | B |
| TFAP2A | 7,044 | 0.83% | 94 | 123 | B |
| NCOR2 | 454 | 1.04% | 96 | 184 | B |
| SIN3A | 3,100 | 0.94% | 41 | 244 | C |
| SPI1 | 13,656 | 0.42% | 48 | 946 | C |
| HCFC1 | 3,316 | 1.98% | 50 | 129 | C |
| SMC1A | 6,434 | 1.25% | 57 | 243 | C |
| POLR2A | 5,825 | 0.83% | 76 | 233 | C |
| SMARCC1 | 1,196 | 1.46% | 77 | 216 | C |
| STAT4 | 1,065 | 1.66% | 97 | 213 | C |
| ZNF263 | 11,945 | 0.62% | 64 | 7,127 | D |

**Note:** Genes listed here are HyperNetWalk-specific candidate genes prioritized at the cohort level in BRCA. GRN out-degree denotes the number of downstream target genes connected to the corresponding gene in the gene regulatory network. Mutation frequency indicates the percentage of samples within the cohort harboring at least one non-silent mutation in the gene. HNW rank denotes the rank assigned by HyperNetWalk in the cohort-level prediction, with smaller values indicating higher priority. Best baseline rank denotes the best (smallest) rank assigned to the same gene by any competing method (DawnRank, DriverMP, and DriverRWH), with smaller values indicating higher priority. Evidence levels (A–D) follow the mutually exclusive categorization based on CGC, NCG, and CancerMine.

**Table S5. HyperNetWalk-specific candidate genes prioritized at the cohort level in LUAD and their evidence levels.**

| Gene | GRN out-degree | Mut. freq. | HNW rank | Best baseline rank | Evidence level |
|---|---|---|---|---|---|
| FLI1 | 14,861 | 3.97% | 13 | 3,956 | B |
| MYH11 | 9,177 | 7.22% | 16 | 108 | B |
| BCOR | 8,466 | 4.51% | 27 | 644 | B |
| BRD4 | 14,508 | 1.26% | 31 | 273 | B |
| CTCF | 16,900 | 1.26% | 35 | 342 | B |
| HCFC1 | 3,316 | 4.69% | 40 | 225 | B |
| HDAC2 | 12,629 | 1.62% | 42 | 128 | B |
| GATA3 | 4,165 | 2.89% | 43 | 226 | B |
| GATA1 | 11,633 | 1.99% | 47 | 516 | B |
| SIN3A | 3,100 | 2.17% | 54 | 196 | B |
| EBF1 | 8,362 | 2.71% | 56 | 844 | B |
| NCOA1 | 97 | 3.97% | 61 | 136 | B |
| AR | 3,773 | 2.89% | 64 | 213 | B |
| MAX | 13,925 | 1.81% | 67 | 875 | B |
| ZEB2 | 19 | 8.30% | 73 | 320 | B |
| NCOA3 | 89 | 3.07% | 74 | 303 | B |
| KLF4 | 11,372 | 1.44% | 75 | 284 | B |
| PRKCB | 0 | 5.78% | 77 | 123 | B |
| DNMT3A | 23 | 3.97% | 78 | 197 | B |
| KDM5B | 7,631 | 2.71% | 82 | 659 | B |
| GSK3B | 0 | 1.44% | 84 | 220 | B |
| RUNX1T1 | 3,104 | 6.50% | 86 | 339 | B |
| EZH2 | 901 | 1.62% | 87 | 142 | B |
| NCOA2 | 79 | 3.43% | 89 | 251 | B |
| TFAP2A | 7,044 | 1.62% | 90 | 2,268 | B |
| SP1 | 8,414 | 1.26% | 91 | 216 | B |
| BRD2 | 11,920 | 1.26% | 93 | 1,181 | B |
| PGR | 3,037 | 3.25% | 96 | 217 | B |
| JAK2 | 0 | 2.89% | 98 | 110 | B |
| HNF4A | 3,928 | 4.15% | 48 | 134 | C |
| POLR2A | 5,825 | 2.89% | 51 | 174 | C |
| CHD8 | 7 | 5.23% | 59 | 189 | C |
| SMC3 | 5,718 | 3.07% | 60 | 345 | C |
| GABPA | 14,018 | 1.26% | 62 | 2,016 | C |
| BPTF | 439 | 5.60% | 66 | 367 | C |
| KDM4C | 6,686 | 3.61% | 92 | 2,443 | C |
| TAF3 | 4,950 | 4.15% | 81 | 472 | D |

**Note:** Genes listed here are HyperNetWalk-specific candidate genes prioritized at the cohort level in LUAD. GRN out-degree denotes the number of downstream target genes connected to the corresponding gene in the gene regulatory network. Mutation frequency indicates the percentage of samples within the cohort harboring at least one non-silent mutation in the gene. HNW rank denotes the rank assigned by HyperNetWalk in the cohort-level prediction, with smaller values indicating higher priority. Best baseline rank denotes the best (smallest) rank assigned to the same gene by any competing method (DawnRank, DriverMP, and DriverRWH), with smaller values indicating higher priority. Evidence levels (A–D) follow the mutually exclusive categorization based on CGC, NCG, and CancerMine.